\documentclass[prodmode]{acmsmall}

\usepackage[final]{fixme}

\usepackage{ifthen}
\newboolean{full}
\setboolean{full}{true}
\mathcode`;="003B  
\mathcode`?="003F  
\mathcode`|="026A  
\mathcode`<="4268  
\mathcode`>="5269  

\mathchardef\ls="213C    
\mathchardef\gr="213E    
\mathchardef\uparrow="0222  
\mathchardef\downarrow="0223  

\usepackage{style_acm}

\usepackage{color}

\begin{document}
%
%
\FXRegisterAuthor{mb}{amb}{Mar}
\FXRegisterAuthor{sm}{asm}{Ste}
\FXRegisterAuthor{as}{aas}{Alex}

%
%

\markboth{Bonsangue, Milius, Silva}{Sound and complete axiomatisations of coalgebraic language equivalence}
\title{Sound and complete axiomatisations \\ of coalgebraic language equivalence${}^\ast$}

\author{
  MARCELLO M.~BONSANGUE
  \affil{Leiden Institute of Advanced Computer Science${}^1$}
  STEFAN MILIUS
  \affil{Technische Universit\"at Braunschweig}
  ALEXANDRA SILVA
  \affil{Radboud University Nijmegen${}^2$}
}

\begin{abstract}
  Coalgebras provide a uniform framework to study dynamical systems,
  including several types of automata. In this paper, we make use of
  the coalgebraic view on systems to investigate, in a uniform way,
  under which conditions calculi that are sound and complete with respect to
  behavioral equivalence can be extended to a coarser coalgebraic
  language equivalence,
  which arises from a generalised powerset construction that
  determinises coalgebras. We show that
  soundness and completeness are established by proving that expressions
  modulo axioms of a calculus form the rational fixpoint of the given
  type functor. Our main result is that the rational
  fixpoint of the functor $FT$, where $T$ is a monad describing the
  branching of the systems (e.g.~non-determinism, weights, probability
  etc.), has as a quotient the rational fixpoint of the
  ``determinised'' type functor $\bar F$, a lifting of $F$ to
  the category of $T$-algebras.
  We apply our framework to the concrete example of weighted automata, for
  which we present a new 
  sound and complete calculus for weighted
  language equivalence. As a special case, we obtain non-deterministic
  automata, where we recover Rabinovich's sound and complete calculus for
  language equivalence.
  \takeout{
  Coalgebras provide a uniform framework to study dynamical systems,
  including several types of automata. In this paper, we make use of
  the coalgebraic view on systems to investigate, in a uniform way,
  under which conditions calculi that are sound and complete with respect to
  behavioral equivalence can be extended to a coarser coalgebraic
  language equivalence,
  which arises from a generalised powerset construction. We illustrate
  the framework with two examples: non-deterministic automata, for
  which we recover Rabinovich's sound and complete calculus for
  language equivalence, and weighted automata, for
  which we present the first sound and complete calculus for weighted
  language equivalence.
  }
\end{abstract}

\category{X.n.n}{ACM-class}{ACM-class}

\terms{Theory}


\acmformat{Bonsangue, M.~M., Milius, S., and Silva, A. 2011. Sound and complete axiomatisations of coalgebraic language equivalence.}

\begin{bottomstuff}
  Authors' addresses: M.~M.~Bonsangue, Leiden Institute of Advanced Computer Science (LIACS),Leiden
  University, The Netherlands;
  S.~Milius, Institut f\"ur Theoretische Informatik, Technische
  Universit\"at Braunschweig, Germany;
  A.~Silva, Intelligent systems section, Radboud University Nijmegen, Nijmegen,
  The Netherlands.\\
${}^1$Also affiliated to Centrum Wiskunde \&
Informatica, Amsterdam, The Netherlands.\\
${}^2$Also affiliated to Centrum Wiskunde \&
Informatica, Amsterdam, The Netherlands  \& HASLab / INESC TEC,
Universidade do Minho,
Braga, Portugal.\\
${}^\ast$This version corrects a few small mistakes and slightly improves 
the published journal paper.
\end{bottomstuff}


\maketitle

%
%
\def\o{\cdot}

\section{Introduction}

State-based structures of various kinds are used to model the behavior
of phenomena in many different
fields of computer science, mathematics, and more recently of biology
and physics. So the theories of state based systems, their
specification, semantics and logical descriptions are topics
at the heart of theoretical computer science.

A major step forward, in the last years was the realisation that a vast majority of state-based systems can
be uniformly described as instances of the general notion of
coalgebra.
\takeout{ 
This has led to a rapid
development of the new field of {\em Universal Coalgebra} in which not only the (co)algebraic
structure theory but also appropriate calculi and logics have been developed on
an abstract level. This yields methods and mathematical tools that are
generic and can be applied uniformly to many different types of
systems. Typical examples of coalgebras are (non-)deterministic and weighted automata,
probabilistic systems, labelled transition systems etc.}
For an endofunctor $F$ on a category $\A$, an
$F$-coalgebra is a pair $(X, f)$, where $X$ is an object of $\A$
representing the state space and $f \colon X \to FX$ is an arrow of
$\A$ defining the observations and transitions of the states. The
strength of coalgebraic modelling lies in the fact that the
type $F$ of the system determines a standard notion of equivalence
called $F$-behavioral equivalence and a canonical domain of
behavior, the so-called \emph{final} coalgebra, into which any $F$-coalgebra
is mapped by a unique homomorphism that identifies all equivalent
states.

The coalgebraic perspective on state-based systems has recently been
proved very relevant by the development of a number of generalised
calculi of \emph{regular expressions} admitting Kleene theorems and
equipped with sound and complete equational logics, which are
expressive enough to characterise the behavioral equivalence of
\emph{all} finite state coalgebras uniformly for an inductively
defined class of type functors on sets. This includes Mealy automata~\cite{brs_fossacs08}, automata
whose type is given by Kripke polynomial functors~\cite{brs_lmcs},
automata for the so-called quantitative functors~\cite{bbrs_ic}
(e.\,g., weighted automata, Segala systems and Pn\"ueli-Zuck systems)
and closed stream circuits~\cite{m_linexp}. This line of work generalises
Kleene's classical theorem~\cite{kleene} as well as work on sound and
complete expression calculi (e.\,g.~\cite{salomaa}, see also~\cite{Kozen}).
A key result for the generalization is that soundness and completeness
is equivalent to proving that generalized regular expressions for
$F$-coalgebras modulo the axioms and rules of the calculus form a final
locally finite coalgebra $\rho F$; equivalently, this is the initial
iterative $F$-algebra of~\cite{amv_atwork}.

The above calculi axiomatize $F$-behavioral equivalence, which, for a functor
$F$ preserving weak pullbacks, is equivalent to bisimilarity. However,
bisimilarity is a very fine grained equivalence, and one is often
interested in a coarser trace or language equivalence. In this paper
we will present a general methodology to extend sound and complete
calculi with respect to behavioral equivalence to sound and complete
calculi for a new \emph{coalgebraic language equivalence}.

As one approach to this equivalence it has recently been
shown~\cite{bbrs_fsttcs} that the classical powerset construction,
which transforms a non-deterministic automaton into a deterministic
one, providing language semantics to the former, can be extended to a
large class of systems, coalgebras for a given type functor, which
includes probabilistic and weighted automata.  The aforementioned
paper models systems as the composite of a functor type $F$ and a
monad $T$, which encodes the non-determinism or probabilities that one
wants to \emph{determinise}.  The determinised coalgebra is actually a
coalgebra in the category of Eilenberg-Moore algebras for the monad
$T$. We will call the equivalence obtained by this construction, that
is, the $F$-behavioral equivalence in the category of $T$-algebras,
coalgebraic language equivalence. For example, the construction
above applied to non-deterministic automata yields a deterministic
automaton in the category of join-semilattices.  Coalgebraic language
equivalence corresponds to ordinary language equivalence, while
$FT$-behavioral equivalence is just ordinary bisimilarity. More
interestingly, the construction also applies to weighted automata, in
which case the determinisation is an automaton in the category of
vector spaces (assuming the weights are elements of a
field). Coalgebraic language equivalence corresponds to weighted
language equivalence, while $FT$-behavioral equivalence is just
weighted bisimilarity of~\cite{buchholz}.

The contributions of our paper are twofold and we explain them in
the following subsections.

\subsection{Finitary Coinduction}

Firstly, we will develop a mathematical theory of finitary coinduction
w.r.t.~coalgebraic language equivalence. Our theory builds
on~\cite{amv_atwork} which provides the foundations for a theory of finitary coinduction
w.r.t.~ordinary behavioral equivalence, and we extend here the first
steps for the new theory taken (in a very special case) by the second
author in~\cite{m_linexp}.

We start by systematically studying coalgebras for endofunctors $F$
having a lifting $\bar F$ to the category of (Eilenberg-Moore) algebras
for a monad $T$. To begin with, in Section~\ref{sec:algcoalg}, we study
the relationship between the final coalgebras $\nu F$ and $\nu(FT)$ as well
as the between the \emph{rational fixpoints} $\rho(FT)$ for $FT$ and $\rho \bar F$
for the lifting of $F$ to $T$-algebras.

Intuitively, one should think of $\nu(FT)$ and $\nu F$ as the
behaviors of \emph{all} systems modulo bisimilarity and language
equivalence, respectively. Furthermore, $\rho(FT)$ and $\rho\bar F$ are the
behaviors of all \emph{finite state} systems modulo bisimilarity and
language equivalence, respectively.

We prove that for every finitary endofunctor $H$, $\rho H$ is the final locally
finitely presentable coalgebra. It is also a fixpoint of $H$, and
the inverse of its structure map yields the initial iterative
algebra for $H$ in the sense of~\cite{amv_atwork}. The latter characterisation gives a
precise connection of the work in~\cite{brs_lmcs,bbrs_ic,m_linexp}
with the classical work on iterative algebras by Nelson~\shortcite{nelson}
  and Elgot's iterative theories~\cite{elgot} (see also~\cite{be}).
In our setting we use a well-known coalgebraic construction of $\rho H$: it is the colimit
of all finite $H$-coalgebras. Here we consider $H = FT$ as above, and we prove
(see Corollary~\ref{cor:colim}) that the rational fixpoint
$\rho\bar F$ of the lifting $\bar F$ of $F$ to $T$-algebras is also a colimit of
finite $FT$-coalgebras but with a different set of connecting morphisms in the
corresponding diagram: in lieu of homomorphisms between coalgebras $X \to FTX$ one uses $\bar F$-coalgebra
homomorphisms on the corresponding \emph{determinisations} $TX \to
FTX$. As our main result we establish the relationships between the four mentioned
fixpoints of $F$ and $FT$ as summarised by the following commutative square
among $T$-algebras (see Theorem~\ref{thm:ratquot}):
\begin{equation}
\label{eq:square}
\vcenter{
\xymatrix{
  \rho(FT)
  \ar@{ >->}[r]
  \ar@{->>}[d]
  &
  \nu(FT)
  \ar@{->>}[d]
  \\
  \rho\bar F
  \ar@{ >->}[r]
  &
  \nu F
  }
  }
\end{equation}
This diagram shows that the rational fixpoints of $FT$ and $\bar F$
are, as expected, subcoalgebras of the respective final coalgebras
(horizontal maps). This makes precise the above intuition that the
rational fixpoints are the behaviors of finite state systems; in fact,
behavioral equivalent states of finite state systems are identified by
the unique coalgebra homomorphisms into $\rho(FT)$ and $\rho\bar F$,
respectively. Furthermore, the final coalgebra for $F$ is a quotient
of the final coalgebra for $FT$ and this quotient restricts to the
respective rational fixpoints (vertical maps). This means that
$FT$-behavioral equivalence implies coalgebraic language equivalence
(i.\,e.~$F$-behavioral equivalence), our abstract version of the
well-known fact that bisimilarity implies language (or trace)
equivalence. All these results hold whenever (a) finitely generated
$T$-algebras are closed under kernel pairs, (b) $T$ is a finitary
monad and (c) $F$ is a finitary functor preserving weak pullbacks and
having a lifting to the category of $T$-algebras. Examples of algebras
satisfying the above condition (a) include join-semilattices, Abelian
groups, vector spaces, semimodules for Noetherian semirings, but
e.g. not groups.


\subsection{Expression Calculi for Coalgebraic Language Equivalence}

In Section~\ref{sec:general} we apply our results from
Section~\ref{sec:algcoalg} to obtain an abstract Kleene's theorem
(Theorem~\ref{thm:genkleene}) and soundness and completeness
results (Theorems~\ref{thm:gensound} and~\ref{thm:gencomp}),
and we show that in our setting it is possible to extend a given calculus for
behavioral equivalence to one for coalgebraic language
equivalence. Here we work without concrete syntax; the results collect
those parts of the soundness and completeness proofs that are generic,
so that for concrete calculi one saves work. We prove that showing
soundness and completeness of a concrete calculus is equivalent to
proving that the syntactic expressions modulo the axioms and rules of
the calculus form the rational fixpoint $\rho \bar F$ for
the lifting of the functor $F$ to $T$-algebras.

Then we apply our abstract results to the monad $V$ of free semimodules
for a Noetherian semiring $\S$ and the functor $FX = \S \times X^A$, where $A$ is a
finite input alphabet, and we show how to obtain a sound and complete
calculus for the language equivalence of weighted automata in Section~\ref{sec:lin},
and, as a special case, of non-deterministic automata in Section~\ref{sec:nondet}.

\newcommand\mtiny[1]{\text{\footnotesize #1}}
Weighted automata were introduced by
Sch\"utzenberger~\shortcite{schuetzenberger}, see also~\cite{dkv}.
For example, take the following two weighted automata over the alphabet $A = \{ a,b,c,d \}$
with weight over the semiring of integers (output values in the states are represented
with a double arrow, when omitted they are zero):
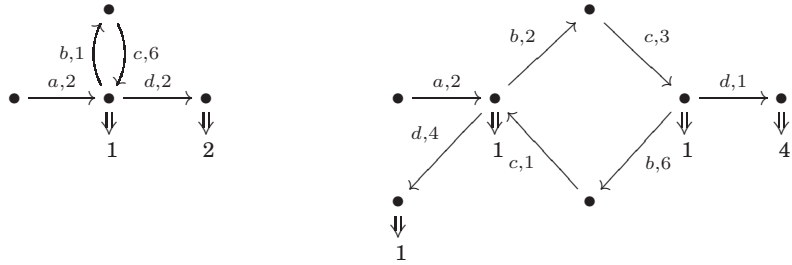
\[
\xymatrix@R=.3cm{ & \bullet  \ar@/^/[dd]^-{c,6}&\\\\
\bullet \ar[r]^-{a,2} & \bullet\ar@{=>}[d] \ar@/^/[uu]^-{b,1}\ar[r]^-{d,2}& \bullet \ar@{=>}[d] && \\
&\mtiny 1 &\mtiny 2
}
\xymatrix@R=.3cm{ & &\bullet\ar[ddr]^{c,3}& &\\\\
\bullet\ar[r]^-{a,2} & \bullet \ar[ddl]_{d,4}\ar[uur]^{b,2}\ar@{=>}[d]& & \bullet \ar[ddl]^{b,6}\ar[r]^{d,1}\ar@{=>}[d]&\bullet\ar@{=>}[d]\\
&\mtiny 1&& \mtiny 1& \mtiny 4\\
\bullet \ar@{=>}[d]& &\bullet\ar[uul]^{c,1}& &\\
\mtiny 1
}
\]
We will see in Section~\ref{sec:coalg} that they are coalgebras for the composition
of the functor functor $FX = \S \times X^A$ with the the monad $V$ of free semimodules
for the semiring $\S$ of integers. What is interesting is that the leftmost
states of these automata are not bisimilar, but they recognise the same weighted language.
Namely, the language that associates with each word $a(bc)^n$ the weight $2\cdot 6^n$,  with $a(bc)^nd$ the
weight $2\cdot 6^n \cdot 4$ and with any other word weight zero. We will provide an
algebraic proof of this equivalence in the sequel. To give upfront the reader a
feeling for how intricate it can get to reason about weighted language equivalence,
we show another example
of two weighted automata over
the singleton alphabet $A = \{ a\}$ but with weight over the field of real numbers.
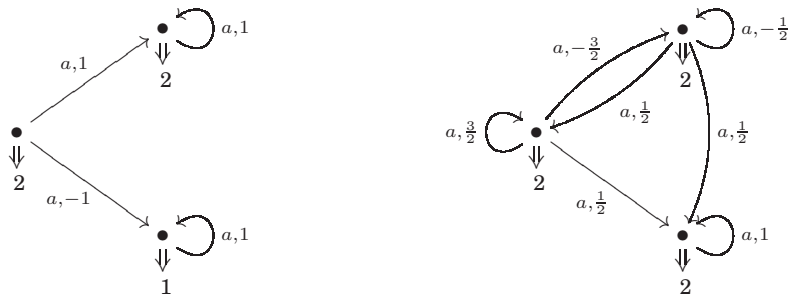
\[
\xymatrix@R=.3cm@C=1.5cm{ &\bullet \ar@{=>}[d]\ar@(dr,ur)_{a,1}&\\
&\mtiny 2\\
 \bullet  \ar[uur]^{a,1}\ar[ddr]_{a,-1}\ar@{=>}[d]& \\
\mtiny 2\\
  &\bullet \ar@{=>}[d] \ar@(dr,ur)_{a,1}&\\
  &\mtiny 1& &
}
\xymatrix@R=.3cm@C=1.5cm{  &\bullet \ar@{=>}[d]\ar@(dr,ur)_{a,-\frac {1} 2}\ar@/^/[ldd]^{a,\frac 1 2}\ar@/^{10pt}/[dddd]^{a,\frac 1 2}&\\
&\mtiny 2\\
 \bullet \ar@(dl,ul)^{a,\frac 3 2} \ar@/^/[uur]^{a,-\frac 3 2}\ar[ddr]_{a,\frac 1 2}\ar@{=>}[d]& \\
\mtiny 2\\
  &\bullet \ar@{=>}[d]\ar@(dr,ur)_{a,1}&\\
  &\mtiny 2&
}
\]
The leftmost states of these automata recognize the weighted language
that assigns to the empty word weight $2$ and to any word $a^n$ ($n\geq
1$) the weight $1$. In the left-hand automaton, it is still relatively
easy to convince oneself that this is the case, whereas for the
right-hand automaton one needs some more ingenuity. We shall see that
the algebraic proof is rather simple and instructive.

We start with the calculus for weighted bisimilarity obtained
from the generic expression calculus of~\cite{bbrs_ic}, and we extend
this by adding three canonical equational axioms. More detailed, the syntactic expressions
of our calculus are defined by the grammar\footnote{Note that we shall
  require a guardedness condition: every variable bound by $\mu$ must
  be within the scope of a $a.(r \dot (-))$ operator, see Section~\ref{sec:lin}.}
$$
E ::= x \mid \zero \mid E \oplus E \mid \ul{r} \mid a.(r \dot E) \mid \mu
x.E,
$$
where $x$ ranges over a finite set of syntactic variables, $a$ over a finite
(input) alphabet, and $r$ over a Noetherian semiring.
We show that each expression denotes a weighted language
(cf.~\refeq{eq:sem}): for example $\zero$ denotes the empty weighted
language, $\oplus$ is union of weighted languages, $\ul r$ denotes the
language that assigns to the empty word the weight $r$ (and $0$ to all
other words), $a.(r\dot E)$ denotes a language that assigns to a word
$aw$ the weight $r\cdot r_w$, where $r_w$ is the weight assigned to $w$
in the language denoted by the expression $E$, and $\mu$ is a fixpoint operator.
From our abstract Kleene theorem (Theorem~\ref{thm:genkleene}) we then obtain
that for every state of a finite weighted automaton there exists an expression
denoting the weighted language accepted by the given automaton starting from the
given state $s$.

For our new axiomatisation of weighted language equivalence we consider
first the following rules:
$$
\begin{array}{rcl@{\qquad}rcl@{\qquad}rcl}
  \multicolumn{3}{l}{E_1 \equiv E_2 [E_1/x] \implies E_1 \equiv \mu x. E_2}
  &
  a.(0 \dot E) & \equiv & \zero &
  \zero \oplus E & \equiv & E \\
  (E_1 \oplus E_2) \oplus E_3 & \equiv & E \oplus (E_2 \oplus E_3) &
  E_1 \oplus E_2 & \equiv & E_2 \oplus E_1 &
  \ul{r} \oplus \ul{s} & \equiv & \ul{r+s} \\
  a.(r \dot E) \oplus a.(s\dot E) & \equiv & a.((r + s) \dot E)
  &
  \mu x. E & \equiv & E[\mu x. E/ x] &
  \ul{0} & \equiv & \zero
\end{array}
$$
As proved in~\cite{bbrs_ic} those axioms and rules together with
$\alpha$-equivalence (i.\,e., renaming of variables bound by $\mu$
does not matter) and the replacement rule
$$
E_1 \equiv E_2 \implies E[E_1/x] = E[E_2/x]
$$
are sound and complete with respect to \emph{weighted bisimilarity}.

Now we add the following three equational axioms to the above calculus:
$$
\begin{array}{rcl}
  a.(r \dot (E_1 \oplus E_2)) & \equiv & a.(r \dot E_1) \oplus a.(r \dot E_2)
  \\
  a.(r \dot \ul{s}) & \equiv & a.(1 \dot \ul{rs})
  \\
  a.(r \dot b.(s \dot E)) & \equiv & a.((rs) \dot b.(1 \dot E))
\end{array}
$$
Here $1$ is the multiplicative unit of the semiring. Our main result
in Section~\ref{sec:lin} is that this augmented calculus is sound and
complete with respect to \emph{weighted language equivalence}.

In Section~\ref{sec:nondet} we mention the special case of non-deterministic automata.
In this case the syntactic expressions simplify to
$$
E ::= x \mid \zero \mid E \oplus E \mid \ul{1} \mid a.E \mid \mu x.E.
$$
and the equational axioms and rules for bisimilarity of expressions
derived from the work in~\cite{bbrs_ic} are
$$
\begin{array}{rcl@{\qquad}rcl@{\qquad}rcl}
  \multicolumn{3}{l}{E_1 \equiv E_2[E_1/x] \implies E_1 \equiv \mu x. E_2}
  &
  \mu x. E & \equiv & E[\mu x. E/ x] &
  \zero \oplus E & \equiv & E \\
  (E_1 \oplus E_2) \oplus E_3 & \equiv & E_1 \oplus (E_2 \oplus E_3) &
  E_1 \oplus E_2 & \equiv & E_2 \oplus E_1 &
  E \oplus E  & \equiv & E
\end{array}
$$
plus $\alpha$-equivalence and the replacement rule. Here we add the
following two axioms
$$
a.(E_1 \oplus E_2) \equiv a.E_1 \oplus a.E_2
\qquad
\text{and}
\qquad
a.\zero \equiv \zero
$$
to obtain a sound and complete calculus for \emph{language equivalence} of
non-deterministic automata.

Notice that the latter calculus coincides with Rabinovich's result for trace
equivalence of finite state labelled transition
systems~\cite{rabinovich}.
\takeout{Another axiomatization of weighted language
equivalence was independently (and almost simultaneously) developed by
\'Esik  and Kuich~\shortcite{ek_2012} who build on an axiomatization of
  rational weighted languages over the semiring of natural numbers by
  Bloom and \'Esik~\shortcite{be_2009}. These axiomatizations use
  a $\ast$-operations and are purely equational as in the axiomatizations
  of regular languages presented in~\cite{Kro91}
  and~\cite{be93}. However, one has an infinite set of equational
  axioms---in fact, it is well-known that there is no finite axiomatization (see e.\,g.~\cite{conway71}).
  Our calculus using a $\mu$-operator and the unique fixpoint rule is
  simpler but not equational.}
Two axiomatizations for weighted language equivalence were recently
developed by \'Esik and Kuich~\shortcite{ek_2012} who build on
axiomatizations of rational weighted languages over the semiring of
natural numbers by Bloom and \'Esik~\shortcite{be_2009}. These
axiomatizations use a $\ast$-operations as in the axiomatizations of
regular languages presented in~\cite{Kro91} and~\cite{be93}. \'Esik
and Kuich's work provides one purely equational axiomatization,
necessarily with an infinite set of equational axioms, and a simpler
axiomatization which is not equational and in which $*$ is a unique
fixpoint operator. Our calculus using a $\mu$-operator and the unique
fixpoint rule is similar to the latter axiomatization. The idea to extend a sound
  and complete calculus for weighted bisimilarity with additional
  axioms as well as our proof method for soundness and completeness are new.
Our result
can also be seen as an extension of the second author's calculus for closed
stream circuits~\cite{m_linexp} to weighted automata over alphabets of
arbitrary size and from weights in a field to weights in a semiring.

We restrict our application to the above two concrete calculi in the
present paper. But our results on finitary coinduction can be applied to
different combinations of monads $T$ and functors $F$. For example, for the
monad of free semimodules for a semiring $\S$ used above, our method can be
applied to calculi for an inductively defined class of functors $F$ in a uniform
way. However, working out these details is non-trivial because the resulting
generic calculus is syntactically more involved as it will be parametric in $F$.
We therefore decided to treat this generic calculus in a subsequent paper.

\section{Preliminaries}

We assume that readers are familiar with basic concepts and notions from
category theory. Here we present some additional basic material needed
throughout the paper. We denote by $\Set$ the category of sets and functions.

\subsection{Semirings and semimodules}
\label{sec:semi}

In our applications we will consider semimodules for a semiring. A
\emph{semiring} is a tuple $(\S, +, \cdot, 0,1)$ where $(\S, + , 0)$ and
$(\S,\cdot, 1)$ are monoids, the former of which is commutative, and
multiplication distributes over finite sums (i.\,e., $r \o 0 = 0 = 0
\o r$, $r\o (s + t) = r \o s + r \o t$ and $(r + s)\o t = r \o t + s
\o t$). We just write $\S$ to denote a semiring. An
\emph{$\S$-semimodule} is a commutative monoid $(M, +, 0)$ with an
action $\S \times M \to M$ denoted by juxtaposition $rm$ for $r \in \S$
and $m \in M$, such that for every $r,s \in \S$ and every $m, n \in M$
the following laws hold:
$$
\begin{array}{rcl@{\qquad}rcl}
  (r+s)m & = & rm + sm & r(m+n) & = & rm + rn \\
  0m &  = & 0 & r0 & = & 0 \\
  1m &  = & m & r(sm) & = & (r \o s) m
\end{array}
$$
A $\S$-semimodule homomorphism is a monoid homomorphism $h\colon M_1 \to M_2$ such that
$h(rm) = rh(m)$ for each $r \in \S$ and $m \in M_1$. We denote by
$$
\SMod
$$
the category of $\S$-semimodules and their homomorphisms.

A semimodule $M$ is \emph{finitely generated} if there is a finite set
$G \subseteq M$ such that every element of $M$ can be written as a finite
linear combination of elements from $G$. Equivalently, there exists a
surjective homomorphism $\S^n \to M$ for some natural number $n$. A semimodule
$M$ is called \emph{finitely presentable} if it can be presented by
finitely many generators and relations. Equivalently, $M$ is a
coequalizer of some parallel pair of semimodule homomorphisms
$\S^m \parallel \S^n$, where $m$ and $n$ are natural numbers.

\begin{definition}[\cite{em_2011}]
  \label{def:Noether}
A semiring $\S$ is called \emph{Noetherian} if every subsemimodule of a
finitely generated $\S$-semimodule is itself finitely generated.
\end{definition}

Examples of Noetherian semirings are: every finite semiring, every
field, every principal ideal domain such as the ring of integers and
therefore every finitely generated commutative ring by Hilbert's Basis
theorem. As recently proved by \'Esik and
Maletti~\shortcite{em_2010}, the tropical semiring $(\Nat \cup
\{\infty\}, \min, +, \infty, 0)$ is not Noetherian. Also the semiring
$(\Nat,+, \cdot, 0, 1)$ of natural numbers is not Noetherian, as shown in the following example.

\begin{example}
  The $\Nat$-semimodule $\Nat \times \Nat$ (with the pointwise
  action) is finitely generated. But its subsemimodule generated by
  the infinitely many elements
  \[
  (2,1), (3,2), (4,3), \ldots
  \]
  is not finitely generated.
\end{example}

\begin{remark}
  In the literature (see e.\,g.~\cite{golan}) a semiring is sometimes
  called Noetherian if every of its ideals is finitely generated. This
  is the same notion that is considered in ordinary ring theory (see
  e.\,g.~\cite[VI, Proposition~1.5]{lang}), and, in fact, for a ring $\S$
  this notion is equivalent to the one in Definition~\ref{def:Noether}.
  However, in general, this is not the case for semirings: while every ideal of
  the semiring $(\Nat, + ,0, \cdot, 1)$ is finitely generated, we have seen in
  the above example that this semiring is not Noetherian according to
  Definition~\ref{def:Noether}.
\end{remark}

\begin{lemma}
  \label{lem:prodmod}
  For every semiring $\S$, finitely generated $\S$-semimodules are closed under finite products.
\end{lemma}
\begin{proof}
  Clearly the terminal semimodule $\{0\}$ is finitely generated. Given two
  finitely generated semimodules $M$ and $N$ with the corresponding
  quotients $p\colon \S^m \to M$ and $q\colon \S^n \to N$ we have the quotient
  $$
  \xymatrix@1{\S^{m+n} = \S^m \times \S^n \ar[r]^-{p \times q} & M \times N.}
  $$
  \qed
\end{proof}

The following proposition gives a slightly more easy criterion to
verify Noetherianess of a semiring.

\begin{proposition}
  \label{prop:noether}
  For a semiring $\S$ the following are equivalent:

  \smallskip
  (1)~$\S$ is Noetherian,

  \smallskip
  (2)~every subsemimodule of a free finitely generated semimodule
  $\S^n$ is finitely generated.
\end{proposition}
\begin{proof}
  (1) $\Rightarrow$ (2) trivially hold.

  (2) $\Rightarrow$ (1). Suppose that $N$ is a subsemimodule of the
  finitely generated $\S$-semimodule $M$ via $m\colon N \to M$. Take a
  quotient $q\colon \S^n \to M$ and form the pullback of $m$ along $q$:
  $$
  \xymatrix{
    N'
    \ar@{ >->}[r]^{m'}
    \ar@{->>}[d]_{q'}
    &
    \S^n
    \ar@{->>}[d]^q
    \\
    N
    \ar@{ >->}[r]_{m}
    &
    M
    }
  $$
  Since surjective and injective homomorphism are stable under
  pullback, we see that $N'$ is a submodule of $\S^n$ and $N$ is a
  quotient of $N'$. So $N'$ is finitely generated by assumption, and,
  hence, so is its quotient $N$.
  \qed
\end{proof}

We will use the following properties of Noetherian semirings.

\begin{proposition}
  \label{prop:lfpmodules}
  If $\S$ is a Noetherian semiring, then the following hold:

  \smallskip
  (1)~every finitely generated semimodule is finitely
  presentable.

  \smallskip
  (2)~finitely generated $\S$-semimodules are closed under finite
  limits.
\end{proposition}
\begin{proof}
  Ad~(1). Let $M$ be a finitely generated $\S$-semimodule, and take a surjective
  homomorphism $h\colon \S^n \to M$. Since $h$ is a regular epimorphism, it
  follows that $h$ is the coequalizer of its kernel pair. So we form
  the kernel pair $p, q\colon K \parallel \S^n$ of $h$. Then $K$ is a
  subsemimodule of the free finitely generated module
  $\S^{n+n}$. Hence, since $\S$ is Noetherian, $K$ is a finitely
  generated semimodule, too. So we have a surjective homomorphism $g\colon
  \S^m \to K$. This implies that $h$ is a coequalizer of the parallel
  pair $p \o g, q \o g\colon \S^m \to \S^n$, which shows that $M$ is
  finitely presentable.

  Ad~(2).  It suffices to prove closedness under finite
  products and subsemimodules. The former was established in
  the Lemma~\ref{lem:prodmod} and the latter is by hypothesis.
\end{proof}

\begin{example}
  \label{ex:fgnonfpmodule}
  For general (semi)rings finitely generated modules need not be
  finitely presentable. For a counterexample consider the ring
  $\S = (\Int_2)^\Nat$ and its ideal $I$ formed by all functions $f\colon \Nat
  \to \Int_2$ with finite support. Then the quotient $\S/I$ is clearly
  finitely generated as an $\S$-module (since there is a surjective
  homomorphism $q\colon \S \to \S/I$). But $\S/I$ is not finitely presented;
   it is easy to show that the kernel $I$ of $q$ is not
  finitely generated as an $\S$-module.
\end{example}

\begin{remark}
  For a ring $\S$ the item~(2) of Proposition~\ref{prop:lfpmodules} is
  actually equivalent to $\S$ being Noetherian. To see this recall that the
  ring $\S$ is Noetherian if and only if every of its ideals is finitely
  generated (see~\cite[Chapter~VI, Proposition~1.5]{lang}).

  Now suppose that finitely generated $\S$-modules are closed under
  finite limits, and let $I$ be any ideal of $\S$. Form the quotient ring
  $\S/I$, i.\,e., the quotient homomorphism $c\colon  \S \to \S/I$ is the
  coequalizer of the inclusion $i\colon I \subto \S$ and the $0$-morphism $I
  \to \S$. Now notice that $I$ is a split quotient of the domain $K =
  \{\,(x,y) \mid cx = cy\,\}$ of the kernel pair of $c$ via $q\colon K \to
  I$ with $q(x,y) = x-y$.

  The quotient $\S/I$ is of course finitely generated (with one
  generator). Since the free $\S$-module $\S$ is also finitely
  presented, so are $K$ (by assumption) and $I$ (since finitely
  generated objects are closed under quotients).
\end{remark}

Let us mention a few special cases of the category $\SMod$ of $\S$-semimodules: for the Boolean
semiring $\S = (\{\,0,1\,\}, \vee, \wedge, 0, 1)$, $\SMod$ is the category $\Jsl$ of
(bounded) join-semilattices and join-preserving maps\footnote{We consider
join-semilattices with a least element $0$. So a join-semilattice
is, equivalently, a commutative idempotent monoid.}. If $\S$ is a field,
then $\SMod$ is the category $\SVec$ of vector spaces over $\S$ and
linear maps; for $\S$ the ring of integers we get the category of Abelian
groups and for $\S$ the natural numbers $\SMod$ is the category of commutative monoids.

\subsection{Coalgebras}
\label{sec:coalg}

Let $\A$ be a category, and let $F\colon \A \to \A$ be an endofunctor. A
\emph{coalgebra} for $F$ is a pair $(C,c)$ consisting of an object $C$
and a structure morphism $c\colon  C \to FC$. For example, if $\A = \Set$,
then we can understand coalgebras as systems, where the set $C$
consists of all states of the system and where the map $c$ provides
the transitions whose type is described by the endofunctor $F$. Concrete examples of coalgebras for set
endofunctors include various kinds of automata (deterministic,
non-deterministic, Mealy, Moore), stream systems, probabilistic
automata, weighted ones, labelled transition systems and many
others. We now mention two leading examples that we will consider in
our applications in Sections~\ref{sec:lin} and~\ref{sec:nondet}; for more
examples see e.\,g.~\cite{rutten,bbrs_ic}.

\takeout{ 
Firstly, image finite labelled transition systems are coalgebras for the
functor $FX = (\powf X)^A$ where $A$ is the set of
actions. To give a coalgebra $c\colon  C \to (\powf C)^A$ is
the same as to give a set $C$ of states and an image finite transition
relation $\to \subseteq C \times A \times C$. }

Firstly, non-deterministic automata are coalgebras for the set functor $FX
= 2 \times (\powf X)^A$, where $A$ is the finite input alphabet, and $\powf$
is the finite powerset functor. A coalgebra $c\colon  C \to 2 \times
(\powf C)^A$ is precisely the same as a set $C$ of states together
with an image finite transition relation $\delta \subseteq C \times A
\times C$ and a subset $C'\subseteq C$ of final states.

Our second leading example is weighted automata~\cite{schuetzenberger,dkv}. Let $\S$ be a
semiring. We consider the functor $V_\S\colon \Set \to \Set$ defined on sets
$X$ and maps $h\colon X \to Y$ as follows:
\begin{equation}
  \label{eq:V}
  V_\S X = \{\, f\colon X \to \S \mid \text{$f$ has finite support}\,\}, \qquad
  V_\S h(f) = \big(y \mapsto \sum_{x \in h^{-1}(y)} f(x)\big),
\end{equation}
where a function $f\colon X \to \S$ is said to have finite support if $f(x)
\neq 0$ holds only for finitely many elements $x \in X$. In the sequel we will omit the subscript $\S$ from the
above functor as we will always work with a fixed semiring $\S$. One can
think of $VX$  as consisting of all formal linear combinations on elements of $X$;
in other words, $VX$ is the free $\S$-semimodule on $X$. A weighted
automaton with finite input alphabet $A$ is simply a coalgebra for the
functor $FX = \S \times (VX)^A$. In more detail, a coalgebra $c\colon  C
\to \S \times (VX)^A$ is given by a set $C$ of states, a map $o\colon C \to \S$
associating an output weight with every state and a map $t\colon C \to (VX)^A$
encoding the transition relation in the following way: the state $s
\in C$ can make a transition to $s' \in C$ with input $a \in A$ and
weight $w \in \S$ if and only if $t(s)(a)(s') = w$.

Notice that taking $\S$ to be the Boolean semiring weighted automata are precisely the classical
non-deterministic ones as $V$ and $\powf$ are naturally isomorphic. So
the first example is actually a special case of the second one.

For $F$-coalgebras to form a category we need morphisms: a
\emph{coalgebra homomorphism} from a coalgebra $(C, c)$ to a coalgebra
$(D,d)$ is a morphism $h\colon C \to D$ preserving the transition
structure, i.\,e., such that $d \o h = Fh \o c$. We write
$$
\Coalg(F)
$$
for the category of $F$-coalgebras and their homomorphisms.

An important concept in the theory of coalgebras is that of a final
coalgebra. An $F$-coalgebra $(T, t)$ is said to be \emph{final}
if for every $F$-coalgebra $(C, c)$ there exists
a unique coalgebra homomorphism $\fin c$ from $(C,c)$ to $(T,t)$:
\[
\xymatrix{
  C
  \ar[r]^-c
  \ar[d]_{\fin c}
  &
  FC
  \ar[d]^-{F\fin c}
  \\
  T
  \ar[r]_-t
  &
  FT
}
\]
We will write
\[
\nu F
\]
for the final coalgebra $T$, if it exists.\footnote{Existence of a final coalgebra can be guaranteed by mild
  assumptions on $F$, e.\,g., every bounded (or, equivalently,
  accessible) endofunctor on $\Set$ has a final coalgebra.} The final
coalgebra is uniquely determined up to isomorphism. Moreover, the
structure map $t\colon \nu F \to F(\nu F)$ of a final coalgebra is an
isomorphism by Lambek's Lemma~\cite{lambek}. So $\nu F$
is a fixpoint of the endo\-functor $F$. More generally, any coalgebra
$(C,c)$ with $c$ an isomorphism is said to be a \emph{fixpoint} of
$F$. For an endofunctor on $\Set$, the elements of the final coalgebra
provide semantics for the behavior of the states of a system regarded as
$F$-coalgebra $(C,c)$ via the unique coalgebra homomorphism $\fin c$.

Let us note that finality also provides the basis for semantic equivalence. Let $(C,
c)$ and $(D, d)$ be two coalgebras for an endofunctor $F$ on $\Set$
with the final coalgebra $(\nu F, t)$.
Then two states
$x \in C$ and $y \in D$ are called \emph{behavioral equivalent} if
$\fin c(x) = \fin d(y)$. If $F$
preserves weak pullbacks then behavioral equivalence
coincides with the well-known notion of bisimilarity. The states $x$
and $y$ are called bisimilar if they are in a special relation
called a \emph{bisimulation}~\cite{aczel+mendler}. We shall not define
that concept here as it is not needed in the present paper; for
details see~\cite{rutten}. Let us just remark that the coalgebraic
notion of bisimulation generalises the concepts known, under the same name
for concrete classes of systems, e.\,g., for labelled transition systems,
where coalgebraic bisimulation coincides with Milner's strong bisimulation.
The requirement that $F$ preserve weak pullbacks is not very restrictive; many
functors of interest in coalgebra theory do indeed preserve weak pullbacks.
We list some examples of interest in this paper.
\begin{examples}%
  \label{ex:weak}%
  \begin{enumerate}[(1)]
  \item Let $\Sigma$ be a signature of operations symbols with
    prescribed finite arities, i.\,e.~a sequence $(\Sigma_n)_{n \ls \omega}$ of
    sets. The associated polynomial functor $F_\Sigma$ is defined by
    the object assignment
    \[
    \label{eq:FSigma}
    F_\Sigma X = \coprod\limits_{n \ls \omega} \Sigma_n \times X^n.
    \]
    All polynomial set functors preserve weak pullbacks.
  \item The finite powerset functor preserves weak pullbacks.
  \item Composites, products and coproducts of weak pullback
    preserving sets functors preserve weak pullbacks.
  \item It follows from~(1)--(3) that the functors $X\mapsto 2 \times X^A$,
      $X\mapsto 2 \times (\powf X)^A$ and
      $X \mapsto B \times X^A$ of deterministic, non-deterministic, and Moore automata, respectively, preserve weak pullbacks.
  \item The functor $V$ from~\refeq{eq:V} preserves weak pullbacks if and only if
    the monoid $(\S , + , 0)$ is
        \begin{enumerate}[(a)]
    \item \emph{positive}, i.\,e., $a + b = 0$ implies $a = 0 = b$ and
    \item \emph{refinable}, i.\,e., whenever $a_1 + a_2 = b_1 + b_2$
      then there exists a $2 \times 2$-matrix with row sums $a_1$ and
      $a_2$ and column sums $b_1$ and $b_2$, respectively,
    \end{enumerate}
    see~\cite{gs01} and the discussion
    in~\cite{amms11}. So if $(\S, +,0)$ is positive and refinable the
    type functor $X \mapsto\S \times (VX)^A$ of weighted automata preserves
    weak pullbacks.
  \item Giry's probability monad~\cite{giry} on the category of
    measurable spaces does not preserve weak pullbacks (see~\cite{vig05}).
  \end{enumerate}
\end{examples}

We now mention some examples of final coalgebras in more detail.

\begin{examples}
  \label{ex:finalcoalg}
  \begin{enumerate}[(1)]
  \item For a polynomial set endofunctor $F_\Sigma$, the final
    coalgebra consists of all (finite and infinite) $\Sigma$-trees,
    i.\,e.~rooted and ordered trees labelled in $\Sigma$ such that a
    node with $n$ children is labelled by an $n$-ary operation
    symbol. The coalgebra structure is given by the inverse of tree
    tupling.

  \item Classical deterministic automata with input alphabet $A$
    are coalgebras for the functor $FX = 2 \times X^A$, where $2 = \{\,0,1\,\}$, and the
    final $F$-coalgebra is carried by the set $\Pow(A^*)$ of all formal
    languages on $A$; its coalgebra structure is given by the two maps
    $o\colon \Pow(A^*) \to 2$ and $t\colon \Pow(A^*) \to \Pow(A^*)^A$ where for
    a formal language $L \subseteq A^*$ we have
    $$
    o(L) = 1 \iff \eps \in L
    \quad
    \text{and}
    \quad
    t(L)(a) = L_a = \{w \mid aw \in L\}.
    $$
    Moreover, for a deterministic automaton presented as
    an $F$-coalgebra $(C,c)$ the unique homomorphism $\fin c\colon C \to
    \Pow(A^*)$ assigns to every state $s \in C$ the formal language it
    accepts.

  \item Deterministic Moore automata with input alphabet $A$ and
    outputs in the set $B$ are coalgebras for the functor $FX = B \times X^A$.
    The final $F$-coalgebra is carried by the set $B^{A^*}$. The coalgebra
    structure on $B^{A^*}$ is given by the two maps $o\colon B^{A^*} \to B$ and
    $t\colon B^{A^*} \to (B^{A^*})^A$ with
    $$
    o(L) = L(\eps)
    \quad
    \text{and}
    \quad
    t(L)(a) = \lambda w.L(aw).
    $$
    We shall be interested in the case where $B = \S$ is a semiring, so
    $\S^{A^*}$ are weighted languages (or formal power series).

  \item In the example of non-deterministic automata as coalgebras for
    $FX = 2 \times (\powf X)^A$ the elements of
     the final coalgebra can be thought of as representatives of all
    finitely branching processes with outputs in $2$ modulo strong
    bisimilarity. A more concrete description follows from the
    result on the final coalgebra for $\powf$ given by
    Worrell~\shortcite{worrell} (see also~\cite{amms11}):

    Consider all (rooted) finitely
    branching trees with edges labelled in $A$ and nodes labelled in
    $2$. Every such tree can be considered as an $F$-coalgebra in a
    canonical way (with the coalgebra structure assigning to a node
    $x$ of a tree the pair $(o,t)$, where $o$ is the node label of $x$
    and $f$ is the function mapping an input symbol $a \in A$ to the finite set of
    child nodes of $x$ reachable by $a$-labelled edges).
    A \emph{tree bisimulation} between a tree $t$ and a tree $s$ is a
    bisimulation $R$ between the corresponding coalgebras such
    that~(a) the roots of $s$ and $t$ are in $R$, (b)~whenever two
    nodes are related, then their parents are related and (c)~only
    nodes of the same depth are related. A tree is
    said to be \emph{strongly extensional} if there is no non-trivial
    tree bisimulation on the coalgebra induced by the tree. The final
    coalgebra consists of all finitely branching strongly extensional
    trees with nodes labelled in $2$ and edge labels from $A$ with the coalgebra structure given
    by the inverse of tree tupling.

  \item Finally, for weighted automata considered as coalgebras for
    the functor $FX = \S \times (V X)^A$ a final coalgebra exists since
    the functor is finitary. However, an explicit description of its
    elements does not seem to be known in general. In the following
    special case an explicit description easily follows from the
    recent work of Ad\'amek et al.~\shortcite{amms11}: let $(\S, +, 0)$ be a
    positive and refinable monoid (equivalently, $V$ preserves weak pullbacks,
    cf.~Example~\ref{ex:weak}(5)). Then the final coalgebra for $V$ is
    carried by the set of all strongly extensional, finitely
    branching, $\S$-labelled trees with the coalgebra structure given
    by the inverse of tree tupling. Similarly, it is not difficult to
    prove that the final coalgebra for $F$ is carried by the set
    of all strongly extensional, finitely branching, $(\S, A)$-labelled
    trees (i.\,e.~each edge is labelled by a weight from $\S$ and an
    input symbol from $A$) with all nodes labelled in $\S$.

    The latter trees are precisely the behaviors of weighted
    automata modulo weighted bisimilarity; in fact, weighted
    bisimilarity~\cite{buchholz} is precisely the behavioral equivalence
    for the above functor $F$ (see~\cite[Proposition~3.4]{bbrs_ic}).
  \end{enumerate}
\end{examples}

\begin{remark}
  \begin{enumerate}[(1)]
  \item We shall need to work with quotients of coalgebras. In
    general, a (strong) \emph{quotient} of an object $X$ in a category
    $\A$ is represented (up to isomorphism) by a strong epimorphism
    $q\colon \xymatrix@1@C-1pc{X \ar@{->>}[r] & Y}$; we shall simply call $Y$ a quotient of $X$. Similarly, a
    \emph{quotient coalgebra} is represented by a coalgebra
    homomorphism $q\colon \xymatrix@1@C-1pc{(X, x) \ar@{->>}[r] & (Y, y)}$
    such that
    $q\colon \xymatrix@1@C-1pc{X \ar@{->>}[r] & Y}$ is a strong
    epimorphism in $\A$.

  \item Our choice of strong epimorphisms to represent
  quotients stems from the fact that in an
  Eilenberg-Moore category $\Set^T$ of a monad $T$ on $\Set$ the strong
  epimorphisms are precisely the surjective $T$-algebra
  homomorphisms. In general, epimorphisms may not be surjective in
  $\Set^T$, e.\,g.~the embedding $\Int \to \Rat$ is an epimorphism in
  the categories of rings and semigroups
  (see~\cite[Example~7.40(5)]{ahs09}), which are both isomorphic to
  $\Set^T$ for appropriate monads $T$.
 \end{enumerate}
\end{remark}

\subsection{Eilenberg-Moore-Algebras and the generalised powerset
  construction}
\label{sec:monad}

The recent paper~\cite{bbrs_fsttcs} provides a coalgebraic version of
the powerset construction applicable to many different system types
expressed as coalgebras for a set endofunctor. One considers an
endofunctor $H$, giving the transition type of a class of systems, that
is obtained
as the composition of two functors $F$ and $T$ on $\Set$, i.e. $H =
FT$. Intuitively, $F$ gives the ``behavior type'' and $T$ the ``branching behavior''
of that class of systems.
We already saw this in our two leading examples above: non-deterministic
automata are $FT$-coalgebras where $FX = 2 \times X^A$ and $T = \powf$ is the finite
powerset functor, and weighted automata are $FT$-coalgebras for $FX = \S \times X^A$
and $T = V$.

To apply the generalised powerset construction to a coalgebra $c\colon  C
\to FTC$ it is important that $T$ is the functor part of a monad and
that $FTC$ is an Eilenberg-Moore algebra for $T$. We now briefly
recall these concepts (see e.\,g.~\cite{maclane} for a detailed
introduction).

A \emph{monad} is a triple $(T, \eta, \mu)$, where
$\eta\colon \Id \to T$ and $\mu\colon TT \to T$ are natural transformations such
that $\mu \o \eta T = \id_T = \mu \o T\eta$ and $\mu \o T\mu = \mu \o
\mu T$. An \emph{Eilenberg-Moore}
algebra for a monad $T$ (or \emph{$T$-algebra}, for short) is a
pair $(A, \alpha)$ consisting of an object $A$ and a structure
morphism $\alpha\colon TA \to A$ such that $\alpha \o \eta_A = \id_A$ and
$\alpha \o \mu_A = \alpha \o T\alpha$. A $T$-algebra
homomorphism from $(A, \alpha)$ to $(B, \beta)$ is a morphism $h\colon A
\to B$ such that $h \o \alpha = \beta \o Th$. Eilenberg-Moore
algebras for a monad $T$ on $\Set$ form the category denoted by
$\Set^T$. Clearly, for every set $X$, $(TX, \mu_X)$ is an
Eilenberg-Moore algebra for $T$. Moreover, this $T$-algebra is
\emph{free} on $X$, i.\,e., for every $T$-algebra $(A,\alpha)$ and
every map $f\colon X \to A$ there is a unique $T$-algebra homomorphism
$\ext f\colon TX \to A$ such that $\ext f \o \eta_X = f$:
\begin{equation}
  \label{not:ext}
  \vcenter{
    \xymatrix{
      TTX
      \ar[r]^-{\mu_X}
      \ar[d]_{T\ext f}
      &
      TX
      \ar[d]^{\ext f}
      &
      X
      \ar[l]_-{\eta_X}
      \ar[ld]^-f
      \\
      TA
      \ar[r]_-\alpha
      &
      A
    }
  }
\end{equation}
Notice also that we have $\ext f = \alpha \o Tf$.

Now we are ready to recall the generalised powerset construction
from~\cite{bbrs_fsttcs}. Let $F$ be an endofunctor on $\Set$ with the
final coalgebra $\nu F$ and let $T$ be a monad. Suppose we are given an
$FT$-coalgebra $(C, c)$ such that $FTC$ carries some $T$-algebra
structure. Then we can form the $F$-coalgebra $\ext c\colon TC \to
FTC$ and consider the unique
$F$-coalgebra homomorphism $\fin{(\ext c)}$ into the
final coalgebra $\nu F$ as summarised by the following diagram:
\begin{equation}
  \label{diag:genpow}
  \vcenter{
\xymatrix@R-.5pc{
  C
  \ar[r]^-c
  \ar[d]_{\eta_C}
  &
  FTC
  \ar[dd]^{F\fin{(\ext c)}}
  \\
  TC
  \ar[ru]_{\ext c}
  \ar[d]_{\fin{(\ext c)}}
  \\
  \nu F
  \ar[r]_-t
  &
  F(\nu F).
}}
\end{equation}

\begin{notation}
  \label{not:ddagger}
  For every $FT$-coalgebra $(C,c)$ we denote the map $\fin{(\ext c)} \o \eta_C$
  arising from the generalized powerset construction by
  \[
  \lang c\colon C \to \nu F.
  \]
  and we call $\lang c$ the \emph{coalgebraic language map} of
  $(C,c)$.
\end{notation}

\begin{definition}
  Let $(C,c)$ and $(D,d)$ be $FT$-coalgebras and let $x \in C$ and $y
  \in D$. The states $x$ and $y$ are called \emph{(coalgebraically)
    language equivalent} if $\lang c(x) = \lang d(y)$ holds.
\end{definition}

In concrete instances, the construction of the coalgebra $(TC, \ext c)$
is determinisation and the map $\lang c \colon C \to \nu F$ assigns to
states of the coalgebra $C$ their language or set of traces.

For example, as we saw previously, non-deterministic automata are $FT$-coalgebras where the
functor is $FX = 2\times X^A$ and the monad is $T = \powf$. The construction extending the coalgebra
structure $c\colon  C \to 2 \times (\powf X)^A$ to $\ext c\colon \powf C \to 2
\times (\powf C)^A$ is precisely the usual powerset construction
determinising the given non-deterministic automaton.  Moreover, the final coalgebra for $F$ consists of all formal
languages, and the map $\lang c$ provides the usual
language semantics of a non-deterministic automaton. In contrast, as
we saw in Example~\ref{ex:finalcoalg}(4), the
final coalgebra for $FT$ provides the bisimilarity semantics taking
into account the non-deterministic branching of automata (thus, for
example, a non-deterministic automaton and its determinisation are in
general not equivalent in this semantics).

In our second leading example of weighted automata we consider
$FT$-coalgebras for the functor $FX = \S \times X^A$ and the monad $T = V$. The construction
extending a coalgebra $c\colon C \to \S \times (VX)^A$ to $\ext c$ can
be understood as determinisation of the given weighted automaton
again. Moreover, we saw in Example~\ref{ex:finalcoalg}(2) that the
final coalgebra for $F$ is carried by the set $\S^{A^*}$ of weighted
languages, and so the map $\lang c\colon C \to \S^{A^*}$ assigns to a
state of a weighted automaton the weighted language it accepts. To
summarise: behavioral equivalence of $FT$-coalgebras coincides with
weighted bisimilarity, while behavioral equivalence of
$F$-coalgebras yields weighted language
equivalence.

\subsection{Liftings of functors to algebras}

We have seen that the category of Eilenberg-Moore algebras for a set
monad $T$ plays an important r\^ole for the generalised powerset
construction presented in the previous section. For our work in the
present paper we make use of functors $F$ that lift to the category
$\Set^T$ and we shall study fixpoints of $F$ and its lifting. We
now briefly recall the necessary background material.

Let $F\colon \Set \to \Set$ be a functor and let $(T, \eta, \mu)$ be a
monad on $\Set$. We denote by $U\colon \Set^T \to \Set$ the forgetful functor mapping
a $T$-algebra to its underlying set. A \emph{lifting} of $F$ to
$\Set^T$ is a functor $\bar F\colon \Set^T \to \Set^T$ such that the square
below commutes:
$$
\xymatrix{
  \Set^T
  \ar[r]^-{\bar F}
  \ar[d]_U
  &
  \Set^T
  \ar[d]^U
  \\
  \Set
  \ar[r]_F
  &
  \Set
}
$$
In general, a lifting of $F$ need not be unique.
It is well-known that to have a lifting of $F$ to $\Set^T$ is the same as to
have a distributive law of the monad $T$ over the functor $F$  (see~\cite{applegate,johnstone_lift}).
Recall from {\em loc.~cit.} that a \emph{distributive law} of $T$ over $F$ is a natural transformation
$\lambda\colon TF \to FT$ such that the following two laws hold:
\begin{equation}
  \label{eq:lambda}
  \lambda \o \eta F = F\eta
  \qquad\textrm{and}\qquad
  \lambda \o \mu F = F \mu \o \lambda T \o T \lambda.
\end{equation}

\begin{remark}
  Suppose that $F$ has a lifting to $\Set^T$. Then $FTC$
  carries a $T$-algebra structure for every object $C$: apply
  the lifting $\bar F$ to the free $T$-algebra $(TC, \mu_C)$.
  Thus, the generalised powerset construction described
  in~\refeq{diag:genpow} can be applied to every coalgebra $c\colon C \to FTC$.
\end{remark}

The functors in our leading examples have liftings to
the respective Eilenberg-Moore categories. For the case of non-deterministic
automata  recall that we have $FX = 2 \times X^A$ and $T = \powf$ and notice that
$\Set^{\powf}$ is (equivalent to)  the category $\Jsl$ of
join-semilattices. The lifting $\bar F$ maps a join-semilattice
$X$ to $2 \times X^A$, where $2$ carries the join-semilattice
structure with $0 \leq 1$, and on the product and the power to the set
$A$ one takes the join-semilattice structure componentwise.
More generally, every non-deterministic functor as defined
in~\cite{brs_lmcs} canonically lifts to $\Jsl$.

For the case of weighted automata we have $FX = \S \times X^A$ and $T=V$.
Then $\Set^{V}$ is (equivalent to) the category $\SMod$ of
$\S$-semimodules. The lifting $\bar F$ maps a semimodule $X$ to $\S
\times X^A$, again with the obvious componentwise structure.

We leave it to the reader to work out the distributive laws corresponding to the liftings.

\takeout{
\begin{remark}
  \label{rem:Fgrammar}
  For every monad $T$, it is not difficult to verify by an induction
  argument that  every endofunctor on $\Set$  defined by the following
  grammar has a canonical lifting to $\Set^T$:
  $$F ::= B \mid \Id \mid F \times F \mid F^A \mid F \o T \mid T \o G,$$
 where $A$ ranges over finite sets, $B$ ranges over $T$-algebras and
 $G$ over finitary endofunctors on $\Set$.
\end{remark}
}

\section{Coalgebras over Algebras}
\label{sec:algcoalg}

For the results in the current paper we will study the move from
coalgebras for a functor $F$ to coalgebras for the lifted functor $\bar F$ more
thoroughly. In this section we develop the necessary mathematical
theory of finitary coinduction that we later use to
obtain desired general soundness and completeness theorems. The main
contributions of this section are:
in subsection~\ref{sec:lfp}, the proof that locally finitely
presentable coalgebras are closed under quotients (Lemma~\ref{lem:lfpquot}); in
subsection~\ref{sec:term}, the proof that the final $FT$-coalgebra
also carries a $T$-algebra structure (Lemma~\ref{lem:fix}) and the relation between
the final $FT$-coalgebra and the final $\bar F$-coalgebra
(Proposition~\ref{prop:quot}); and in subsection~\ref{sec:coalg_on_alg},
the relation between the rational fixpoints of $FT$ and $\bar F$
(recall~\refeq{eq:square} and see Theorem~\ref{thm:ratquot}).

\subsection{Locally finitely presentable coalgebras}
\label{sec:lfp}

For the soundness and completeness proofs of the expression calculi presented
in~\cite{brs_lmcs}, locally finite coalgebras play an important
r\^ole, and for the sound and complete calculus for linear systems
given in~\cite{m_linexp} one uses locally finite dimensional coalgebras.
More precisely, expressions modulo the equations and rules of the calculus form
a final locally finite (or, locally finite dimensional, respectively) coalgebra.
In~\cite{m_linexp}, \emph{locally finitely presentable coalgebras}
were introduced as a common generalization of locally finite and locally
finite dimensional coalgebras.
Next, we recall the necessary material and further extend the theory
so as to be able to relate the final locally finitely presentable coalgebras
for $FT$ and $\bar F$.

For a general category, local finiteness of coalgebras is based on a notion
of finiteness of objects of the category, and the latter is captured
by locally finitely presentable categories; we now briefly recall the
basics from~\cite{ar}. A functor is \emph{finitary} if it preserves
filtered colimits, and an object $X$ of a category $\A$ is called
\emph{finitely presentable} if its hom-functor $\A(X,-)$ is
finitary. A category $\A$ is called \emph{locally finitely
  presentable} (\emph{lfp}, for short) if
\begin{enumerate}[(1)]
\item it is cocomplete and
\item has a set of finitely presentable
  objects such that every object of $\A$ is a filtered colimit of
  objects from that set.
\end{enumerate}
We write $\Afp$ for the full subcategory of $\A$
given by all finitely presentable objects.

Our categories of interest, $\Set$ and $\SMod$, are locally
finitely presentable with the expected notion of finitely presentable
objects: finite sets, and finitely presentable $\S$-semimodules, respectively.
In the special instances of $\Jsl$ and vector spaces over a field $\S$ the
finitely presentable objects are finite join-semilattices and finite dimensional
vector spaces, respectively. Other examples of lfp categories are the
categories of posets, graphs, groups and, in fact, every finitary
variety of algebras is lfp. The corresponding notions of finitely
presentable objects are: finite posets or graphs and those groups or
algebras presented by finitely many generators and relations. Notice
that finitary varieties are precisely the Eilenberg-Moore categories
for finitary set monads, so $\Set^T$ is lfp for every
finitary monad $T$ on $\Set$ (here we call a monad finitary if
its the underlying functor is finitary).
In contrast, the category of complete partial orders (cpo's) and
continuous maps is not lfp; there are no non-trivial finitely
presentable objects.

\begin{assumption}
  For the rest of this section we assume that $\A$ is an lfp category
  and that $F\colon \A \to \A$ is a finitary functor on $\A$.
\end{assumption}

\begin{examples}
  There are many examples of finitary functors on lfp categories. We
  mention only those two of interest in the current paper.

  \noindent
  (1)~Every non-deterministic functor on $\Set$ as defined in~\cite{brs_lmcs}
  is finitary. All these functors lift to finitary functors on $\Jsl$
  (e.\,g.~the functor $FX = 2 \times X^A$).

  \noindent
  (2)~The functor $FX = \S \times X^A$ is finitary on $\Set$ and it lifts to a finitary functor of $\SMod$.
\end{examples}

\begin{remark}
  \label{rem:fs}
  (1)~We shall need the following property of lfp categories, and we
  recall this from~\cite{ar}:  Every morphism $f$ in an lfp category $\A$ can be factorized as a
  strong epi $e$ followed by a monomorphism $m$, i.e. $f = m \o e$. This
  factorisation has the following diagonalisation property: for
  every commutative square
  $$
  \xymatrix{
    A
    \ar[d]_f
    \ar@{->>}[r]^e
    &
    B
    \ar[d]^g
    \ar@{-->}[ld]_d
    \\
    C
    \ar@{ >->}[r]_m
    &
    D
    }
  $$
  with $m$ a monomorphism and $e$ a strong epimorphism there exists a
  unique morphism $d\colon  B \to C$ such that $m \o d = g$ and $d\o e =
  f$.

  \medskip
  \noindent
  (2)~It follows that $\Coalg(F)$ also has factorisations whenever $F$ preserves monomorphisms.
   Given the coalgebra homomorphism $f\colon (C,c) \to (D, d)$ we take its strong epi-mono
  factorisation $f = m \o e$ in $\A$. By diagonalisation, we
  obtain a unique $F$-coalgebra structure on the codomain of $e$ such that $e$ and
  $m$ are coalgebra homomorphisms:
  $$
  \xymatrix{
    C
    \ar[r]^-c
    \ar@{->>}[d]_e
    &
    FC
    \ar[d]^{Fe}
    \\
    E
    \ar@{-->}[r]
    \ar@{>->}[d]_m
    &
    FE
    \ar@{>->}[d]^{Fm}
    \\
    D
    \ar[r]_-d
    &
    FD
    }
  $$
  Notice that we do not claim that $e$ is a strong epimorphism in
  $\Coalg(F)$ (and, in general, this claim is false). Also observe
  that for $\A = \Set$ the above argument works for all endofunctors
  since set endofunctors preserve all non-empty monomorphisms $m$ and
  the case of $m\colon \emptyset \to A$ is trivial.
\end{remark}

\begin{notation}
  We denote by $\Coalgf(F)$ the category of all coalgebras $p\colon P \to
  FP$ with a finitely presentable carrier $P$.
\end{notation}

In the current setting, local finiteness of coalgebras is captured by
the following notion introduced in~\cite{m_linexp}.
\begin{definition}
  An $F$-coalgebra $(S, s)$ is called \emph{locally finitely
    presentable} if the canonical forgetful functor $\Coalgf(F)/(S,s)
  \to \Afp / S$ is cofinal.
\end{definition}

\begin{remark}
  More explicitly $(S, s)$ is locally finitely presentable if and only if the
  following two conditions are satisfied:
  \begin{enumerate}[(1)]
  \item for every $f\colon X \to S$ where $X$ is a finitely presentable
    object of $\A$ there exists a coalgebra $(P,p)$ from $\Coalgf(F)$,
    a coalgebra homomorphism $h\colon (P,p) \to (S,s)$ and a morphism $f'\colon X
    \to P$ such that $h \o f' = f$.

  \item The factorisation in~(1) is essentially unique in the sense that
    for every $f''\colon X \to P$ with $h \o f'' = f$ there exists a
    homomorphism $\ell\colon (P,p) \to (Q,q)$ in $\Coalgf(F)$ and a coalgebra
    homomorphism $h'\colon (Q, q) \to (S, s)$ such that $\ell \o f' = \ell \o
    f''$.
  \end{enumerate}
\end{remark}

\begin{example}
  (1)~For $\A = \Set$ an $F$-coalgebra is locally finitely presentable
  if and only if every finite subset of its carrier is contained in a finite
  subcoalgebra. As discussed in~\cite{m_linexp}, if $F$
  preserves weak pullbacks the above notion coincides with that
  of local finiteness considered in~\cite{brs_lmcs}.

  \medskip\noindent
  (2)~Analogously for $\A = \Jsl$, an $F$-coalgebra is locally
  finitely presentable if and only if every finite sub-join-semilattice of its carrier is
  contained in a finite subcoalgebra.

  \medskip\noindent
  (3)~For $\A=\FVec$, the category of vector spaces over a field $\F$, an $F$-coalgebra
  is locally finitely presentable if and only if every finite dimensional subspace of its
  carrier is contained in a finite dimensional subcoalgebra, i.\,e., the given
  coalgebra is locally finite dimensional.
\end{example}

The following theorem gives an easier characterisation of locally finitely presentable
coalgebras, and in particular of the \emph{final} locally finitely presentable
coalgebra, which can be described by considering \emph{only} those coalgebras
with a finitely presentable carrier.

\begin{theorem}[\cite{m_linexp}]
  \label{thm:lfpcoalg}
  (1)~A coalgebra is locally finitely presentable if and only if it is a filtered colimit
  of a diagram of coalgebras from $\Coalgf(F)$.

  \medskip\noindent
  (2)~A locally finitely presentable coalgebra $(R,r)$ is final in
  the category of all locally finitely presentable coalgebras if and only if for
  every coalgebra $(P,p)$ from $\Coalgf(F)$ there exists a unique
  homomorphism from $(P,p)$ to $(R,r)$.
\end{theorem}

An immediate consequence of point~(1) in the previous theorem is that
the final locally finitely presentable coalgebra for a finitary functor $F$
always exists.

\begin{corollary}
  \label{cor:colimrho}
  (1)~The final locally finitely presentable $F$-coalgebra $\rho F$ of a
  finitary functor $F$ exists and is  constructed as the colimit of $\Coalgf(F)$; in symbols:
  $$
  \rho F = \colim(\Coalgf(F) \subto \Coalg(F)).
  $$

  \medskip\noindent
  (2)~Furthermore, $\rho F$ is a fixpoint of $F$.
\end{corollary}

For the proof of point~(2) in the above Corollary~\ref{cor:colimrho}, see~\cite[Theorem~3.3]{amv_atwork}.
The colimit construction in point~(1) is
exactly the construction given in {\em loc.~cit.} of the initial iterative
algebra for $F$. We shall not recall the notion of iterative
algebras here as this plays no r\^ole in the present paper,
but just mention the following result to make an explicit connection
of the work here and in~\cite{brs_lmcs,bbrs_ic,m_linexp} to iterative
theories of~\cite{elgot}.

\begin{corollary}
  The final locally finitely presentable coalgebra for $F$ is
  equivalently characterised as the initial iterative algebra for $F$.
\end{corollary}

Continuing on the above connection, in~\cite{amv_atwork} it was proved that
the monad of free iterative algebras for $F$ is the free iterative
monad $\mathcal R$ on $F$. Thus, our Corollary~\ref{cor:finalexp}
below (page~\pageref{cor:finalexp}) and the
corresponding theorem in~\cite{brs_lmcs,bbrs_ic} provide a new syntactic
characterisation of the closed terms in the free iterative theory
(i.\,e., $\mathcal R0$, where $0$ denotes the initial object).

Next we return to our study of locally finitely presentable coalgebras. We shall
continue to use the notation
\[
\rho F
\]
for the final locally finitely presentable coalgebra for $F$ in analogy to the
notation $\nu F$ for the final $F$-coalgebra, and we will call  $\rho F$ the
\emph{rational fixpoint} of $F$.

\begin{example}
  \label{ex:final}
  We mention a number of examples of rational fixpoints $\rho F$ to illustrate that they
  capture finite system behavior; further examples can be found
  in~\cite{amv_atwork,amv_horps}.

  (1)~For a polynomial endofunctor $F = F_\Sigma$ on $\Set$ (see
  Example~\ref{ex:weak}(1)), recall that the final coalgebra is
  carried by the set of all $\Sigma$-trees, and $\rho F$ consists of
  all \emph{rational} $\Sigma$-trees, i.\,e., $\Sigma$-trees having, up to
  isomorphism, only finitely many subtrees (see~\cite{ginali}).

  \medskip\noindent
  (2)~For the special case $FX = 2 \times X^A$ on
  $\Set$, recall that a coalgebra is a deterministic automaton, and
  the final coalgebra is carried by the set $\mathcal{P}(A^*)$ of all
  formal languages on $A$. Here $\rho F$ is the subcoalgebra given by
  all regular languages.

  \medskip\noindent
  (3)~Let $\F$ be a field. For the functor $FX = \F \times X$ on $\Set$, $\rho F$ consists of
  all streams $\sigma$ that are eventually periodic, i.\,e., $\sigma =
  uvvv\cdots$ where $u$ and $v$ are finite words on $\F$. However, for
  the lifting $\bar F$ to $\FVec$, $\rho \bar F$ is the subcoalgebra of
  $\F^\omega$ given by all rational streams (see~\cite{m_linexp} for
  details).

  \medskip\noindent
  (4)~Similarly, for the lifted functor $\bar FX = \S \times X^A$ on
  the category of $\SMod$ for a semiring $\S$, the
  final coalgebra is carried by the set $\S^{A^*}$ of formal power series
  (or weighted languages) on $\S$. We will see later in this section
  that, whenever $\S$ is Noetherian, $\rho \bar F$ can be characterised by those coalgebras
  with a carrier freely generated by a finite set $X$; equivalently, a
  weighted automaton with the finite state set $X$. By the
  Kleene-Sch\"utzenberger theorem~\cite{schuetzenberger} (see
  also~\cite{br88}) it follows that
  $\rho \bar F$ is the subcoalgebra of all rational formal power series.
  Our sound and complete calculus for language equivalence of weighted
  automata in Section~\ref{sec:lin} is based on this example. Note that
  in the special case that $\S$ is a field and $A$ is the singleton set,
  one can use a different (but equivalent as it coincides with $\rho
  \bar F$) definition of rational formal power series~\cite{rutten_behdiff}.
\end{example}

In all the examples above, the rational fixpoint $\rho F$ always
occurs as a subcoalgebra of $\nu F$. This is no coincidence as we
will now prove.

Recall from~\cite{ar} that a \emph{finitely generated} object is an
object $X$ such that its covariant hom-functor $\A(X,-)$ preserves
directed unions (i.\,e., colimits of directed diagrams of
monomorphisms). Clearly, every finitely presentable object is finitely
generated, but in general the converse does not hold. In fact, finitely
generated objects are closed under quotients (whereas finitely presentable
objects are not, in general), and an object is finitely generated if and only
if it is a quotient of a finitely presentable object. Therefore, to say that
finitely generated and finitely presentable objects coincide
(cf.~Proposition~\ref{prop:sub} below) is equivalent
to the statement that finitely presentable objects are closed under quotients.
The following proposition follows from~\cite[Proposition~4.6 and Remark~4.3]{amv_62}. We
include a proof for the convenience of the reader.

\begin{proposition}
  \label{prop:sub}
  Suppose that in an lfp category $\A$ finitely generated objects are
  finitely presentable, and that $F$ preserves monomorphisms.
  Then $\rho F$ is the subcoalgebra of $\nu F$ given by the union of
  images of all coalgebra homomorphisms $(P,p) \to (\nu F, t)$ where
  $(P,p)$ ranges over $\Coalgf(F)$.
\end{proposition}
\begin{proof}
  Recall that for every coalgebra $p\colon P \to FP$, $\fin p\colon P \to \nu F$
  denotes the unique coalgebra homomorphism. Let $R$ be the union from
  the statement of the proposition:
  $$
  R = \bigcup \mathsf{im}(\fin p)
  \qquad
  \textrm{where $p\colon P \to FP$ ranges over $\Coalgf(F)$.}
  $$
  More precisely, for every $(P,p)$ in $\Coalgf(F)$, let $I =
  \mathsf{im}(\fin p)$ be the subobject of $\nu F$ given by
  factorizing $\fin p$ as a strong epimorphism $e\colon P \to I$ followed
  by a monomorphism $m\colon I \to \nu F$. Since $\Coalgf(F)$ is a filtered category it follows
  that the subobjects $\mathsf{im}(\fin p)$ and their inclusions form
  a directed diagram $\D$, and $R$ is the colimit of this diagram.
  In addition, from Remark~\ref{rem:fs}(2) we see, since $F$ preserves
  monomorphisms, that $I$ carries a coalgebra $i\colon I \to FI$ such
  that $(I,i)$ is a quotient coalgebra of $(P,p)$ via $e$ and a subcoalgebra of $\nu
  F$ via $m$. Thus, the union $R$ is a subcoalgebra of $\nu F$:
  indeed, being a colimit of a diagram of coalgebras, $R$ carries a
  canonical coalgebra structure, and, in addition, the cocone given by
  all monomorphisms $m\colon I \to \nu F$ factors through a monomorphism $R
  \to \nu F$ (see~\cite{ar}).

  Furthermore, by assumption, we have that the quotient $I$ of the
  finitely presentable object $P$ is finitely presentable, too. So
  $\D$ is actually a full subcategory of $\Coalgf(F)$. Since
  we have the morphism $e\colon (P,p) \to (I,i)$, we see that the inclusion
  of $\D$ into $\Coalgf(F)$ is cofinal. It follows that the colimits
  of $\D$ and $\Coalgf(F)$ are the same, in symbols: $R \cong \rho
  F$, which completes the proof. \qed
\end{proof}

\begin{example}
  \label{ex:fg=fp}
  Let us list some examples of categories in which our first assumption
  of Proposition~\ref{prop:sub} holds, i.\,e., finitely generated and
  finitely presentable objects coincide.

  \smallskip
  \noindent
  (1)~The categories of sets, of posets and of graphs obviously have
  the desired property since finitely presentable objects are just
  finite sets (or posets or graphs, respectively).

  \smallskip
  \noindent
  (2)~The categories $\Jsl$ of join-semilattices, of vector
  spaces over a field and of Abelian groups satisfy the property.
  More generally, the category $\SMod$ satisfies this assumption
  whenever $\S$ is a Noetherian semiring (see Proposition~\ref{prop:lfpmodules}).

  \smallskip
  \noindent
  (3)~A \emph{locally finite variety} is a finitary variety in which free
  algebras on finite sets are themselves finite (e.\,g., Boolean
  algebras, distributive lattices or join-semilattices). It is not difficult to see that in such
  a category finitely presentable and finitely generated objects coincide and
  are precisely the finite algebras.

  \smallskip
  \noindent
  (4)~In the categories of commutative monoids and commutative
  semigroups finitely presentable and finitely generated objects
  coincide as proved by R\`edei~\shortcite{redei} (see
  also~\cite{rosales&sanchez} and, for a rather short proof, \cite{freyd68}).
  Notice that commutative monoids are $\SMod$ for $\S$ the natural numbers,
  which we have already seen do \emph{not} form a Noetherian semiring.
  So the proof is different than what we saw in
  Proposition~\ref{prop:lfpmodules}.

  \smallskip
  \noindent
  (5)~The category of presheaves on finite sets (equivalently, finitary
  endofunctors of $\Set$) (see~\cite{amv_horps}).
\end{example}

We have seen that for many interesting categories it holds that finitely generated and finitely presentable objects
coincide. However there are many other relevant
categories in which this fails:

\begin{example}
  \label{ex:fg!=fp}
  \begin{enumerate}[(1)]
  \item In the category of groups, finitely generated objects are
    precisely those groups having a presentation by finitely many
    generators, and finitely presentable groups are precisely those
    groups with a presentation by finitely many generators and finitely
    many relations. It is well-known that there exist finitely generated
    groups that are not finitely presented.

  \item Similarly, in the category of all (not necessarily commutative)
    monoids finitely presentable and finitely generated objects do not
    coincide (see e.\,g.~\cite{ruskuc99} for a finitely generated monoid that is
    not finitely presentable)

  \item In the category of $\S$-modules for the ring $\S =
    (\Int_2)^\Nat$, finitely presentable and finitely generated
    objects do not coincide as shown in
    Example~\ref{ex:fgnonfpmodule}.
 \end{enumerate}
\end{example}

With the next example we show that Proposition~\ref{prop:sub} does not hold
without the assumption that finitely presentable and finitely
generated objects coincide.

\begin{example}\label{ex:counter}
  We take as $\A$ the category of algebras for the signature $\Sigma$
  with a unary and a binary operation symbol. Then the natural numbers
  $\Nat$ with the operations of addition and $n \mapsto 2\o n$ is an
  object of $\A$. Thus, we have an endofunctor $FX = \Nat \times X$ on
  $\A$, and its final coalgebra is carried by the set $\Nat^\omega$ of all streams of natural
  numbers with the obvious componentwise algebra structure. Now consider the $F$-coalgebra $\alpha\colon A \to FA$, where $A$ is the
  free (term) algebra on one generator $x$ and $\alpha$ is uniquely
  determined by the assignment
  $\alpha(x) = (1, 2\o x)$. The unique $F$-coalgebra homomorphism $h\colon
  A \to \nu F$ maps $x$ to the stream $(1,2,4,8,\cdots)$ of powers of
  $2$, and we have
  $$
  h(2\o x) = h(x + x) = (2, 4, 8, 16, \ldots).
  $$
  Now notice that $(A,\alpha)$ lies in $\Coalgf(F)$, and so there is also
  a unique $F$-coalgebra homomorphism $h_0\colon A \to \rho F$. However,
  we will now prove that
  \begin{equation}
    \label{eq:neq}
    h_0(2\o x) \neq h_0(x + x),
  \end{equation}
  and this implies that $\rho F$ is not a subcoalgebra of $\nu F$.
  
%
  We prove~\refeq{eq:neq} by contradiction. Suppose that
  $h_0(2 \o x) = h_0(x+x)$. By the construction of $\rho F$ (see
  Corollary~\ref{cor:colimrho}), we know that there is a coalgebra
  $\beta: B \to FB$ in $\Coalgf(F)$ and an $F$-coalgebra homomorphism
  $g: A \to B$ in $\A$ with $g(2 \o x) = g(x + x)$. Since $B$ is
  finitely presented it is the quotient in $\A$ of a free algebra $A'$
  on a finite set $Y$ of generators modulo finitely many relations,
  via a quotient homomorphism $q: A' \to B$, say. Next, observe that
  there is an $F$-coalgebra structure $\alpha': A' \to FA'$ such that
  $q$ is an $F$-coalgebra morphism: indeed, choose some map
  $s: B \to A'$ with $q \o s = \id$; then extending
  $Fs \o \beta \o q \o \eta_Y$, where $\eta_Y: Y \to A'$ is the
  embedding of generators yields the desired $F$-coalgebra
  $A' \to FA'$.

  Now choose a term $t_x$ in $A'$ with $q(t_x) = g(x)$. This implies
  that $q(2 \o t_x) = q(t_x + t_x)$ since $q$ is a coalgebra
  homomorphism. Since $g$ is a coalgebra homomorphism it merges the
  right-hand component of $\alpha(2 \o x)$ and $\alpha(x+x)$, in
  symbols: $g(2 \o (2 \o x)) = g ((2\o x) + (2\o x))$. It follows that
  $q$ satisfies: $q(2 \o (2 \o t_x)) = q ((2\o t_x) + 2 \o t_x))$.

  Continuing to use that $g$ and $q$ are homomorphisms, we obtain the
  following infinite list of elements (terms) of $A'$ that are merged
  by $q$ (we write these pairs as equations):
  \begin{eqnarray}
    2 \o t_x & = & t_x + t_x \nonumber \\
    2 \o (2 \o t_x) & = & (2\o t_x) + (2\o t_x) \label{eq:eqnlist} \\
    2 \o (2 \o (2\o t_x)) & = & (2\o(2\o t_x)) + (2\o (2\o t_x)) \nonumber \\
    & \vdots \nonumber
  \end{eqnarray}

  We need to prove that there exists no finite set of relations
  $E \subseteq A' \times A'$ generating the above congruence $q\colon A' \to B$.

  Suppose the contrary, let $A'_0$ be the $\Sigma$-subalgebra of $A'$
  generated by $\{t_x\}$, and let $T \subseteq A_0'$ be the subset of
  those terms (or finite $\Sigma$-trees on~$Y$) such that every path
  in $t$ from the root to the generator $t_x$ has the same length. For
  example, of the following finite $\Sigma$-trees
  $$
  x \qquad
  \vcenter
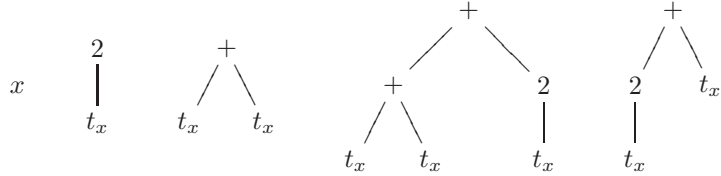
{
    \xy
    \POS   (000,000) *+{2} = "2"
    ,      (000,-10) *+{t_x} = "x"
    \ar@{-} "2";"x"
    \endxy
  }
  \qquad
  \vcenter{
    \xy
    \POS   (000,000) *+{+} = "+"
    ,      (-05,-10) *+{t_x} = "x1"
    ,      (005,-10) *+{t_x} = "x2"
    \ar@{-} "+";"x1"
    \ar@{-} "+";"x2"
    \endxy
  }
  \qquad
  \vcenter{
    \xy
    \POS   (000,000) *+{+} = "+1"
    ,      (-10,-10) *+{+} = "+2"
    ,      (010,-10) *+{2} = "2"
    ,      (-15,-20) *+{t_x} = "x1"
    ,      (-05,-20) *+{t_x} = "x2"
    ,      (010,-20) *+{t_x} = "x3"
    \ar@{-} "+1";"+2"
    \ar@{-} "+1";"2"
    \ar@{-} "+2";"x1"
    \ar@{-} "+2";"x2"
    \ar@{-} "2";"x3"
    \endxy
  }
  \qquad
  \vcenter{
    \xy
    \POS   (000,000) *+{+} = "+"
    ,      (-05,-10) *+{2} = "2"
    ,      (-05,-20) *+{t_x} = "x1"
    ,      (005,-10) *+{t_x} = "x2"
    \ar@{-} "+";"2"
    \ar@{-} "2";"x1"
    \ar@{-} "+";"x2"
    \endxy
  }
  $$
  the first four are in $T$ but not the fifth one. For every
  $\Sigma$-tree in $t$ we call the maximal length of a path from the
  root to the generator $t_x$ the \emph{height} of that tree. Now let
  $t$ and $s$ be $\Sigma$-trees of different height in $T$. Then we
  clearly have $k(t) \neq k(s)$, where $k: B \to \nu F$ is the unique
  $F$-coalgebra homomorphism; this follows from the fact that for a
  tree $t$ of height $n$ in $T$ we have
  $$
  k(t) = (2^n, 2^{n+1}, 2^{n+2}, \ldots).
  $$
  Thus, the equation $t = s$ is not in the congruence generated by $E$
  (otherwise, we would have $q(t) = q(s)$ which implies
  $k(t) = k(s)$). Now let $\ell$ be the height of the tallest
  $\Sigma$-tree that occurs in a relation from $E$. Then the
  $\ell+1$-st equation in~\refeq{eq:eqnlist} with $x$ replaced by
  $t_x$ is not generated by $E$ as this equation is of the form
  $2 \o t = t' + t''$ with $t$, $t'$ and $t''$ of height $\ell+1$. If
  $s$ and $s'$ are terms of height greater that $k$ related by the
  smallest congruence generated by $E$, then $s$ and $s'$ must have
  the same head symbol. So $2 \o t$ and $t' + t''$ are not
  related. Thus, we arrive at the desired contradiction.
\end{example}

We already mentioned that one way to say that finitely generated and
finitely presentable objects coincide is to say that finitely presentable objects are closed under
quotients. The following lemma extends the latter property to locally finitely presentable coalgebras.

\begin{lemma}
  \label{lem:lfpquot}
  Under the assumptions of Proposition~\ref{prop:sub} every quotient
  coalgebra of a locally finitely presentable
  coalgebra is itself locally finitely presentable.
\end{lemma}
\begin{proof}
  Let $q\colon (C, c) \to (D, d)$ be a quotient coalgebra, where
  $(C,c)$ is a locally finitely presentable $F$-coalgebra. So $q\colon C
  \to D$ is a strong epimorphism in $\A$. By
  Theorem~\ref{thm:lfpcoalg}(1), $(C,c)$ is a filtered colimit of a
  diagram of coalgebras $(C_i, c_i)$ from $\Coalgf(F)$ with the
  colimit injections $\ini_i\colon (C_i, c_i) \to (C,c)$. For every $i$
  factorize $q \o \ini_i$ as a strong epi- followed by a
  monomorphism in $\Coalg(F)$ (see Remark~\ref{rem:fs}(2)):
  \[
  \xymatrix{
    (C_i, c_i)
    \ar[r]^-{\ini_i}
    \ar@{->>}[d]_{e_i}
    &
    (C,c)
    \ar[d]^-{q}
    \\
    (D_i,d_i)
    \ar@{ >->}[r]_-{m_i}
    &
    (D,d)
    }
  \]
  By assumption, each $(D_i, d_i)$ lies in
  $\Coalgf(F)$. Moreover, each connecting morphism $c_{ij}\colon (C_i, c_i)
  \to (C_j, c_j)$ induces a coalgebra homomorphism $d_{ij}\colon (D_i,
  d_i) \to (D_j, d_j)$ turning the $D_i$ into a filtered diagram (with the same
  diagram scheme as for the $C_i$). To conclude our proof it suffices
  to show that $D$ is a colimit of this new diagram. We shall
  now prove that $D$ is the \emph{union} of its subobjects $m_i\colon D_i
  \to D$, i.\,e., $D$ has no proper subobject containing every
  $m_i$. It then follows that $D = \colim D_i$ (see~\cite{ar}, 1.63). So let $m\colon M \to D$ be a subobject containing all $m_i$,
  i.\,e., for every $i$ we have monomorphisms $n_i\colon D_i \to M$ such that $m \o n_i =
  m_i$. Now the outside of the following square commutes:
  $$
  \xymatrix{
    \coprod_i C_i
    \ar[d]_{[n_i \o e_i]_i}
    \ar@{->>}[r]^-{[\ini_i]_i}
    &
    C
    \ar@{->>}[r]^-q
    &
    D
    \ar@{=}[d]
    \ar@{-->}[lld]_s
    \\
    M
    \ar@{ >->}[rr]_-m
    &&
    D
    }
  $$
For every $i$ we have
  $$
  q \o \ini_i = m_i \o e_i = m \o n_i \o e_i.
  $$
  Moreover, notice that the copairing $[\ini_i]_i$ is a strong
  epimorphism since it is the copairing of all the injections of the
  colimit $C$. Since strong epimorphisms compose, we see that the
  upper edge of the above diagram is a strong epimorphism. Hence, we
  get, by diagonalisation, the morphism $s\colon D \to M$ such that
  $m \o s = \id$ showing $m$ to be a split epimorphism, whence an
  isomorphism. This completes the proof.
  \qed
\end{proof}

\subsection{The overall setting -- algebraic categories}

For our soundness and completeness proofs in Section~\ref{sec:general}
we need to consider coalgebras for a lifted endofunctor on categories
of Eilenberg-Moore algebras. So we will now focus our attention on
\emph{algebraic categories} $\A$, i.\,e., $\A = \Set^T$ for a finitary
monad $T$ on $\Set$.  Recall that a kernel pair of a morphism $f\colon X
\to Y$ is a pair $k_1, k_2 \colon R \to X$ forming a pullback of $f$ with
itself, and that in every algebraic category kernel pairs exist and
are formed in $\Set$, i.\,e.
\[
R = \{\,(x,y) \mid x,y \in X, fx = fy\,\}
\]
where $k_1, k_2$ are the projections.

\begin{assumption}
  \label{ass:setT}
  For the rest of the paper we assume that $\A = \Set^T$ for the
  finitary monad $(T, \eta, \mu)$, and we also assume that in $\Set^T$
  finitely generated algebras are closed under taking kernel pairs. In addition
  we require that $F\colon\Set \to \Set$ is a finitary endofunctor weakly preserving
  pullbacks and having a lifting $\bar F\colon \Set^T \to \Set^T$.
\end{assumption}

From the above assumptions that $F$ and $T$ are finitary endofunctors
on $\Set$ we know that the final coalgebras $\nu F$ and $\nu(FT)$
exist, see e.\,g.~\cite{barr_coalg}. We also know that the rational
fixpoint $\rho(FT)$ of $FT$ exists since $FT$ is a finitary functor of
$\Set$. Recall that $\Set^T$ is an lfp category, and notice that the
lifting $\bar F$ is finitary because $F$ is finitary and filtered
colimits in $\Set^T$ are formed on the level of $\Set$. Thus, the
final $\bar F$-coalgebra and the rational fixpoint $\rho\bar F$ also
exists. Notice that, in general, $\rho\bar F$ is different from the rational
fixpoint of $F\colon \Set \to \Set$ as demonstrated by
Example~\ref{ex:final}(3).

For some results in the previous section we assumed finitely presentable objects to be closed
under quotients (or, equivalently, finitely generated objects to be finitely
presentable). In this section, we restrict our attention to algebras
for the finitary monad $T$ and we assume that finitely generated $T$-algebras are closed under
kernel pairs. The kernel pair of a morphism gives always a congruence,
and, conversely every congruence relation $\sim$ on an algebra $A$ is
the kernel of the corresponding quotient homomorphism $A \to
A/\mathord{\sim}$. Our assumption above is thus requiring
finitely generated algebras to be \emph{closed under congruences};
more precisely, every congruence of a finitely generated algebra $A$
is itself finitely generated (as a subalgebra of $A \times A$). After
a few examples below we will see that this assumption implies finitely presentable
algebras to be closed under quotients.

\begin{example}
  \label{ex:ass}
  Let us come back to the categories in
  Example~\ref{ex:fg=fp} and see whether they satisfy our
  assumptions.

  \medskip\noindent
  (a)~We have seen that finitely generated commutative monoids and semigroups
  are also finitely presentable (cf. Example~\ref{ex:fg=fp}(4)). However
  congruences of finitely generated commutative monoids need not be
  finitely generated as a monoid. Consider the following example
  from~\cite{Chap2006}: let $R$ be the congruence on $\Nat$, the free
  commutative monoid on one generator, defined by
   \[
   (x,y) \in R  \;\;\mbox{iff}\;\;   (x \geq 1 \;\mbox{and}\; y \geq 1) \;\mbox{or}\; x = y.
  \]
  It is easy to see that $\{(x, 1) \mid x \geq 1\} $ is contained in $R$. But
  the elements of this set cannot be expressed as a sum of two other
  nontrivial elements of $R$. Therefore $R$ cannot be finitely generated
  as a monoid.

  \medskip\noindent
  (b)~All other categories from Example~\ref{ex:fg=fp} satisfy the condition that
  finitely generated objects are closed under taking kernel
  pairs: indeed, for sets, posets, graphs (cf.~\ref{ex:fg=fp}(1)) and
  locally finite varieties (cf.~\ref{ex:fg=fp}(3)) this clearly holds,
  and for semimodules of a Noetherian semiring (cf.~\ref{ex:fg=fp}(2)) see
  Proposition~\ref{prop:lfpmodules}.

  \medskip\noindent
  (c)~The following categories from Example~\ref{ex:fg=fp} are
  categories of algebras for a monad on $\Set$ (in each case we list
  the monad):

  \begin{center}
    \begin{tabular}{p{5cm}||p{6cm}}
      {\bf category} & {\bf is $\Set^T$ for \dots} \\
      \hline\hline
      $\Set$ & $T = \Id$ \\
      \hline
      $\Jsl$ & $T = \powf$ \\
      \hline
      $\FVec$ & $T = V$, where $\S = \F$ is a field\\
      \hline
      abelian groups & $T = V$, $\S = \Int$ the ring of integers \\
      \hline
      $\SMod$ & $T = V$, $\S$ is a (Noetherian) semiring\\
      \hline
      commutative semigroups & $TX = \text{non-empty bags on $X$}$\footnotemark \\
      \hline
      commutative monoids &  $TX = \text{bags on $X$}$\\
    \end{tabular}
    \footnotetext{A \emph{bag} is a finite multiset.}
  \end{center}


  The categories of posets, graphs and
  finitary endofunctors of $\Set$ are not (equivalent to) $\Set^T$ for
  any finitary monad $T$ on $\Set$.

  \medskip\noindent
  (d)~None of the categories from Example~\ref{ex:fg!=fp} has finitely
  generated objects closed under kernel pairs; this follows from the
  next lemma.
\end{example}

From the counterexample in Example~\ref{ex:ass}(a) and the lemma below, it
follows that our Assumption~\ref{ass:setT} is
strictly stronger than the one used in Proposition~\ref{prop:sub}.

\begin{lemma}
In an algebraic category, if finitely generated algebras are closed
under kernel pairs then they are finitely presentable.
\end{lemma}
\begin{proof}
  Let $A$ be a finitely generated algebra. So $A$ is the quotient of
  some finitely presentable algebra $B$ via the surjective
  homomorphism $q\colon B \to A$. Then $q$ is the coequalizer of its
  kernel pair $f,g\colon K \parallel B$. Since $A$ and $B$ are
  finitely generated so is $K$. Hence, $K$ is a quotient of the
  finitely presentable algebra $L$ via $p\colon L \to K$. As $p$ is an
  epimorphism it follows that $q$ is the coequalizer of $f\o p$ and
  $g\o p$. Since $L$ and $B$ are finitely presentable, and finitely
  presentable objects are closed under finite colimits, also $A$ is
  finitely presentable.
\end{proof}

\begin{remark}
  \label{rem:barFmono}
  \begin{enumerate}[(1)]
  \item
    The lifted functor $\bar F$ on $\Set^T$ preserves monomorphisms
    since monomorphisms are just injective $T$-algebra homomorphisms and
    since $F$ preserves all injective maps (it even preserves weak
    pullbacks by assumption). Thus, the previous lemma ensures that
    $\rho\bar F$ is a subcoalgebra of $\nu \bar F$ (see
    Proposition~\ref{prop:sub})
    and that locally finitely presentable $\bar F$-coalgebras are closed
    under quotients (see Lemma~\ref{lem:lfpquot}).

  \item Actually, the previous lemma holds
    more generally: in any lfp category, where strong epimorphisms
    are regular and finitely generated objects are closed under kernel
    pairs, we have that finitely generated objects are finitely
    presentable.
  \end{enumerate}
\end{remark}

\subsection{Final coalgebras over algebras}
\label{sec:term}

In this subsection we show that the final coalgebra $\nu F$ lifts to a
final coalgebra of $\bar F$, and we prove that $\nu(FT)$ also carries
the structure map of a $\bar F$-coalgebra and that $\nu F$ is a quotient
coalgebra of $\nu(FT)$.

\begin{remark}
  For the results in this subsection rational fixpoints are not
  necessary.  Our results here hold, more generally, for any monad $T$ and any
  endofunctor $F$ on an arbitrary category such that $\nu(FT)$ and
  $\nu F$ exist and $F$ has a lifting to the category of $T$-algebras.
\end{remark}

\begin{notation}
  From now on we write
  \[
  \cFT\colon\nu(FT) \to FT(\nu(FT))
  \qquad
  \text{and}
  \qquad
  \cF \colon \nu F \to F(\nu F),
  \]
  for the structure maps of the final $FT$ and $F$-coalgebra,
  respectively.

  We also write
  \[
  \lambda\colon TF \to FT
  \]
  for the distributive law that (uniquely) corresponds to the lifting $\bar F\colon \Set^T \to \Set^T$.
\end{notation}

First, let us recall that in our setting the final coalgebra for $F$
lifts to a final coalgebra for $\bar F$. This result essentially
follows from the work in~\cite{bartels_thesis} (see Theorem~3.2.3) and
also cf.~\cite{tp}. More explicitly, one obtains the unique coalgebra
homomorphism $\aF\colon T(\nu F) \to \nu F$ as displayed below:
$$
\xymatrix{
  T(\nu F)
  \ar[r]^-{T\cF}
  \ar[d]_\aF
  &
  TF(\nu F)
  \ar[r]^-{\lambda_{\nu F}}
  &
  FT(\nu F)
  \ar[d]^{F\aF}
  \\
  \nu F
  \ar[rr]_-{\cF}
  &&
  F(\nu F)
}
$$
It is then easy to prove that $(\nu F, \aF)$ is an Eilenberg-Moore
algebra for $T$ such that $\cF\colon \nu F \to \bar F(\nu F)$ is a
$T$-algebra homomorphism, and, moreover, $(\nu F, \cF)$ is a final
$\bar F$-coalgebra. So we shall write $\nu F$ for both the
final coalgebras for $F$ and its lifting $\bar F$.

\begin{example}
  We only mention how the terminal coalgebras lifts for the two concrete examples for which we discuss a
  sound and complete expression calculus in Sections~\ref{sec:lin}
  and~\ref{sec:nondet}.  Some
  further examples of the setting of this subsection may be found
  in~\cite{bbrs_fsttcs}.

  \begin{enumerate}[(1)]
  \item In the case of non-deterministic automata we saw that the
    functor $FX = 2 \times X^A$ lifts to $\Set^{\powf}$, and so, the
    final coalgebra for the lifting $\bar F$ is carried by the set of
    formal languages with the join-semilattice structure given
    by union of formal languages.

  \item For the case of weighted automata we saw that the functor $FX
    = \S\times X^A$ lifts to the category $\Set^{V}$ of
    $\S$-semimodules. Hence, the final coalgebra for the lifting $\bar
    F$ is carried by the set $\S^{A^*}$ of weighted languages with the
    canonical (pointwise) structure of a semimodule.
  \end{enumerate}
\end{example}

Next, we want to relate the final coalgebras for $F$ and $FT$.
As a first step we show in the following lemma that the isomorphism
of every fixpoint of $FT$ (and whence the structural map of the final
$FT$ coalgebra), say $c\colon C \stackrel\cong\to FTC$, is a $T$-algebra
homomorphism. In other words, $(C, c)$ can be regarded as an $\bar
F$-coalgebra. Recall that the the generalized powerset constructions
turns $(C,c)$ into the $\bar F$-coalgebra $(TC, \ext c)$ (see
diagram~\refeq{diag:genpow}).

\begin{lemma}
  \label{lem:fix}
  Every fixpoint $(C, c)$ of $FT$ has a unique
  $T$-algebra structure $\gamma\colon TC \to C$ such that $c\colon  C \to FTC$ a $T$-algebra homomorphism.
  Furthermore,
  \[
  \gamma\colon (TC, \ext c) \to (C, F\gamma\o c)
  \]
  is an $\bar F$-coalgebra homomorphism.
\end{lemma}

\begin{proof}
  On $FTC$ we have the $T$-algebra structure
  $$
  \bar F(TC, \mu_C) = (
  \xymatrix@1{
    TFTC
    \ar[r]^-{\lambda_{TC}}
    &
    FTTC
    \ar[r]^-{F\mu_C}
    &
    FTC
    })
  $$
  Since the forgetful functor $U\colon \Set^T \to \Set$ creates
  isomorphisms we have that
  $$
  \gamma = (
    \xymatrix@1{
      TC
      \ar[r]^-{Tc}
      &
      TFTC
      \ar[rr]^-{(F\mu \o \lambda T)_C}
      &&
      FTC
      \ar[r]^-{c^{-1}}
      &
      C
    }),
  $$
  is the unique $T$-algebra structure on $C$ such that $c$ is a
  $T$-algebra homomorphism.
  To see that $\gamma$ is an coalgebra homomorphism for the set
  functor $F$ consider the commutative diagram below:
  \begin{equation}
    \label{diag:Talg}
    \vcenter{
      \xymatrix{
        TC
        \ar[d]_\gamma
        \ar[r]^-{Tc}
        &
        TFTC
        \ar[r]^-{\lambda_{TC}}
        &
        FTTC
        \ar[d]_{F\mu_C}
        \ar[r]^-{F\mu_C}
        &
        FTC
        \ar[d]^-{F\gamma}
        \\
        C
        \ar[rr]_-{c}
        &&
        FTC
        \ar[r]_-{F\gamma}
        &
        FC
      }
    }
  \end{equation}
  Now we will show that the coalgebras in the upper and lower rows are actually $\bar
  F$-coalgebras. Firstly, as just proved, $c$ is a $T$-algebra homomorphism and
  so is the $T$-algebra structure $\gamma$ (by one of the axioms of
  Eilenberg-Moore algebras) whence $F\gamma = \bar F \gamma$ is a
  $T$-algebra homomorphism. This proves $(C, F\gamma\o c)$ to be an
  $\bar F$-coalgebra. Secondly, $Tc$ is clearly a $T$-algebra homomorphism and
  $F\mu_C \o \lambda_{TC}$ is the structure of the $T$-algebra $\bar
  F(TC, \mu_C)$ whence a $T$-algebra homomorphism. This proves that
  the coalgebra in the upper row is an $\bar F$-coalgebra. So $\gamma$
  is an $\bar F$-coalgebra homomorphism.

  We still need to show that the coalgebra structure in the top row
  of~\refeq{diag:Talg} is $\ext c$. This follows from the universal
  property of the free algebra $TC$ by showing that the $T$-algebra
  homomorphism $F\mu_C \o \lambda_{TC} \o Tc$ extends $c$:
  $$
  \begin{array}{rcl@{\qquad}p{3.2cm}}
    F\mu_C \o \lambda_{TC} \o Tc \o \eta_C & = & F\mu_C \o
    \lambda_{TC} \o \eta_{FTC} \o c & naturality of $\eta$, \\
    & = & F\mu_C \o F\eta_{TC} \o c & $\lambda$ a distributive law,
    \\
    & = & c & since $\mu \o \eta T = \id$.
  \end{array}
  $$
  \qed
\end{proof}

From the above lemma we have $T$-algebra structures on the final
coalgebra and on the rational fixpoint for $FT$, and hence, both
can be given structures of $\bar F$-coalgebras. We now fix notation for
these structures for the rest of paper.

\begin{notation}
  \label{not:structs}
  We already fixed the notation $\cFT$ and $\cF$ for the structures of
  the final coalgebras $\nu(FT)$ and $\nu F$, respectively. We will
  henceforth write
  \[
  \rFT\colon \rho(FT) \to FT(\rho(FT))
  \]
  for the structure map of the rational fixpoint of $FT$.
  We also denote the $T$-algebra structures on $\nu(FT)$ and $\rho(FT)$
  obtained from the previous lemma by
  \[
  \aFT\colon T(\nu(FT)) \to \nu (FT)
  \qquad
  \text{and}
  \qquad
  \bFT\colon T(\rho(FT)) \to \rho(FT).
  \]
  Thus, both $\nu(FT)$ and $\rho(FT)$ are $\bar F$-coalgebras with the
  structure maps
  \[
  \xymatrix@1{
    \nu(FT) \ar[r]^-{\cFT} & FT(\nu(FT)) \ar[r]^-{F\aFT} & F(\nu(FT))
  }
  \quad
  \text{and}
  \quad
  \xymatrix@1{
    \rho(FT) \ar[r]^-{\rFT} & FT(\rho(FT)) \ar[r]^-{F\bFT} & F(\rho(FT))
  }
  \]
  Notice also that $\aFT$ and $\bFT$ are $\bar F$-coalgebra
  homomorphisms:
  \[
  \aFT\colon (T\nu(FT), \ext \cFT) \to (\nu(FT), F\aFT \o \cFT)
  \quad
  \text{and}
  \quad
  \bFT\colon (T(\rho(FT), \ext \rFT) \to (\rho(FT), F\bFT \o \rFT).
  \]
\end{notation}

\takeout{
As a corollary of the above lemma, we now have that the final $FT$-coalgebra has a $T$-algebra structure
and hence it is an $\bar F$-coalgebra.

\begin{corollary}
  \label{cor:alg}
  The final coalgebra $\nu(FT)$ has a canonical $T$-algebra
  structure
  \[
  \aFT \colon T(\nu (FT)) \to \nu(FT),
  \]
  and hence, it is an $\bar F$-coalgebra with the structure
  $F\aFT \o \cFT\colon \nu(FT) \to F(\nu FT)$.
\end{corollary}
}

Taking a step further into deepening the understanding of the relation
between final $FT$-coalgebra and the final $\bar F$-coalgebra, we
now prove that the latter is actually a quotient of the former,
meaning it is the codomain of a surjective coalgebra homomorphism.

\begin{proposition}
  \label{prop:quot}
  The final $\bar F$-coalgebra is a quotient coalgebra of the final $FT$-coalgebra.
\end{proposition}
\begin{proof}
  Consider the following $\bar F$-coalgebra homomorphism obtained by
  using the universal property of $\nu F$ (just within this proof we abuse notation and write
  $F$ instead of $\bar F$, and we also write $Z$ for  $\nu F$ and
  $Z_0$ for $ \nu FT$):
  \begin{equation}
    \label{diag:p}
    \vcenter{
      \xymatrix{
        Z_0
        \ar[d]_p
        \ar[r]^-{\cFT}
        &
        FTZ_0
        \ar[r]^-{F\aFT}
        &
        FZ_0
        \ar[d]^{Fp}
        \\
        Z
        \ar[rr]_-{\cF}
        &&
        FZ
      }
    }
  \end{equation}
  Since all horizontal morphisms are $T$-algebra homomorphisms, then so is
  $p\colon Z_0 \to Z$. To see that $p$ is surjective we show it has a
  splitting $s\colon Z \to Z_0$ in $\Set$. To obtain $s$ we use the
  universal property of $Z_0$; there is a unique $FT$-coalgebra
  homomorphism $s$ such that the diagram below commutes:
  \begin{equation}
    \label{diag:s}
    \vcenter{
      \xymatrix{
        Z
        \ar[r]^-{\cF}
        \ar[d]_s
        &
        FZ
        \ar[r]^-{F\eta_Z}
        &
        FTZ
        \ar[d]^{FTs}
        \\
        Z_0
        \ar[rr]_-{\cFT}
        &&
        FTZ_0
      }
    }
  \end{equation}
  To see that $p \o s = \id$ holds, we verify that the following
  diagram commutes:
  $$
  \xymatrix{
    Z
    \ar[d]_s
    \ar[r]^-\cF
    \ar@{}[drr]|{\refeq{diag:s}}
    &
    FZ
    \ar[r]^-{F\eta_Z}
    &
    FTZ
    \ar[r]^-{F\aF}
    \ar[d]_{FTs}
    \ar@{}[dr]|{(*)}
    &
    FZ
    \ar[d]^{Fs}
    \\
    Z_0
    \ar@{}[drrr]|{\refeq{diag:p}}
    \ar[d]_p
    \ar[rr]_-{\cFT}
    &&
    FTZ_0
    \ar[r]_-{F\aFT}
    &
    FZ_0
    \ar[d]^{Fp}
    \\
    Z
    \ar[rrr]_-\cF
    &&&
    FZ
    }
  $$
  Indeed, the upper left-hand and lower parts commute as indicated,
  but we do not claim that part $(*)$ commutes. This part commutes
  when precomposed with $F\eta_Z$; to see this remove $F$ and consider
  $$
  \xymatrix{
    Z
    \ar[r]_-{\eta_Z}
    \ar[d]_s
    &
    TZ
    \ar[r]_-\aF
    \ar[d]_{Ts}
    &
    Z
    \ar[d]^s
    \ar@{<-} `u[l] `[ll]_{\id} [ll]
    \\
    Z_0
    \ar[r]^-{\eta_{Z_0}}
    &
    TZ_0
    \ar[r]^-{\aFT}
    &
    Z_0
    \ar@{<-} `d[l] `[ll]^{\id} [ll]
    }
  $$
  where the left-hand square commutes by the naturality of $\eta$ and
  the upper and lower triangle by the unit law of $T$-algebras. \qed
\end{proof}

From the previous theorem we obtain an alternative way to define, for
a given $FT$-coalgebra $(C, c)$, a coalgebraic language map $C \to \nu
F$, namely as
\[
\xymatrix@1{C \ar[r]^-{\fin c} & \nu(FT) \ar@{->>}[r]^-p & \nu F},
\]
where $\fin c$ is the unique $FT$-coalgebra homomorphism. We shall now
prove that this map coincides with the coalgebraic language map
$\lang c\colon C \to \nu F$ from Notation~\ref{not:ddagger}. We first
prove that $FT$-coalgebra homomorphisms preserve coalgebraic language
equivalence:
\begin{lemma}
  \label{lem:ddagger}
  Let $h\colon (C,c) \to (D,d)$ be an $FT$-coalgebra
  homomorphism. Then we have
  \[
  \lang c = (\xymatrix@1{C \ar[r]^-h & D \ar[r]^-{\lang d} & \nu F}).
  \]
\end{lemma}
\begin{proof}
  Given $h\colon (C,c) \to (D,d)$ we prove that $Th\colon TC \to TD$ is an
  $F$-coalgebra homomorphism. To see this consider the following
  diagram:
  \[
  \xymatrix{
    TC
    \ar[ddd]_{Th}
    \ar[rr]^-{\ext c}
    &&
    FTC
    \ar[ddd]^{FTh}
    \\
    &
    C
    \ar[ru]^-c
    \ar[lu]_-{\eta_C}
    \ar[d]_h
    \\
    &
    D
    \ar[rd]_-d
    \ar[ld]^-{\eta_D}
    \\
    TD
    \ar[rr]_-{\ext d}
    &&
    FTD
    }
  \]
  The outside of the diagram consists of $T$-algebra
  homomorphisms. By the freeness of $TC$ it suffices to show that it
  commutes when extended by $\eta_C$, and this follows from the
  commutativity of the inner parts; to see this, use naturality of $\eta$ and
  the fact that $h$ is an $FT$-coalgebra homomorphism.

  Now by the uniqueness of coalgebra homomorphisms into $\nu F$ we
  have $\fin{(\ext d)} \o Th = \fin{(\ext c)}$, and so we conclude
  \[
  \begin{array}{rclp{4cm}}
    \lang d \o h & = & \fin{(\ext d)} \o \eta_D \o h & definition
    of $\lang d$ \\
    & = & \fin{(\ext d)} \o Th \o \eta_C & naturality of $\eta$ \\
    & = & \fin{(\ext c)} \o \eta_C & from above \\
    & = & \lang c & definition of $\lang c$.
  \end{array}
  \]
\qed
\end{proof}

\begin{proposition}
  \label{prop:power}
  Let $(C,c)$ be an $FT$-coalgebra. Then we have
  \[
  \xymatrix{
    C \ar[r]^-{\fin c}
    \ar[rd]_{\lang c}
    &
    \nu(FT)
    \ar@{->>}[d]^p
    \\
    &
    \nu F
    }
  \]
  where $\fin c$ is the unique $FT$-coalgebra homomorphism
\end{proposition}
\begin{proof}
  (1)~We first prove that for the final $FT$-coalgebra $(\nu(FT), \cFT)$ we
  have
  \[
  \lang\cFT = p\colon \nu(FT) \to \nu F.
  \]
  The desired equation follows from the following one by precomposing
  with $\eta_{\nu(FT)}$:
  \[
  \fin{(\ext \cFT)} = (\xymatrix@1{T(\nu(FT))\ar[r]^-{\aFT} & \nu(FT) \ar[r]^-p &
    \nu F}),
  \]
  where $\aFT$ is the $T$-algebra structure of $\nu(FT)$ and $\fin{(\ext \cFT)} \colon
  (\nu(FT), \ext \cFT) \to (\nu F, \cF)$ the unique $F$-coalgebra
  homomorphism (cf.~Diagram~\refeq{diag:genpow}). Now to establish the latter
  equation one proves that $p \o \aFT$ is an $F$-coalgebra
  homomorphism:
  \[
  \xymatrix{
   T(\nu(FT))
   \ar[d]_{\aFT}
   \ar[r]^-{\ext \cFT}
   &
   FT(\nu(FT))
   \ar[d]^{F\aFT}
   \\
   \nu(FT)
   \ar[ru]_-{\cFT}
   \ar[d]_p
   &
   F(\nu(FT))
   \ar[d]^{Fp}
   \\
   \nu F
   \ar[r]_-{\cF}
   &
   F(\nu F)
  }
  \]
  The lower part commutes by the definition of $p$ in~\refeq{diag:p}, and
  for the upper triangle one uses that both $\ext \cFT$ and $\cFT \o
  \aFT$ are $T$-algebra homomorphisms that are equal when composed
  with $\eta_{\nu(FT)}$, in symbols:
  \[
  \ext \cFT \o \eta_{\nu(FT)} = \cFT = \cFT \o \aFT \o \eta_{\nu(FT)}.
  \]
  It follows that the outside of the above diagram
  commutes as desired.

  (2)~We are now ready to prove the statement of the proposition. Given
  the $FT$-coalgebra $(C,c)$ we use item~(1) above and Lemma~\ref{lem:ddagger} to conclude
  that
  \[
  \lang c = \lang \cFT \o \fin c = p \o \fin c,
  \]
  which completes the proof.
\qed
\end{proof}

As a corollary we get the main result of~\cite{bbrs_fsttcs} that
behavioral equivalence implies coalgebraic language equivalence.
\begin{corollary}
  Let $(C,c)$ and $(D,d)$ be $FT$-coalgebras. Then for every $x \in C$
  and $y \in D$ we have
  \[
  \fin c(x) = \fin d(y) \quad \Longrightarrow \quad \lang c(x) =
  \lang d(y).
  \]
\end{corollary}

We have now a formal relation between the final coalgebras for $FT$
and for $\bar F$. These final coalgebras contain, respectively,
canonical representatives for bisimilarity and language
equivalence. As demonstrated by the previous corollary, the abstract
result that the final coalgebra for $\bar F$ is a quotient of the
final coalgebra for $FT$ instantiates, for non-deterministic automata
and labelled transition systems, to the well-known fact that language
(or trace) equivalence is coarser than bisimilarity. Similarly, in the
case of weighted automata, we have that weighted bisimilarity implies
weighted language equivalence.

\takeout{
In the next two sections, we will lay down the foundations to help us
in the main quest of this paper: find a sound and complete
axiomatisation of weighted language equivalence and later
 of (coalgebraic) language equivalence for a large
class of systems in a uniform manner.
}

\subsection{Locally Finitely Presentable Coalgebras over Algebras}
\label{sec:coalg_on_alg}

The aim of this subsection is to establish our main result
concerning final coalgebras and rational fixpoints as explained in
the introduction (see~\refeq{eq:square}). In fact, we will show that
the rational fixpoint $\rho \bar F$ is a quotient of the rational
fixpoint $\rho(FT)$, thus these rational fixpoints share the same relationship that
we saw for the corresponding final coalgebras in Proposition~\ref{prop:quot}.
In addition, we will use the fact that we work with algebras to improve
on the finality criterium for $\rho \bar F$ from Theorem~\ref{thm:lfpcoalg}(2),
and we show that $\rho \bar F$ can be constructed from only those coalgebras
with a {\em free} finitely presentable carrier.

These results lay down the foundation for the work in the subsequent
sections: establishing that soundness and completeness proofs amount to
proving that expressions modulo axioms of a calculus are isomorphic to
$\rho\bar F$ (Section~\ref{sec:general}) and the application of this
to our calculus for weighted language equivalence in Section~\ref{sec:lin}.

\begin{remark}
Let us collect some facts which are true for every finitary monad $T$ on $\Set$ and
every functor $F$ having a lifting to $\Set^T$ and that we will subsequently
need in our proofs.
  \label{rem:fp}

 \noindent (1)~Every free algebra $TX$ is
  \emph{projective}:   for every (strong) epimorphism $q\colon A \to
  B$ in $\Set^T$ (i.\,e., $q$ is a surjective homomorphism) and every
  $T$-algebra homomorphism $f\colon TX \to B$ there exists a homomorphism
  $g\colon TX \to A$ such that $q \o g = f$:
  $$
  \xymatrix{
    TX
    \ar[rd]_{f}
    \ar[r]^g & A
    \ar@{->>}[d]^q
    \\
    &
    B
    }
  $$
 Since $q$ is surjective we have a (not necessarily
  homomorphic) map $s\colon B \to A$ with
  $q \o s = \id$. Then we use the freeness of $TX$ to extend the map $s
  \o f \o \eta_X\colon X \to A$ to the homomorphism $g\colon TX \to A$, which has
  the desired property.

  \medskip \noindent
  (2)~As we mentioned already, finitely presentable algebras are precisely
  those algebras that are presentable by finitely many generators and
  relations. In category theoretic terms, an algebra $A$ is finitely
  presentable if and only if it is the (reflexive) coequalizer of a parallel pair $f, g\colon TX
  \to TY$ of homomorphism between \emph{free} finitely presentable
  algebras, i.\,e., free algebras on the finite sets
  $X$ and $Y$ (cf.~\cite[Proposition~5.17]{arv}).

  \medskip\noindent
  (3)~The monad $T$ yields a functor $T'\colon \Coalg(FT) \to
  \Coalg(\bar F)$; it assigns to every $FT$-coalgebra $c\colon  X \to FTX$
  the coalgebra $\ext c\colon TX \to FTX$ obtained by the generalized
  powerset construction, and
  on morphisms $T'$ acts like $T$. It is easy to see that
  $T'$ is finitary; this follows essentially from the fact that
  the filtered colimits in $\Coalg(FT)$ and $\Coalg(\bar F)$ are
  formed on the level of $\Set$ (since the forgetful functors of
  $\Coalg(FT)$, $\Coalg(\bar F)$ and $\Set^T$ create filtered
  colimits).
\end{remark}

\begin{notation}
  \label{not:D}
  We denote by
  \[
  \Coalgfree(\bar F)
  \]
  the full subcategory of $\Coalgf(\bar F)$ given by
  coalgebras with a \emph{free} finitely presentable carrier. That means
  that the objects of $\Coalgfree(\bar F)$ are of the form $TX \to FTX$ with $X$ a
  finite set.
\end{notation}

\begin{remark}
  Observe that the objects of $\Coalgfree(\bar F)$ are precisely the
  results of applying the generalized powerset construction in
  Section~\ref{sec:monad} to every finite coalgebra $c\colon X \to
  FTX$. Indeed, $T'(X,c) = (TX, \ext c)$ lies in $\Coalgfree(\bar F)$,
  and, conversely, for every $d\colon TX \to FTX$ in $\Coalgfree(\bar F)$
  it is easy to see that $d = \ext{(d \o \eta_X)}$.
\end{remark}

\begin{lemma}
  \label{lem:coprod}
  The category $\Coalgfree(\bar F)$ is closed in $\Coalg(F)$ under finite coproducts.
\end{lemma}
\begin{proof}
  The empty $FT$-coalgebra $0 \to FT0$ extends uniquely to an $\bar
  F$-coalgebra $T0 \to FT0$, and this is the initial object of $\Coalgfree(\bar F)$.

  Let $\ext c\colon TX \to FTX$ and $\ext d\colon TY \to FTY$ be objects of
  $\Coalgfree(\bar F)$ with the corresponding $FT$-coalgebras $c\colon  X \to FTX$ and $d\colon
  Y \to FTY$. Now form
  $$
  k = (\xymatrix@1{
    X + Y \ar[r]^-{c + d} & FTX + FTY \ar[r]^-{\can} & FT(X+Y)
    }),
  $$
  where $\can = [FT\inl, FT\inr]$,
  and extend $k$ to the $T$-algebra homomorphism
  \[\ext k\colon T(X+Y) \to FT(X+Y).\]
  It is not difficult to verify that this $\bar F$-coalgebra
  is the coproduct of $(TX, \ext c)$ and $(TY, \ext d)$ in
  $\Coalgfree(\bar F)$. To see this, first verify that $T\inl\colon TX \to T(X+Y)$ and
  $T\inr\colon TY \to T(X+Y)$ are $\bar F$-coalgebra homomorphisms. Next we
  show that they serve as the
  coproduct injections. Suppose we have two $\bar F$-coalgebra
  homomorphisms $f\colon (TX, \ext c) \to (A, a)$ and $g\colon (TY, \ext d)
  \to (A, a)$. Let $f_0 = f \o \eta_X$ and $g_0 = g \o
  \eta_Y$. Now extend the morphism $h_0 = [f_0, g_0]\colon X + Y \to A$ to
  a $T$-algebra homomorphism $h\colon T(X+Y) \to A$. Then one readily
  verifies using the universal properties of free $T$-algebras that
  $h$ is the unique $\bar F$-coalgebra homomorphism from $(T(X+Y), \ext k)$ to
  $(A,a)$ such that $h \o T\inl = f$ and $h \o T\inr = g$. \qed
\end{proof}

The next proposition is the key to the main results of this
section. It uses the full strength of our
Assumption~\ref{ass:setT}, in particular that finitely generated
algebras are closed under taking kernel pairs.

\begin{proposition}
  \label{prop:coeq}
  Every coalgebra in $\Coalgf(\bar F)$ is the coequalizer of a pair of
  morphisms in $\Coalgfree(\bar F)$.
\end{proposition}
\begin{proof}
  Let $a\colon A \to \bar F A$ be a coalgebra from $\Coalgf(\bar F)$,
  so $A$ is a finitely presentable $T$-algebra. From
  Remark~\ref{rem:fp}(2) we recall that $A$ is the coequalizer of some
  pair $TX' \parallel TX$ of $T$-algebra homomorphisms with $X'$ and
  $X$ finite sets via some $q\colon TX \to
  A$. Being a functor on $\Set$, $F$ preserves epimorphisms. Thus,
  $\bar F q$ is a strong epimorphism in $\Set^T$. Now we use that $TX$
  is projective to obtain a coalgebra structure $c\colon  TX \to FTX$
  as displayed below:
  \begin{equation}
    \label{diag:proj}
    \vcenter{
      \xymatrix{
        TX
        \ar@{-->}[r]^-{c}
        \ar[d]_q
        &
        F TX
        \ar@{->>}[d]^{Fq}
        \\
        A
        \ar[r]_-a
        &
        FA
      }
    }
  \end{equation}
  Now since $\Set^T$ is a category with pullbacks we know that every
  coequalizer in that category is the coequalizer of its kernel
  pair. So let $f,g\colon K \to TX$ be the kernel pair of $q$ in $\Set^T$.
  Notice that since $TX$ and
  $A$ are finitely presentable $T$-algebras, so is $K$ because
  finitely presentable (equivalently, finitely generated) $T$-algebras
  are closed under taking kernel pairs by Assumption~\ref{ass:setT}.
  Since the forgetful functor $\Set^T \to \Set$ preserves limits we have a pullback in $\Set$, and
  since $F$ weakly preserves pullbacks $Ff, Fg$ form a weak pullback
  of $Fq$ along itself in $\Set$. Thus, we have a map $k\colon K \to FK$ such that
  the diagram below commutes:
  \begin{equation}
    \label{diag:kappa}
    \vcenter{
    \xymatrix{
      K
      \ar@<-.5ex>[d]_f
      \ar@<.5ex>[d]^g
      \ar@{-->}[r]^-k
      &
      FK
      \ar@<.5ex>[d]^{Fg}
      \ar@<-.5ex>[d]_{Ff}
      \\
      TX
      \ar[d]_q
      \ar[r]^-c
      &
      FTX
      \ar[d]^{Fq}
      \\
      A
      \ar[r]_-a
      &
      FA
      }}
  \end{equation}
  Notice that we do not claim that $k$ is a $T$-algebra
  homomorphism. However, since $K$ is a finitely presentable
  $T$-algebra it is the coequalizer of some pair $TY' \parallel TY$ of
  $T$-algebra homomorphisms, $Y'$ and $Y$ finite, via
  $p\colon TY \to K$. Now we choose some splitting $s\colon K \to TY$ of $p$ in
  $\Set$, i.\,e., $s$ is a map such that $p \o s = \id$. Next we extend the map $d_0 =
  Fs \o k \o p \o \eta_Y$ to a $T$-algebra homomorphism $d\colon  TY
  \to FTY$:
  \begin{equation}
    \label{diag:d}
    \vcenter{
      \xymatrix{
        Y
        \ar[d]_{\eta_Y}
        \ar[rd]^{d_0}
        \\
        TY
        \ar@{-->}[r]^-d
        \ar[d]_p
        &
        FTY
        \ar[d]^{Fp}
        \\
        K
        \ar[r]_-k
        &
        FK
      }
    }
  \end{equation}
  (Notice that to obtain $d$ we cannot simply use projectivity of $TY$
  similarly as in~\refeq{diag:proj} since $k$ is not necessarily a
  $T$-algebra homomorphism.)

  We do not claim that this makes $p$ a coalgebra homomorphism
  (i.\,e., we do not claim the lower square in~\refeq{diag:d} commutes). However, $f\o p$
  and $g \o p$ are $\bar F$-coalgebra homomorphisms from $(TY, d)$ to
  $(TX, c)$. To see that
  $$
  c \o (f \o p) =  F(f \o p) \o d
  $$
  it suffices that this equation of $T$-algebra homomorphisms holds
  when both sides are precomposed with $\eta_Y$. To see this we compute
  $$
  \begin{array}{rcl@{\qquad}p{3cm}}
    c \o f \o p \o \eta_Y & = & Ff \o k \o p \o \eta_Y &
    see~\refeq{diag:kappa}, \\
    & = & Ff \o Fp \o d_0 & outside of~\refeq{diag:d},
    \\
    & = & Ff \o Fp \o d \o \eta_Y & definition of $d$.
  \end{array}
  $$
  Similarly, $g \o p$ is a coalgebra homomorphism. Since $p$ is an
  epimorphism in $\Set^T$ it follows that $q$ is a coequalizer of $f
  \o p$ and $g \o p$. Thus $f\o p$ and $g \o p$ form the desired pair
  of morphisms in $\Coalgfree(\bar F)$ such that $(A, a)$ is a coequalizer of
  them, which completes the proof. \qed
\end{proof}

As a consequence of the previous proposition we obtain that the
rational fixpoint $\rho \bar F$ can be constructed just using those
coalgebras obtained by applying the generalized powerset construction
to finite $FT$-coalgebras.

\begin{corollary}
  \label{cor:colim}
  The rational fixpoint of $\bar F$ is the colimit of all
  coalgebras in $\Coalgfree(\bar F)$; in symbols:
  \[
  \rho \bar F = \colim(\Coalgfree(\bar F) \subto  \Coalg(\bar F)).
  \]
\end{corollary}
\begin{proof}
  We first show that $\Coalgf(\bar F)$ is the closure of $\Coalgfree(\bar F)$ under
  coequalizers in the category $\Coalg(\bar F)$. Since finitely
  presentable algebras are closed under finite colimits and finite colimits
  in $\Coalgf(\bar F)$ are formed on the level of $\Set^T$ we clearly
  see that the closure of $\Coalgfree(\bar F)$ under coequalizers is a subcategory of
  $\Coalgf(\bar F)$. But, by the previous proposition, each object of
  $\Coalgf(\bar F)$ is a coequalizer of some parallel pair of morphisms
  from $\Coalgfree(\bar F)$, which establishes the desired statement.

  It is easy to prove that the colimit of $\Coalgfree(\bar F)$ and
  the filtered colimit of its closure under coequalizers coincide. But
  the latter is $\rho\bar F$ by Corollary~\ref{cor:colimrho}.\qed
\end{proof}

Furthermore, and playing a crucial r\^ole in simplifying our proof
burden for completeness later, we have that a locally finitely presentable coalgebra
$r\colon R\to \bar FR$ is final for \emph{all} locally finitely
presentable coalgebras if there is a unique homomorphism from those
coalgebras whose carrier is {\em free on a finite set} to  $(R,r)$. This means that
when proving finality of $(R,r)$ one does not need to show the existence
of a unique homomorphism for all coalgebras but only for the much
smaller class of coalgebras from $\Coalgfree(\bar F)$.
\begin{corollary}
  \label{cor:final}
  A locally finitely presentable $\bar F$-coalgebra $(R, r)$ is
  final in the category of all locally finitely presentable $\bar
  F$-coalgebras if and only if for every coalgebra $(TX, \ext c)$ from $\Coalgfree(\bar F)$ there
  exists a unique coalgebra homomorphism from $(TX, \ext c)$ to
  $(R,r)$.
\end{corollary}
\begin{proof}
  Necessity of a unique coalgebra homomorphism from each $(TX, \ext
  c)$ to $(R,r)$ is clear. For sufficiency let $a\colon A \to FA$ be a
  coalgebra in $\Coalgf(\bar F)$. By Proposition~\ref{prop:coeq}, we
  have a coequalizer diagram
  \[
  \xymatrix{
    (TX, \ext c)
    \ar@<3pt>[r]^-f
    \ar@<-3pt>[r]_-g
    &
    (TY, \ext d)
    \ar[r]^-{q}
    &
    (A,a),
    }
  \]
  with $(TX, \ext c)$ and $(TY, \ext d)$ in $\Coalgfree(\bar F)$. The unique coalgebra
  homomorphism $h\colon (TY, \ext d) \to (R,r)$ satisfies $h \o f = h \o g$ since
  both of these are coalgebra homomorphisms from $(TX, \ext c)$ to
  $(R,r)$. So by the universal property of the coequalizer we get a
  unique coalgebra homomorphism $k\colon (A,a) \to (R,r)$. The desired
  result now follows from Theorem~\ref{thm:lfpcoalg}(2).
\end{proof}

We are now ready to relate the rational fixpoints of $FT$ and $\bar
F$. Recall the congruence quotient $p\colon\nu(FT) \to \nu F$ from
Proposition~\ref{prop:quot} and notice that the rational fixpoint
$\rho(FT)$ is a subcoalgebra of $\nu(FT)$ (see
Proposition~\ref{prop:sub}). From our assumptions we also know that
$\rho\bar F$ is a subcoalgebra of $\nu F$ (recall from
Section~\ref{sec:term} that $\nu F$ denotes the final $\bar
F$-coalgebra).

\begin{notation}
  \label{not:i}
  We denote the corresponding inclusion homomorphisms by
  \[
  i\colon \rho(FT) \to \nu(FT)\qquad\text{and}\qquad j\colon \rho\bar F \to \nu
  F.
  \]
\end{notation}

Furthermore, recall from Notation~\ref{not:structs} that $\rho(FT)$ is an
$\bar F$-coalgebra with the structure $\rcomp$, where $\rFT$ is the coalgebra map
of the rational fixpoint $\rho(FT)$, and $\bFT$ its $T$-algebra structure.

\begin{lemma}
  \label{lem:lfp}
  The coalgebra
  \[
  \xymatrix@1{
    \rho(FT) \ar[r]^-{\rFT} & FT(\rho(FT)) \ar[r]^-{F\bFT} & F(\rho(FT))
  }
  \]
  is a locally finitely presentable $\bar F$-coalgebra.
\end{lemma}
\begin{proof}
  By Theorem~\ref{thm:lfpcoalg}(1) the coalgebra $(\rho(FT), \rFT)$ is the
  filtered colimit of the inclusion functor $I\colon \Coalgf(FT) \subto
  \Coalg(FT)$. The finitary functor $T'\colon \Coalg(FT) \to
  \Coalg(\bar F)$ from Remark~\ref{rem:fp}(3) preserves this colimit,
  and so the coalgebra $T'(\rho(FT), \rFT) = (T(\rho(FT), \ext{\rFT})$ is the filtered colimit of the
  diagram of all $\bar F$-coalgebras $T'(C, c) = (TC, \ext c)$.
  (Notice that the corresponding diagram scheme contains the same objects but fewer
  connecting morphisms than $\Coalgfree(\bar F)$ from
  Notation~\ref{not:D}---here we consider only the morphisms $Th$ for
  $h$ an $FT$-coalgebra homomorphism.)

  Thus, since the carrier of every object in this diagram is a finitely presentable
  algebra, we can apply Theorem~\ref{thm:lfpcoalg} to conclude
  that $(T(\rho(FT)), \ext{\rFT})$ is a locally finitely presentable
  coalgebra. We also know that
  \[
  \bFT\colon (T(\rho(FT)), \ext{\rFT}) \to (\rho(FT), F\bFT \o \rFT)
  \]
  is a homomorphism of $\bar F$-coalgebras (see
  Notation~\ref{not:structs}). This is a strong epimorphism in
  $\Set^T$ (because $\bFT \o  \eta_{\rho(FT)} = \id$). Hence, being a
  quotient of a coalgebra that is locally finitely presentable,
  $(\rho(FT), F\bFT \o \rFT)$ also has that property (see Lemma~\ref{lem:lfpquot}). \qed
\end{proof}

Next, we show that the coalgebra $(\rho(FT), \rcomp)$  is weakly
final\footnote{A \emph{weakly final} object in a category is an object
  $W$ such that for every object $X$ there exists a (not necessarily
  unique) morphism $X \to W$.}
among the locally finitely presentable $\bar F$-coalgebras.

\begin{lemma}
  \label{lem:weak}
  For every locally finitely presentable $\bar F$-coalgebra there
  exists a canonical homomorphism into the coalgebra $(\rho(FT),
  F\bFT\o \rFT)$.
\end{lemma}
\begin{proof}
  It suffices to show the statement for every coalgebra from $\Coalgfree(\bar F)$. It
  then follows that every coalgebra from $\Coalgf(\bar F)$ (being a
  coequalizer of a pair of morphisms in $\Coalgfree(\bar F)$) admits a
  homomorphism into $\rho(FT)$. Hence, every filtered colimit of
  coalgebras from $\Coalgf(\bar F)$ admits a homomorphism into
  $\rho(FT)$.

  Now suppose we are given $\ext c\colon TX \to FTX$ from $\Coalgfree(\bar F)$. Consider the corresponding
  $FT$-coalgebra $c\colon  X \to FTX$. Since $X$ is
  a finite set we obtain a unique $FT$-coalgebra homomorphism $h$ from
  $(X,c)$ to the final locally finite coalgebra $\rho(FT)$:
  $$
  \xymatrix{
    X
    \ar[r]^-{c}
    \ar[d]_{h}
    &
    FTX
    \ar[d]^{FTh}
    \\
    \rho(FT)
    \ar[r]_-{\rFT}
    &
    FT(\rho(FT))
  }
  $$
  We apply the functor $T'\colon \Coalg(FT) \to \Coalg(\bar F)$ to obtain
  an $\bar F$-coalgebra homomorphism $Th$ from $(TX, \ext c)$ to $(TR,
  \ext{\rFT})$. Then compose with the $\bar F$-coalgebra homomorphisms
  $\bFT\colon T(\rho(FT) \to \rho(FT)$ to obtain the desired
  coalgebra homomorphism from $(TX, \ext c)$ to $(\rho(FT), \bFT \o \rFT)$.
  \qed
\end{proof}

As a consequence of the previous lemma, also every quotient
of $(\rho(FT), \rcomp)$ is weakly final among the locally finitely
presentable $\bar F$-coalgebras:

\begin{corollary}
  \label{cor:weak}
  Every quotient coalgebra of $(\rho(FT),F\bFT\o \rFT)$ admits a homomorphism
  from every locally finitely presentable coalgebra for $\bar F$.
\end{corollary}

At last, we can state the formal relation between the rational fixpoints of $FT$ and $\bar F$:

\begin{theorem}
  \label{thm:ratquot}
  The rational fixpoint of $\bar F$ is the image of $\rho(FT)$ under the
  quotient $p\colon \nu(FT) \to \nu F$ from Proposition~\ref{prop:quot}, that is,
  there is a surjective $\bar F$-coalgebra homomorphism $q\colon \rho(FT) \to \rho\bar F$ such
  that the following square commutes (using Notation~\ref{not:i}):
$$
\xymatrix{
  \rho(FT)
  \ar@{ >->}[r]^-i
  \ar@{->>}[d]_q
  &
  \nu(FT)
  \ar@{->>}[d]^p \\
  \rho \bar F
  \ar@{ >->}[r]_-j
  &
  \nu F
}
$$
\end{theorem}
\begin{proof}
We first need to verify that $i\colon \rho(FT) \to \nu(FT)$ is a
homomorphism of $\bar F$-coalgebras. By definition we have $\cFT \o i
= FTi \o \rFT$, and since $\cFT$ is invertible we get:
\[
i = \cFT^{-1} \o FTi \o \rFT.
\]
Thus, $i$ is a $T$-algebra homomorphism since all three morphisms on
the right-hand side of the above equation are. Now the following
diagram commutes as desired:
\[
\xymatrix{
  \rho(FT)
  \ar[d]_i
  \ar[r]^-{\rFT}
  &
  FT(\rho(FT))
  \ar[r]^-{F\bFT}
  \ar[d]^{FTi}
  &
  F(\rho(FT))
  \ar[d]^{Fi}
  \\
  \nu(FT)
  \ar[r]_-{\cFT}
  &
  FT(\nu(FT))
  \ar[r]_-{F\aFT}
  &
  F(\nu(FT))
}
\]

Let $I$ be the image in $\nu F$ of $\rho(FT)$ under $p$, i.\,e., we
take the image factorisation $m \o e$ of $p \o i$. Then $I$ is a
sub-$T$-algebra of $\nu F$.
Since $\bar F$ preserves monomorphisms (cf.~Remark~\ref{rem:barFmono}), it follows that
$I$ carries the structure $z\colon I \to \bar F I$ of an $\bar F$-coalgebra making it a
subcoalgebra of $\nu F$ (see Remark~\ref{rem:fs}(2)). We will prove that $I$ is the final
locally finitely presentable $\bar F$-coalgebra and a quotient
coalgebra of $(\rho(FT), F\bFT \o \rFT)$.

Firstly, by an application of Lemma~\ref{lem:lfpquot} we see that the
quotient $(I,z)$ is locally finitely presentable since the coalgebra
$(\rho(FT), F\bFT \o \rFT)$ also has this property (see
Lemma~\ref{lem:lfp}). Thus, by Corollary~\ref{cor:final}, we only need to prove that for every $\bar
F$-coalgebra $\ext c\colon TX \to FTX$ from the category $\Coalgfree(\bar F)$ there exists
a unique coalgebra homomorphism from $(TX, \ext c)$ to $(I,z)$.
Since $(I,z)$ is a subcoalgebra of
the final $\bar F$-coalgebra $\nu F$ the uniqueness of a
homomorphism is clear, and the existence of a homomorphism is clear
since $(I,z)$ is a quotient coalgebra of $(\rho(FT), \rcomp)$ (use
Corollary~\ref{cor:weak}).

This proves that $I \cong \rho\bar F$ and composing this isomorphism
with $m$ and $e$ yields $q$ and $j$ as displayed in the square above.
\qed
\end{proof}

Let us summarize the four fixpoints from the previous theorem and
their coalgebra structures in one picture for future reference:
\[
\xymatrix{
  &
  F\rho(FT)
  \ar[rr]^-{Fi}
  \ar'[d]_(.8){Fq}[dd]
  &&
  F\nu(FT)
  \ar[dd]^{Fp}
  \\
  \rho(FT)
  \ar@{ >->}[rr]^(.6)i
  \ar@{->>}[dd]_q
  \ar[ru]^(.4){\rcomp}
  &&
  \nu(FT)
  \ar@{->>}[dd]^(.6)p
  \ar[ru]^(.4){F\aFT \o \cFT}
  \\
  &
  F\rho\bar F
  \ar'[r]^(.8){Fj}[rr]
  &&
  F\nu F
  \\
  \rho \bar F
  \ar@{ >->}[rr]_-j
  \ar[ru]^-{\rF}
  &&
  \nu F
  \ar[ru]_-\cF
}
\]
From Proposition~\ref{prop:power} we also see that the quotient map $p$ is the coalgebraic language map
$\lang\cFT$ for the final $FT$-coalgebra $\cFT\colon \nu(FT) \to
FT(\nu(FT))$ and the diagonal of the front square is
$\lang\rFT$ for the final locally finite $FT$-coalgebra $\rFT\colon
\rho(FT) \to FT(\rho(FT))$:
\[
\xymatrix{
  \rho(FT)
  \ar@{ >->}[r]^-i
  \ar@{->>}[d]_q
  \ar[rd]^-{\lang\rFT}
  &
  \nu(FT)
  \ar@{->>}[d]^p \\
  \rho \bar F
  \ar@{ >->}[r]_-j
  &
  \nu F
}
\]

In this section, we have developed the theory of locally finitely
presentable coalgebras (over algebras). All the abstract work and
results in this section will play a prominent r\^ole in the rest of
the paper; they enable stating and proving a Kleene like theorem and
soundness and completeness of axiomatisation results for coalgebraic
language equivalence, for a large class of systems, uniformly. We will
demonstrate this with our calculus for weighted automata in
Section~\ref{sec:lin}. The first pay-off of this abstract work appears
immediately in the next section, where we will narrow down what proof
obligations one has after extending a sound and complete calculus for
bisimilarity with extra axioms in order to guarantee that the resulting
calculus is sound and complete with respect to (coalgebraic) language
equivalence.

\section{Soundness, Completeness and Kleene's theorem in general}
\label{sec:general}

In this section we obtain a generalisation of Kleene's classical
theorem from automata theory~\cite{kleene} to the setting of
$FT$-coalgebras as presented in Section~\ref{sec:algcoalg}. We also
present generic coalgebraic formulations of soundness and completeness
of an expression calculus
that we will then instantiate in the concrete example of weighted
automata in the next section. The goal is to push as much work as
possible to the present abstract setting and only do the minimal
necessary amount of work in concrete instances.

We still work in the setting as described in
Assumption~\ref{ass:setT}. Thus we consider a finitary endofunctor
$F\colon\Set \to \Set$ that weakly preserves pullbacks and has a
lifting $\bar F\colon \Set^T \to \Set^T$, for a finitary monad $(T, \eta, \mu)$
such that in $\Set^T$ finitely generated algebras are closed under taking
kernel pairs.

Let us first consider our two leading
examples. For the functor $FX = 2 \times X^A$
and the monad $T = \powf$ consider the expression calculus obtained
from (the structure of) the functor $FT$; we recalled the syntax in
the introduction. Let $\Exp$ denote the \emph{closed} syntactic
expressions, i.\,e., those expressions in which every variable is
bound by a $\mu$-operator, and let $\equiv$ be the least equivalence
on $\Exp$ generated by the proof rules of the calculus. Then, as
proved in~\cite{brs_lmcs}, $\Expmod$ is isomorphic to $\rho(F\powf)$.

Similarly, for the semiring $\S$, $FX = \S \times X^A$ and $T = V$ one can
define an expression calculus with closed syntactic expressions $\Exp$, and
proof rules such that $\Expmod$ is isomorphic to $\rho(FV)$ (see~\cite{bbrs_ic}).

In each case we write $q_0\colon \Exp \to \Expmod$ for the canonical
quotient map. This motivates the following definition.

\begin{definition}
  \label{dfn:exprcalc}
  We call a set $\Exp$ with a surjective map $q_0\colon \Exp \to \rho(FT)$
  an (abstract) \emph{expression calculus} (for $FT$). The elements of
  $\Exp$ are referred to as \emph{expressions}. \takeout{We say that two
  expressions $E$ and $F$ are \emph{provably equivalent} if $q_0(E) =
  q_0(F)$.\footnote{Notice that
    these definitions make sense for every finitary
    functor $G$ on a concrete locally finitely presentable category.}}
\end{definition}

Besides the $FT$-bisimilarity semantics from~\cite{brs_lmcs,bbrs_ic}
for which the calculi from the introduction are sound and complete,
there is a different semantics that we now introduce.

Let us fix an expression calculus $q_0\colon \Exp \to \rho(FT)$ for the
rest of this section. Then we see
that every expression $E$ in $\Exp$ denotes an element $\lsem E\rsem$
of the final coalgebra $\nu F$. More precisely, the semantics function
$\lsem -\rsem\colon \Exp \to \nu F$ is defined by
\begin{equation}
  \label{eq:sem}
  \lsem -\rsem = (
  \xymatrix@1{
    \Exp
    \ar@{->>}[r]^-{q_0}
    &
    \rho(FT)
    \ar[r]^-{\lang\rFT}
    &
    \nu F
  }
  ),
\end{equation}
where $\lang\rFT$ is the coalgebraic language map of $\rho(FT)$.

In our leading examples this semantics is the usual language
semantics; for non-deterministic automata $\lsem E\rsem$ is the
formal language the expression $E$ denotes, and, similarly, in the example of weighted
automata $\lsem E\rsem$ is the weighted language denoted by $E$.

\takeout{ 
We will soon prove a Kleene like theorem, but before doing that we need a technical
lemma. It states that for every element $x$ of a locally finitely presentable $\bar
F$-coalgebra $(C,c)$ there exists a finite coalgebra $d\colon  Y \to FTY$
with an $\bar F$-coalgebra homomorphism $f\colon (TY, \ext d) \to (C,c)$
such that $x$ is contained in the image of $Y$ (considered as the set
of generators of $TY$) under $f$. Recall that the category $\Coalgfree(\bar F)$ from
Notation~\ref{not:D} is the full subcategory of $\Coalgf(\bar F)$ given by
coalgebras with a \emph{free} finitely presentable carrier.

\begin{lemma}
  \label{lem:genkleene}
  Let $(C, c)$ be a locally finitely presentable $\bar F$-coalgebra.
  For every map $m\colon X \to C$ with $X$ a finite set there
  exists a coalgebra $TY$ in $\Coalgfree(\bar F)$, a coalgebra homomorphism $f\colon TY \to
  C$ and a map $m'\colon X \to Y$ such that the following triangle
  $$
  \xymatrix{
    &&X
    \ar[lld]_{m'} \ar[d]^m
    \\
    Y
    \ar[r]_-{\eta_Y} & TY \ar[r]_-f & C
    }
  $$
  commutes.
\end{lemma}
\begin{proof}
  Given $m\colon X \to C$ we have, since $C$ is a locally finitely
  presentable coalgebra, some
  coalgebra $(P,p)$ from $\Coalgf(\bar F)$, a coalgebra homomorphism
  $g\colon P \to C$ and a map $n\colon X \to P$ such that $g \o n =
  m$. By Proposition~\ref{prop:coeq}, $P$ is the coequalizer of some
  parallel pair in $\Coalgfree(\bar F)$, and so we have some coalgebra $(TZ, \ext e)$
  in $\Coalgfree(\bar F)$ and a surjective homomorphism $g'\colon TZ \to P$ in $\Coalgf(\bar
  F)$. Choose some map $s\colon P \to TZ$ with $g' \o s = \id$ and let $n' =
  s \o n$. Then $g' \o n' = n$.

  Now let $Y = X + Z$ and consider the $T$-algebra homomorphism
  $\ext{[n', \eta_Z]}\colon TY \to TZ$. This is a split epimorphism in
  $\Set^T$; we have $T\inr\colon TZ \to TY$ with
  $$\ext{[n', \eta_Z]} \o T\inr = \ext{\eta_Z} = \id_{TZ}.$$
  Therefore we have the coalgebra structure
  $$
  \ext d = (\xymatrix@1@C+1pc{
    TY \ar[r]^-{\ext{[n',\eta_Z]}} &
    TZ \ar[r]^-{\ext e} &
    FTZ \ar[r]^-{FT\inr} &
    FTY
    })
  $$
  such that $\ext{[n',\eta_Z]}$ is a $\bar F$-coalgebra
  homomorphism from $(TY, \ext d)$ to $(TZ, \ext e)$. Since $Y$ is a
  finite set we see that $(TY, \ext d)$ is a coalgebra in $\Coalgfree(\bar F)$. So $Y$ together with  $m' = \inl\colon X \to
  Y$ and $f = g \o g' \o \ext{[n', \eta_Z]}$ are the required data;
  the diagram below commutes:
  $$
  \xymatrix@C+1pc{
    &&&&X
    \ar[lllld]_{m'}
    \ar[lld]_(.6){n'}
    \ar[ld]^(.4){n}
    \ar[d]^m
    \\
    Y=X+Z
    \ar[r]_-{\eta_{Y}}
    &
    TY
    \ar[r]_-{\ext{[n',\eta_Z]}}
    &
    TZ
    \ar[r]_-{g'}
    &
    P
    \ar[r]_-{g}
    &
    C
    }
  $$
  This completes the proof.
  \qed
\end{proof}
} 

We now prove a Kleene like theorem that establishes a one-to-one
correspondence between expressions and states of finite
$FT$-coalgebras.
\begin{theorem}
  \label{thm:genkleene}
  Every state of a finite coalgebra for $FT$ can
  equivalently be presented by an expression and vice versa. More precisely, we have:

  \noindent
  (1)~Let $E$ be an expression in $\Exp$, then there exists a finite
  $FT$-coalgebra $(S, g)$ and a state $s \in S$ having
  the behavior $\lsem E \rsem$, i.\,e., $\lang g(s) = \lsem E\rsem$.

  \noindent
  (2)~Conversely, let $(S, g)$ be a finite $FT$-coalgebra and let $s \in S$ be a state. Then there exists an
  expression $E$ such that the behavior of $s$ is $\lsem E \rsem$; in
  symbols: $\lang g(s) = \lsem E \rsem$.
\end{theorem}
\begin{proof}%
  \takeout{ 
  Ad~(1). Given the expression $E$ we have $q\o q_0(E) \in\rho\bar
  F$. Since $\rho\bar F$ is a locally finitely presentable,
  coalgebra we can apply Lemma~\ref{lem:genkleene} to obtain an
  $\bar F$-coalgebra $(TS, \ext g)$ in
  $\Coalgfree(\bar F)$ and a homomorphism $f\colon TS \to \rho\bar F$ such
  that $q \o q_0(E) = f \o \eta_S(s)$ for some $s \in S$. Now compose
  with the homomorphism $j\colon \rho\bar F \to \nu F$ from
  Theorem~\ref{thm:ratquot} to obtain:
  $$
  \lsem E \rsem = \lang\rFT \o q_0(E) = j \o q\o q_0(E) =
  j \o f \o \eta_S(s) = \fin{(\ext g)} \o \eta_S(s) =
  \lang g(s),
  $$
  where the last equation uses the finality of $\nu F$.
  } 
  Ad~(1). Given the expression $E$ we have $q_0(E) \in
  \rho(FT)$. Since $\rho(FT)$ is locally finitely presentable there
  exists a finite $FT$-coalgebra $(S, g)$, a state $s \in S$ and a coalgebra homomorphism
  $h\colon (S,g) \to (\rho(FT), \rFT)$ with $h(s) = q_0(E)$. We
  compose this with the coalgebraic language map $\lang \rFT$ to
  obtain:
  \[
  \lsem E \rsem = \lang \rFT \o q_0(E) = \lang \rFT \o h(s) = \lang
  g(s),
  \]
  where the last equation uses Lemma~\ref{lem:ddagger}.

  Ad~(2). Given the $FT$-coalgebra $(S, g)$ and $s \in S$ form the $\bar
  F$-coalgebra $(TS, \ext g)$ and take the unique $\bar F$-coalgebra
  homomorphism $f$ into the final locally finitely presentable
  coalgebra $\rho \bar F$. Let $E$ be such that $q\o q_0(E) =
  f \o \eta_S(s)$, where $q\colon\rho(FT) \to \rho\bar F$ is the
  quotient homomorphism from Theorem~\ref{thm:ratquot}. Now composing with
  $j\colon\rho\bar F \to \nu F$ yields $\lsem E\rsem =
  \lang g(s)$ as before.
\qed\end{proof}

Next, we will show that, for $F$ and $T$ satisfying our assumptions, it is
always possible to ``add proof rules'' to an existing expression calculus in
order to arrive at a sound and complete calculus w.\,r.\,t.~the language
semantics given by $\lsem -\rsem$ in~\refeq{eq:sem}.

\begin{definition}
  Let $(\E, e)$ be an $\bar F$-coalgebra and let $f\colon \Exp \to \E$ be a
  map. We call $(\E,e,f)$ \emph{sound} if for two expressions $E$ and
  $F$ in $\Exp$, $f(E) = f(F)$ implies that $\lsem E\rsem = \lsem F\rsem$, and $(\E,
  e, f)$ is called \emph{complete} if $\lsem E \rsem = \lsem F \rsem$
  implies $f(E) = f(F)$.
\end{definition}

One should think of $\E$ in the above definition as a quotient
coalgebra of $(\Expmod) = \rho(FT)$ obtained by adding proof rules so
as to obtain a coarser equivalence $\equiv_D$ with $\E = (\Expmodd)$. In
fact, we have the following

\begin{theorem}[Soundness]
  \label{thm:gensound}
  Every quotient coalgebra of the $\bar F$-coalgebra $(\rho(FT), F\bFT
 \o\rFT)$ is sound.
\end{theorem}
\begin{proof}
  Let $\E$ be a quotient coalgebra of $\rho(FT)$ via $q\colon \rho(FT) \to \E$ and let $j\colon \E \to \nu F$ be the unique coalgebra
  homomorphism. We consider the map $q \o q_0\colon \Exp \to \E$ and verify
  the soundness by proving that the diagram below commutes:
  \begin{equation}
    \label{diag:sound}
    \vcenter{
      \xymatrix{
        \Exp
        \ar[r]^-{q_0}
        \ar `d[rd] [rrd]_(.4){\lsem -\rsem}
        &
        \rho(FT)
        \ar[r]^-q
        \ar[rd]_{\lang\rFT}
        &
        \E
        \ar[d]^j
        \\
        &&
        \nu F
      }
    }
  \end{equation}
  The left-hand part commutes by the definition of the
  semantic map $\lsem -\rsem$ (see~\refeq{eq:sem}), and the
  right-hand part commutes since all its arrows are $\bar F$-coalgebra
  homomorphisms and using finality of $\nu F$.

  Now whenever for two expressions $E$ and $F$ in $\Exp$ we have $q\o
  q_0(E) = q \o q_0(F)$ we clearly have $\lsem E \rsem = \lsem F
  \rsem$, and this is the desired soundness.
\qed\end{proof}

In particular, we see that $(\rho(FT), \rcomp, q_0)$ itself is sound. Now recall that
the final locally finitely presentable coalgebra $\rho \bar F$ is the (greatest)
quotient of $(\rho(FT), \rcomp)$ via the homomorphism $q\colon \rho(FT) \to
\rho\bar F$ (see Theorem~\ref{thm:ratquot}). So, in addition we have

\begin{theorem}[Completeness]
  \label{thm:gencomp}
  The final locally finitely presentable coalgebra $\rho\bar F$
  together with the map $q\o q_0\colon \Exp \to \rho\bar F$ is complete.
\end{theorem}
\begin{proof}
  Recall the four $\bar F$-coalgebra homomorphisms from the statement
  of Theorem~\ref{thm:ratquot}. Now consider diagram~\refeq{diag:sound}
  where $\E = \rho \bar F$. If for two expression $E$ and $F$ in
  $\Exp$ we have $\lsem E \rsem = \lsem F \rsem$ then $q\o q_0(E) =
  q\o q_0 (F)$ since $j\colon \rho\bar F \to \nu F$ is
  injective. Therefore we obtain the desired completeness.
\qed\end{proof}

Intuitively, this theorem states that, under our assumptions, it is always
possible to obtain a complete calculus for (coalgebraic) language equivalence
as a quotient of a calculus for bisimilarity (hence by adding new
sound rules). However, the theorem does not give any indication how
the added rules should look like in concrete instances and not even whether
it suffices to add finitely many new rules.

One may wonder at this point about the relevance of the
theorems in this section because we did not introduce any concrete
syntax and proof rules. But we shall see in the next sections that
from the above abstract results we automatically obtain soundness,
completeness and Kleene theorems for concrete syntactic calculi once
we have established that the quotient formed by concrete syntactic
expressions modulo proof rules forms the rational fixpoint $\rho\bar F$.

\section{Expression calculus for weighted automata}
\label{sec:lin}

In~\cite{m_linexp} the second author has presented a sound and
complete expression calculus for linear systems presented in the form
of closed stream circuits, which are equivalent to weighted automata
with unary input alphabet $A = \{*\}$ and weights in a field. In this section we are going to use the
ideas from \emph{loc.~cit.} and apply the results from
Section~\ref{sec:lfp} to provide a sound and complete expression
calculus for the language equivalence of weighted automata. This
extends the previous work to weighted systems with several different
inputs and from weights in a field to weights in a semiring.

As discussed in the introduction, an axiomatization for weighted
language equivalence also follows from~\cite{ek_2012}. Their result
holds for so-called proper commutative semirings, a class of semirings containing
all Noetherian semirings but also the semiring of natural numbers. Our
work here is independent.

\begin{assumption}
  In this section we work with the category $\SMod$ for a semiring $\S$
  such that finitely generated semimodules are closed under kernel
  pairs. We consider the free-semimodule monad $T = V$ (see~\refeq{eq:V})
  and the functor $F = \S \times (-)^A$.
\end{assumption}
The above assumption on $\S$ holds whenever $\S$ is Noetherian
(see~Proposition~\ref{prop:lfpmodules}).
Notice
that the functor $F$ has a canonical lifting $\bar F$ to $\SMod = \Set^V$.
So our assumptions in~\ref{ass:setT} clearly hold.

As we saw in Example~\ref{ex:finalcoalg}, coalgebras for the composite
$FV$ are weighted automata with weights in the semiring $\S$, and the final coalgebra
for $F$ and its lifting is carried by the set $\S^{A^*}$ of all
weighted languages.

The expression calculus one obtains in this particular
instance from the work in~\cite{bbrs_ic} allows one to reason about
the equivalence of weighted automata w.\,r.\,t.~weighted
bisimilarity. We will now recall the syntax and
proof rules of this calculus. The syntactic expressions are
defined by the following grammar
\begin{eqnarray*}
  E & ::=  & x \mid \zero \mid E \oplus E \mid \ul{r} \mid a.(r \dot E) \mid \mu
  x.E^g,\\
  E^g & :: = & \zero \mid E^g \oplus E^g \mid \ul r \mid a.(r \dot E)
  \mid \mu x. E^g.
\end{eqnarray*}
Notice that the variable binding operator
$\mu x.-$ is only applied to \emph{guarded} expressions, i.\,e.,
expressions $E^g$ where each occurrence of $x$ is within the scope of
an operator $a.(r \dot -)$.

We write $\Exp$ for the set of all \emph{closed expressions} defined by
the above grammar. One puts on these
expressions certain rules and equations stating that $\mu$ is a unique
fixpoint operator, that $\oplus$ is a commutative and associative
binary operation with the neutral element $\zero$, etc; here is
the list of rules:
\[
\begin{array}{rcl@{\qquad}rcl@{\qquad}l}
\ul{0} & \equiv & \zero          &
\ul{r} \oplus \ul{s} & \equiv & \ul{r+s} &
\\
\zero \oplus E & \equiv  & E &
E_1 \oplus E_2 & \equiv & E_2 \oplus E_1 &
(E_1 \oplus E_2) \oplus E_3 \equiv E_1 \oplus (E_2 \oplus E_3)
\\
a.(0 \dot E) &\equiv & \zero &
\multicolumn{4}{l}{%
  a.(r \dot E) \oplus a.(s\dot E) \equiv a.((r + s) \dot E)
}
\\
  \mu x. E & \equiv &  E[\mu x. E/ x] &
\multicolumn{4}{l}{
E_1 \equiv E_2 [E_1/x] \implies E_1 \equiv \mu x. E_2}
\end{array}
\]
We call the last two rules pertaining to $\mu$ the \emph{fixpoint
  axiom} (\axFP, for short) and the \emph{uniqueness rule}, respectively.
In addition the rules contain $\alpha$-equivalence, i.\,e., renaming
of bound variables does not matter, and the \emph{replacement rule}
(also called congruence rule):
\begin{equation}
  \label{eq:repl}
  \frac{E_1 \equiv E_2}{E[E_1/x] \equiv E[E_2/x]},
\end{equation}
where $E_1$, $E_2$ and $E$ are expressions and $x$ is a free variable in
$E$. We write $\equiv$ for the least equivalence on $\Exp$
generated by the above rules.

The main result of~\cite{bbrs_ic} is that this calculus is sound
and complete for bisimilarity equivalence of weighted
automata. As previously mentioned, the key fact used in order to prove soundness and
completeness is that the set
$\E = \Expmod$ of closed syntactic expressions modulo the proof rules
above is (isomorphic to) the final locally finite coalgebra
$\rho(FV)$.

Now we will turn to a different semantics of the expressions in
$\Exp$, the weighted languages described by them.   The canonical
quotient map $q_0\colon \Exp \to \E = \rho(FV)$ gives us an expression
calculus in the sense of Definition~\ref{dfn:exprcalc}, and we obtain the
corresponding semantics map from~\refeq{eq:sem}:
\[
\lsem -\rsem\colon
\xymatrix@1{\Exp \ar[r]^-{q_0} & \E = \rho(FV)
  \ar[r]^-{\lang\rFT} & \nu F = \S^{A^*}};
\]
it assigns to every expression the weighted language it denotes.

\begin{remark}
  \label{rem:vecE}
  Before we proceed we gather a number of facts that we will need for
  the subsequent technical development.
  \begin{enumerate}[(1)]
  \item In~\cite{bbrs_ic} a measure of complexity $N(E)$ for guarded
    expressions is defined. For the special instance of the calculus
    we are considering here, $N(E)$ is defined as:
    $$
    \begin{array}{rcl}
      \multicolumn{3}{c}{N(\zero) = N(\ul r) = N(a.(r \dot E)) = 0} \\
      N(E_1 \oplus E_2) & = & 1 + \max\{N(E_1), N(E_2)\} \\
      N(\mu x. E) & = & 1 + N(E).
    \end{array}
    $$
    Notice that for every guarded expression we clearly have
    $N(E_1) = N(E_1[E_2/x])$ for every expression $E_2$.
  \item For every set $X$, every element of $VX$ can be written as
    a formal linear combination
    $$
    \sum\limits_{i=1}^n r_ix_i, \quad \text{with $x_i \in X, r_i \in
      \S$ for $i = 1, \ldots, n$}.
    $$
    This formal linear combination corresponds to $f\colon X \to \S$
    with $f(x_i) = r_i$ for $i = 1, \ldots, n$, and $f(y) = 0$ else.
  \item As usual we denote by $[E]$ the equivalence classes in
  $\E = \rho(FV)$. By Lemma~\ref{lem:lfp}, we see that $\E$
  has a canonical structure
    of a $V$-algebra, i.\,e., $\E$ is an $\S$-semimodule. It is
    straightforward to work out that the semimodule addition is
    $$
    [E_1] + [E_2] = [E_1 \oplus E_2]
    $$
    with the neutral element $[\zero]$
    and that the action of the semiring $\S$ satisfies the following laws:
    \begin{equation}
      \label{eq:scalar}
      \begin{array}{rcl}
        r[\zero] & = & [\zero] \\
        r[E_1 \oplus E_2] & = & r[E_1] +  r[E_2] \\
        r[\mu x. E] & = & r[E[\mu x. E / x]] \\
        r[\ul{s}] & = & [\ul{rs}] \\
        r[a.(s \dot E)] & = & [a.((rs) \dot E)]
      \end{array}
    \end{equation}

    From now on we will omit the square brackets indicating equivalence
    classes w.\,r.\,t.~$\equiv$ and simply write $E$ for elements of
    $\E$. We shall also write $rE$ for any expression in $r[E]$.

  \item Furthermore, since $\E = \rho(FV)$ we have the coalgebra
    structure $\rFT\colon \E \to FV(\E)$ and we have the Eilenberg-Moore
    algebra structure $\bFT\colon V(\E) \to \E$ which gives us an \mbox{$\bar
      F$-coalgebra} structure $\rcomp$ on $\E$ (see
    Notation~\ref{not:structs}).
    For further reference  we note that the coalgebra structure
    $\rFT\colon \E \to \S \times (V\E)^A$ acts, for example, as follows:
    \begin{equation}
      \label{eq:dotE}
      \begin{array}{rcl}
        \rFT(a.(s \dot E)) & = & (0, \lambda b. \left\{
          \begin{array}{lp{1.2cm}}
            sE & if $b = a$ \\
            0 & else
          \end{array}
          \right\}
          ),
        \\
        \rFT(\ul s) & = & (s, \lambda b. \zero),
      \end{array}
    \end{equation}
    (since we omit equivalence classes here, we do have the formal linear
    combination $sE \in V(\E)$).
  \end{enumerate}
\end{remark}

From the generic Kleene theorem~\ref{thm:genkleene} we obtain
immediately a Kleene like theorem stating that that every
state of a weighted automaton can equivalently be specified by an
expression of our calculus.

\begin{corollary}
  (1)~For every expression $E$ in $\Exp$ there exists a finite weighted
  automaton $S$ and a state $s \in S$ such that the weighted language
  accepted by $s$ is $\lsem E\rsem$.

  \noindent
  (2)~For every state $s$ of a finite weighted automaton there exists an
  expression that denotes the same weighted language as the one accepted by the
  state $s$.
\end{corollary}
Indeed, this is just a restatement of Theorem~\ref{thm:genkleene}
noting that finite weighted automata are precisely finite
$FV$-coalgebras.

In classical automata theory one obtains, of course, an algorithmic
construction of an expression for a given state of an automaton. The
above theorem does not provide such a construction. However, in our
theory the respective construction does occur, namely later in the proof of
Theorem~\ref{thm:final}.

\subsection{Axiomatization of weighted language equivalence}
\label{sec:axiom}

We are now going to add the following three additional equational laws
to the calculus from the previous section:
\begin{eqnarray}
  a.(r \dot (E_1 \oplus E_2)) & \equiv_D & a.(r \dot E_1) \oplus a.(r \dot E_2) \label{eq:1} \\
  a.(r \dot b.(s \dot E)) & \equiv_D & a.((rs) \dot b.(1 \dot E)) \label{eq:2} \\
  a.(r \dot \ul{s}) & \equiv_D & a.(1 \dot \ul{rs}) \label{eq:3}
\end{eqnarray}
Notice that we write $\equiv_D$ for the least equivalence generated by
all the above rules (i.\,e., all the rules from the previous section
and the three last ones).

We denote by $\ED = \Expmodd$ the closed expression modulo all these
proof rules. Notice that $\ED$ is a quotient of $\E$ via $q\colon \E \to \ED$, say.

\begin{remark}
  Observe that the following equational law
  \begin{equation}
    a.(r \dot \zero) \equiv_D \zero  \label{eq:4}
  \end{equation}
  is provable from the other laws. Using~\refeq{eq:3} we have
  $$
  a.(r \dot \zero)
  \equiv_D
  a.(r\dot \ul 0)
  \equiv_D
  a.(1\dot \ul{r0})
  \equiv_D
  a.(1\dot \ul{0r})
  \equiv_D
  a.(0\dot \ul r)
  \equiv_D
  \zero.
  $$
\end{remark}

\begin{lemma}
  The quotient $\ED$ is an $\S$-semimodule and $q\colon \E \to \ED$ is a
  homomorphism of semimodules.
\end{lemma}
\begin{proof}
  We only need to prove that the three additional equational laws
  in~\refeq{eq:1}--\refeq{eq:3} respect the semimodule structure of $\E$,
  i.\,e., the semimodule operations on $\ED$ are well-defined on equivalence
  classes.

  For the addition this follows from the replacement
  rule~\refeq{eq:repl}. We verify well-definedness for the action of
  the semiring $\S$ for each of the three equational laws:

  Ad~\refeq{eq:1} we have
  $$
  \begin{array}{rcl@{\qquad}p{1.5cm}}
    s(a.(r \dot E_1) \oplus a.(r \dot E_2))
    & \equiv_D & s(a.(r\dot E_1)) \oplus s(a.(r\dot E_2)) & see~\refeq{eq:scalar} \\
    & \equiv_D & a.((sr) \dot E_1) \oplus a.((sr) \dot E_2) & see~\refeq{eq:scalar} \\
    & \equiv_D & a.((sr)\dot (E_1 \oplus E_2)) & see~\refeq{eq:1} \\
    & \equiv_D & s(a.(r \dot (E_1 \oplus E_2)) & see~\refeq{eq:scalar}.
  \end{array}
  $$

  Ad~\refeq{eq:2} we have
  $$
  \begin{array}{rcl@{\qquad}p{1.5cm}}
    c(a.(r \dot b.(s \dot E)))
    & \equiv_D & a.((cr) \dot b.(s\dot E)) & see~\refeq{eq:scalar} \\
    & \equiv_D & a.((crs) \dot b.(1 \dot E)) & by~\refeq{eq:2} \\
    & \equiv_D & c(a.((rs) \dot b.(1 \dot E))) & see~\refeq{eq:scalar}.
  \end{array}
  $$

  Ad~\refeq{eq:3} we have
  $$
  \begin{array}{rcl@{\qquad}p{1.5cm}}
    c(a.(r \dot \ul s))
    & \equiv_D & a.((cr) \dot \ul s) & see~\refeq{eq:scalar} \\
    & \equiv_D & a.(1 \dot \ul{crs}) & by~\refeq{eq:3} \\
    & \equiv_D & a.(c \dot \ul{rs}) & by~\refeq{eq:3} \\
    & \equiv_D & c(a.(1\dot \ul{rs})) & see~\refeq{eq:scalar}.
  \end{array}
  $$
%
  This completes the proof.
\qed\end{proof}

\begin{lemma}
  \label{lem:scalardot}
  For the action of the semiring $\S$ in $\ED$ we have the following
  provable identity:
  $$
  r(a.(s\dot E)) \equiv_D a.(r \dot sE).
  $$
\end{lemma}
\begin{proof}
  Recall from~\refeq{eq:scalar} that $r(a.(s \dot E)) = a.((rs) \dot E)$. Now
  the proof proceeds by induction on the complexity  $N(E)$ of
  expressions. Here are the different cases (we drop the subscript in $\equiv_D$):

  (1)~For $E = \zero$ we apply~\refeq{eq:4} and get
  $$
  a.((rs) \dot \zero)
  \equiv \zero
  \equiv a.(r \dot \zero)
  \equiv a.(r \dot (s\zero)).
  $$

  (2)~For $E = \ul t$ we use~\refeq{eq:3} and~\refeq{eq:scalar} to obtain
  $$
  a.((rs) \dot \ul t)
  \equiv a.(1 \dot \ul{rst})
  \equiv a.(1 \dot (r \ul{st}))
  \equiv a.(r \dot \ul{st})
  \equiv a.(r \dot (s\ul t)).
  $$

  (3)~For a sum $E = E_1 \oplus E_2$ we compute
  $$
  \begin{array}{rcl@{\quad}p{4cm}}
    a.((rs) \dot (E_1 \oplus E_2))
    & \equiv & a.((rs) \dot E_1) \oplus a.((rs) \dot E_2) & by~\refeq{eq:1} \\
    & \equiv & a.(r \dot sE_1) \oplus a.(r \dot sE_2) & by induction hypothesis \\
    & \equiv & a.(r \dot (sE_1 \oplus sE_2)) & by~\refeq{eq:1} \\
    & \equiv & a.(r \dot s(E_1 \oplus E_2)) & by~\refeq{eq:scalar}.
  \end{array}
  $$

  (4)~For $E = b.(t \dot E')$ we use~\refeq{eq:2} and obtain
  $$
  \begin{array}{rcl}
    a.((rs) \dot b.(t \dot E'))
    & \equiv & a.((rst) \dot b.(1 \dot E'))  \\
    & \equiv & a.(r \dot b.((st) \dot E')) \\
    & \equiv & a.(r \dot s(b.(t \dot E'))).
  \end{array}
  $$

  (5)~Finally, for a $\mu$-term $E = \mu x. E'$ one simply uses the
  induction hypothesis on $E'[\mu x. E' / x]$ to obtain
  $$
  \begin{array}{rcl}
    a.((rs) \dot (\mu x. E'))
    & \equiv & a.((rs) \dot E'[\mu x. E' / x]) \\
    & \equiv & a.(r \dot (sE'[\mu x. E'  / x])) \\
    & \equiv & a.(r \dot s(\mu x. E')).
  \end{array}
  $$
  This completes the proof.
\qed\end{proof}

\begin{remark}
  Note that for a commutative semiring $\S$ we have
  \begin{equation}\label{eq:56c}
    r(a.(s\dot E)) \equiv_D a.(s \dot rE);
  \end{equation}
  indeed, one computes
  $$
  \begin{array}{rcl@{\quad}p{4cm}}
    r(a.(s\dot E)) & \equiv_D & a.((rs) \dot E) & by~\refeq{eq:scalar}
    \\
    & \equiv_D & a.((sr) \dot E) & by commutativity \\
    & \equiv_D & s(a.(r\dot E) & by~\refeq{eq:scalar} \\
    & \equiv_D & a.(s \dot rE) & by Lemma~\ref{lem:scalardot}.
  \end{array}
  $$
\end{remark}

Before we proceed to prove that the axiomatisation above is sound and
complete, let us revisit the examples from the introduction.

\begin{example}
(1)~We start by considering the following two automata:
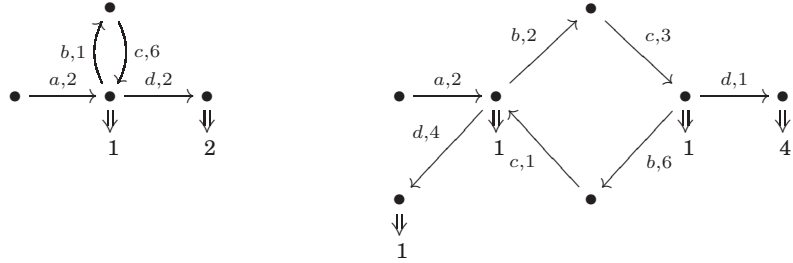
\[
\xymatrix@R=.3cm{ & \bullet  \ar@/^/[dd]^-{c,6}&\\\\
\bullet \ar[r]^-{a,2} & \bullet\ar@{=>}[d] \ar@/^/[uu]^-{b,1}\ar[r]^-{d,2}& \bullet \ar@{=>}[d] && \\
&\mtiny 1 &\mtiny 2
}
\xymatrix@R=.3cm{ & &\bullet\ar[ddr]^{c,3}& &\\\\
\bullet\ar[r]^-{a,2} & \bullet \ar[ddl]_{d,4}\ar[uur]^{b,2}\ar@{=>}[d]& & \bullet \ar[ddl]^{b,6}\ar[r]^{d,1}\ar@{=>}[d]&\bullet\ar@{=>}[d]\\
&\mtiny 1&& \mtiny 1& \mtiny 4\\
\bullet \ar@{=>}[d]& &\bullet\ar[uul]^{c,1}& &\\
\mtiny 1
}
\]
Using the Kleene theorem for weighted automata from~\cite{bbrs_ic}, one can compute
expressions $FV$-equivalent (and thus also $\bar F$-equivalent) to the leftmost states
of the automata above, which we denote by $E_1$ and $E_2$, respectively.
\[
\begin{array}{l@{\qquad}l}
E_1 =  a.(2 \dot E) & E = \mu x. b.(1 \dot c. (6 \dot x))\oplus d.(2\dot \underline 2)\oplus \underline{1}\\\\
E_2 =  a.(2 \dot E')& E' = \mu y. b.(2 \dot c.(3 \dot (b.(6\dot c.(1\dot y))\oplus d.(1\dot \underline 4)\oplus \underline 1)))\oplus d.(4 \dot \underline 1)\oplus\underline{1}
\end{array}
\]
The two expressions are not bisimilar, but they denote the same
weighted language, therefore our goal is to show $E_1\equiv_D E_2$.
Using the congruence rule, we only need to show $E\equiv_D E'$. We will
show that $E \equiv_D b.(2\dot c.(3 \dot (b.(6\dot c.(1\dot E))\oplus d.(1\dot \underline
4)\oplus \underline 1)))\oplus d.(4 \dot \underline
1)\oplus\underline{1}$, which then, using the uniqueness of fixpoints,
yields the desired result. Using the fixpoint rule twice we obtain:
\[
\begin{array}{lcll}
E &\stackrel{(\axFP)}\equiv\hspace*{-6pt}_D& b.(1 \dot c. (6 \dot E))\oplus d.(2\dot \underline 2)\oplus \underline{1}\\
&\stackrel{(\axFP)}\equiv\hspace*{-6pt}_D& b.(1 \dot c. (6 \dot \left(b.(1 \dot c. (6 \dot E))\oplus d.(2\dot \underline 2)\oplus \underline{1}\right)))\oplus d.(2\dot \underline 2)\oplus \underline{1}\\
&\stackrel{\refeq{eq:2}}\equiv\hspace*{-6pt}_D& b.(2 \dot c.(3 \dot (b.(6\dot c.(1\dot E))\oplus d.(2\dot \underline 2)\oplus \underline 1)))\oplus d.(2 \dot \underline 2)\oplus \underline{1}\\
&\stackrel{\refeq{eq:3}}\equiv\hspace*{-6pt}_D& b.(2 \dot c.(3 \dot (b.(6\dot c.(1\dot E))\oplus d.(1\dot \underline 4)\oplus \underline 1)))\oplus d.(4 \dot \underline 1)\oplus \underline{1}\end{array}
\]

\smallskip\noindent
(2)~For another example, consider the automata
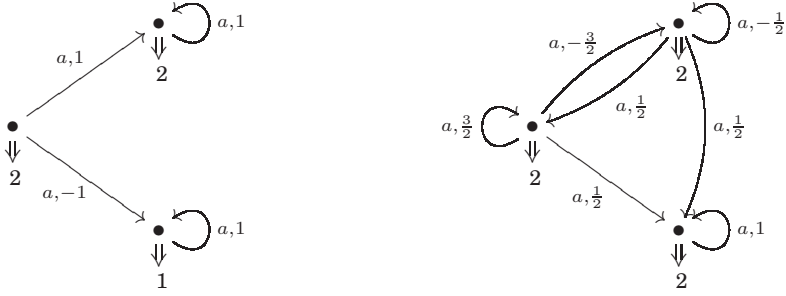
\[
\xymatrix@R=.3cm@C=1.5cm{ &\bullet \ar@{=>}[d]\ar@(dr,ur)_{a,1}&\\
&\mtiny 2\\
 \bullet  \ar[uur]^{a,1}\ar[ddr]_{a,-1}\ar@{=>}[d]& \\
\mtiny 2\\
  &\bullet \ar@{=>}[d] \ar@(dr,ur)_{a,1}&\\
  &\mtiny 1& &
}
\xymatrix@R=.3cm@C=1.5cm{  &\bullet \ar@{=>}[d]\ar@(dr,ur)_{a,-\frac {1} 2}\ar@/^/[ldd]^{a,\frac 1 2}\ar@/^{10pt}/[dddd]^{a,\frac 1 2}&\\
&\mtiny 2\\
 \bullet \ar@(dl,ul)^{a,\frac 3 2} \ar@/^/[uur]^{a,-\frac 3 2}\ar[ddr]_{a,\frac 1 2}\ar@{=>}[d]& \\
\mtiny 2\\
  &\bullet \ar@{=>}[d]\ar@(dr,ur)_{a,1}&\\
  &\mtiny 2&
}
\]
As before, one can compute expressions equivalent to the leftmost states of the automata above,
which we denote by $E_1$ and $E_2$, respectively.
\[
\begin{array}{l@{\qquad\qquad}l}
E_1 =  a.(1 \dot E'') \oplus a.(-1 \dot
E)\oplus \underline{2}& E = \mu y. a.(1 \dot y)\oplus \underline{1}\\[10pt]
E_2 = \mu x. a.(\frac 3 2 \dot x) \oplus a.(-\frac 3 2  \dot E') \oplus a.(\frac 1 2 \dot
E'')\oplus \underline{2}& E'' = \mu y. a.(1 \dot y)\oplus \underline{2}\\[10pt]
E' = \mu z. a.(-\frac 1 2 \dot z) \oplus a.(\frac 1 2  \dot x) \oplus a.(\frac 1 2 \dot
E'')\oplus \underline{2}
\end{array}
\]
The goal is now to show that $E_1 \equiv_D E_2$.  Using the fixpoint
axiom we first observe that
\[
\begin{array}{lcll}
E_2 &\stackrel{(\axFP)}\equiv\hspace*{-6pt}_D& a.(\frac 3 2 \dot E_2) \oplus a.(-\frac 3 2  \dot E'[E_2/x]) \oplus a.(\frac 1 2 \dot
E'')\oplus \underline{2}\\[2ex]
& \stackrel{\refeq{eq:1}}\equiv\hspace*{-6pt}_D&
a.(1 \dot (\frac 3 2 E_2 \oplus -\frac 3 2E'[E_2/x] \oplus \frac 1 2 E'')) \oplus \underline 2 &
\text{using~\refeq{eq:56c},}\\[2ex]
\text{and}&\\[2ex]
E_1
& \stackrel{\refeq{eq:1}}\equiv\hspace*{-6pt}_D&  a.(1 \dot ( E'' \oplus -E)) \oplus \underline 2 & \text{using~\refeq{eq:56c}.}\\
\end{array}
\]
Using the replacement rule, it suffices to prove that $\frac 3 2 E_2
\oplus -\frac 3 2E'[E_2/x] \oplus \frac 1 2 E''\equiv E'' \oplus
-E$. Using the fixpoint axiom we obtain:
\[
\begin{array}{lcll}
&&\frac 3 2 E_2 \oplus -\frac 3 2E'[E_2/x] \oplus \frac 1 2 E'' \\
&\stackrel{(\axFP)}\equiv\hspace*{-6pt}_D& a.(\frac 9 4 \dot E_2) \oplus a.(-\frac 9 4  \dot E'[E_2/x]) \oplus a.(\frac 3 4 \dot
E'')\oplus \underline{3} \\[1.2ex]
&&\oplus\ a.(\frac 3 4 \dot E'[E_2/x]) \oplus a.(-\frac 3 4  \dot E_2) \oplus a.(-\frac 3 4 \dot
E'')\oplus \underline{-3}\\[1.2ex]
&&\oplus\  a.(\frac 1 2 \dot E'')\oplus \underline{1}  \\[2ex]
& \stackrel{\refeq{eq:1}}\equiv\hspace*{-6pt}_D&  a.(1 \dot (\frac 3 2 E_2 \oplus -\frac 3 2E'[E_2/x] \oplus \frac 1 2 E'')) \oplus \underline 1 & \text{using~\refeq{eq:56c}}\\[2ex]
\end{array}
\]
By the unique fixpoint rule we can now conclude that
\[
\frac 3 2 E_2 \oplus -\frac 3 2E'[E_2/x] \oplus \frac 1 2 E'' \equiv_D \mu x . a(1\dot x) \oplus \underline 1.
\]
Now, we use a unique fixpoint argument to verify that $E''
\oplus -E$ is equivalent to  $\mu x . a(1\dot x) \oplus \underline
1$, too:
\[
\begin{array}{lcll}
&&E'' \oplus -E \\
&\stackrel{(\axFP)}\equiv\hspace*{-6pt}_D& a.(1 \dot E'') \oplus \underline{2} \oplus\  a.(-1 \dot E')\oplus \underline{-1}  \\[2ex]
& \stackrel{\refeq{eq:1}}\equiv\hspace*{-6pt}_D&  a.(1 \dot (E'' \oplus - E') \oplus \underline {1} & \text{using~\refeq{eq:56c} and $\underline r \oplus \underline s = \underline {r+s}$}
\end{array}
\]
So by the unique fixpoint rule we obtain $E'' \oplus -E \equiv_D \mu x . a(1\dot x) \oplus \underline 1.$
\end{example}

\subsection{Soundness of the calculus}

Next we show that our calculus is sound for reasoning
about weighted language equivalence.

In order to achieve our goal we will show that $\ED$ is a coalgebra
for the lifting $\bar F\colon \SMod \to \SMod$, and it is a quotient
coalgebra of
\[
\xymatrix@1{
\E = \rho(FV) \ar[r]^-{\rFT} & FV(\rho(FV)) \ar[r]^-{F\bFT} &
F(\rho(FV)) = F(\E).
}
\]
Then we apply the general soundness theorem from Section~\ref{sec:general}.

\begin{lemma}
  \label{lem:quotE}
  The map $Fq \o \rcomp\colon \E \to F(\ED)$ is well-defined on the
  equivalence classes of $\equiv_D$.
\end{lemma}
\begin{proof}
  It is sufficient to show that $\rcomp$ merges both sides of
  the three equations~\refeq{eq:1}--\refeq{eq:3}. We shall use the
  following notation for (certain) elements of $M^A$, where $M$ is some
  semimodule: for $s \in M$ we write $a \asg s$ for the function $f\colon
  A \to M$ with $f(a) = s$ and $f(b) = 0$ for $b \neq a$.
  We also make use of the fact that $\bFT\colon V(\E) \to \E$ is a
  $V$-algebra structure in the following form: for $E_1+E_2$ and
  $rE$ a formal sum and formal scalar product, respectively, in
  $V(\E)$ we have $\bFT(E_1+E_2) = E_1+E_2$ and $\bFT(rE) = rE$, where the
  operations on the right-hand side are the semimodule operations of $\E$.

  Ad~\refeq{eq:1} we compute
  $$
  \begin{array}{rclp{3cm}}
    F\bFT \o \rFT (a.(r\dot E_1) \oplus a.(r \dot E_2))
    & = & F\bFT (0, (a\asg rE_1) + (a \asg rE_2)) & see~\refeq{eq:dotE} \\
    & = & F\bFT (0, a \asg (rE_1 + rE_2)) & semimodule structure on $V(\E)^A$ \\
    & = & (0, a \asg (rE_1\oplus rE_2)) & $\bFT$ is a $V$-algebra\\
    & = & (0, a \asg r(E_1\oplus E_2)) & $\E$ is a semimodule \\
    & = & F\bFT (0, a \asg r(E_1\oplus E_2)) & $\bFT$ is a $V$-algebra \\
    & = & F\bFT \o \rFT (a.(r \dot (E_1 \oplus E_2))) & see~\refeq{eq:dotE}.
  \end{array}
  $$

  Ad~\refeq{eq:2} we compute
  $$
  \begin{array}{rclp{3cm}}
    F\bFT \o \rFT (a.(r \dot b.(s \dot E)))
    & = & F\bFT (0, a \asg r(b.(s \dot E))) & see~\refeq{eq:dotE} \\
    & = & (0, a \asg r(b.(s \dot E))) & $\bFT$ is a $V$-algebra \\
    & = & (0, a \asg b.((rs) \dot E)) & see~\refeq{eq:scalar} \\
    & = & (0, a \asg (rs)(b.(1 \dot E)) & see~\refeq{eq:scalar} \\
    & = & F\bFT (0, a \asg (rs)(b.(1 \dot E))) & $\bFT$ is a $V$-algebra
    \\
    & = & F\bFT \o \rFT (a.((rs) \dot b.(1 \dot E))) & see~\refeq{eq:dotE}.
  \end{array}
  $$

  Ad~\refeq{eq:3} we compute
  $$
  \begin{array}{rclp{4cm}}
    F\bFT \o \rFT (a.(r \dot \ul{s}))
    & = & F\bFT (0, a \asg r\ul{s}) & see~\refeq{eq:dotE} \\
    & = & (0,a\asg r\ul{s}) & $\bFT$ is a $V$-algebra \\
    & = & (0,a \asg \ul{rs}) & see~\refeq{eq:scalar} \\
    & = & (0, a\asg 1\ul{rs}) & $\E$ is a semimodule \\
    & = & F\bFT (0, a\asg 1\ul{rs}) & $\bFT$ is a $V$-algebra\\
    & = & F\bFT \o \rFT (a.(1 \dot \ul{rs})) & see~\refeq{eq:dotE}.
  \end{array}
  $$
%
  This completes the proof.
\qed\end{proof}

\begin{corollary}
  \label{cor:quotE}
  There is a coalgebra structure $c\colon  \ED \to F(\ED)$ such that $q$ is
  a $\bar F$-coalgebra homomorphism from the coalgebra $(\E, \rcomp)$ to $(\ED, c)$.
\end{corollary}
\begin{proof}
  Define $c([E]) = Fq \o F\bFT \o \rFT(E)$. Then $c$ is well-defined
  by Lemma~\ref{lem:quotE}, a semimodule homomorphism since $q$, $\bFT$ and $\rFT$ are
  so, and $c \o q = Fq \o (F\bFT \o \rFT)$ clearly holds. \qed
\end{proof}

\begin{theorem}[Soundness]
  \label{thm:sound}
  The calculus is sound: whenever we have $E_1 \equiv_D E_2$ for two expressions,
  then also $\lsem E_1 \rsem = \lsem E_2 \rsem$.
\end{theorem}

This is just an application of Theorem~\ref{thm:gensound} to
the quotient coalgebra $q\colon (\E, F\bFT \o \rFT) \to (\ED, c)$ for $\bar F$.

\subsection{Completeness}

We are ready to prove the completeness of our calculus
w.\,r.\,t.~weighted language equivalence of expressions. The key
ingredient for our completeness result is the fact that $\ED$ is the
final locally finitely presentable coalgebra for $\bar F\colon \SMod \to
\SMod$.

\begin{lemma}
  \label{lem:iso}
  The map $c\colon \ED \to F(\ED)$ is a semimodule isomorphism.
\end{lemma}
\begin{proof}
  We first define the map $d\colon  F(\ED) \to \ED$ by
  $$
  d(r, \<[E^a]\>_{a \in A}) = [\ul{r} \oplus \bigoplus\limits_{a \in A} a.(1\dot E^a)].
  $$
  By the replacement rule~\refeq{eq:repl}, $d$ is well-defined. We
  first prove that $d$ preserves sums:
  $$
  \begin{array}{r@{\qquad}cl@{\qquad}p{3cm}}
    \multicolumn{4}{l}{d(<r, <[E_1^a]>_{a \in A}> + <s, <[E_2^a]>_{a \in A}>)}
    \\
    & = & d(r+s,
    <[E_1^a \oplus E_2^a]>_{a \in A}) & addition in $\bar F(\E)$
    \\
    & = &
    (r \oplus s) \oplus \bigoplus\limits_{a \in A} a.(1 \dot (E_1^a \oplus
    E_2^a))
    & definition of $d$
    \\
    & = &
    \left(r \oplus \bigoplus\limits_{a \in A} a.(1 \dot E_1^a)\right) \oplus
    \left(s \oplus \bigoplus\limits_{a \in A} a.(1 \dot E_2^a)\right)
    & by~\refeq{eq:1}
    \\
    & = &
    d(r, <[E_1^a]>_{a\in A}) \oplus d(s, <[E_2^a]>_{a \in A})
    &
    definition of $d$.
  \end{array}
  $$

  We now prove that $c$ and $d$ are mutually inverse. It then follows
  that $d$ is a semimodule homomorphism since the forgetful functor
  $\SMod \to \Set$ creates isomorphisms. To see that $c \o d = \id$ we compute:
  $$
  \begin{array}{rcl@{\qquad}p{4.8cm}}
    c \o d (r, \<[E^a]\>_{a \in A})
    & = &
    c\left([\ul{r} \oplus \bigoplus\limits_{a \in A} 1\dot E^a]\right)
    & definition of $d$
    \\
    & = &
    c\left([\ul r]\right) + c\left([\bigoplus\limits_{a \in A} 1\dot
      E^a]\right) & $c$ semimodule homomorphism \\
    & = &
    (r, \lambda b.[\zero]) + \sum\limits_{a\in A}(0, a \mapsto [E^a])

    & see~\ref{lem:quotE}, \ref{cor:quotE} and~\refeq{eq:dotE}
    \\
    & = & (r, \<[E^a]\>_{a \in A})
    & semimodule structure \newline on $\S \times (\ED)^A$.
  \end{array}
  $$

  Finally, we verify that $d \o c = \id$, and we show this by
  induction on the complexity $N(E)$ of expressions $E$:

  For $E = \zero$ we have
  $$
  d \o c([\zero]) = d(0, \<[\zero]\>_{a\in A}) = \left[\ul 0 \oplus
  \bigoplus\limits_{a\in A} a.(1 \dot 0)\right] = [\zero \oplus \zero]
  = [\zero],
  $$
  by the definitions of $c$ and $d$ and using~\refeq{eq:4}.

  For $E = \ul r$ we obtain
  $$
  d \o c([\ul r])
  =
  d(r, \<[\zero]\>_{a\in A})
  =
  \left[\ul r \oplus \bigoplus\limits_{a\in A} a.(1\dot \zero)\right]
  = [\ul r],
  $$
  where the last step uses the semimodule structure on $\ED$
  and~\refeq{eq:4}.

  Next, for $E = E_1 \oplus E_2$ we simply use that $c$ and $d$ preserve
  sums and the induction hypothesis to obtain
  $$
  d \o c([E_1 \oplus E_2]) = d(c(E_1) + c(E_2)) = d(c(E_1)) \oplus d(c(E_2)) = [E_1]
  \oplus [E_2] = [E_1 \oplus E_2].
  $$

  For $E = a.(r \dot E')$ we compute
  $$
  \begin{array}{rclp{4cm}}
    d \o c([a.(r \dot E')])
    & = &
    d(0, a \asg r[E'])
    & see~\refeq{eq:dotE}
    \\
    & = & [\ul 0 \oplus a.(1 \dot (rE'))]
    & definition of $d$ and semimodule structure of $\ED$
    \\
    & = & [a.(1 \dot (rE'))]
    \\
    & = & [r(a.(1\dot E'))]
    & by Lemma~\ref{lem:scalardot}
    \\
    & = & [a.(r\dot E')] & see~\refeq{eq:scalar}.
  \end{array}
  $$

  Finally, for a $\mu$-expression $E = \mu x. E'$ we simply use the
  fixpoint axiom and the induction hypothesis to obtain
  $$
  d \o c([\mu x. E']) = d \o c([E'[\mu x. E' /x]]) = [E'[\mu x. E' / x]] = [\mu x. E'].
  $$
  This completes the proof.
\qed\end{proof}

\begin{notation}
  For expressions $E_1$ and $E_2$ we denote by
  \[
  E_1\{E_2/x\}
  \]
  the \emph{syntactic replacement} of $x$ by $E_2$ in $E_1$, i.\,e., one
  substitutes $E_2$ without first renaming its free variables that are
  bound in $E_1$.

  For example, for $E_1 = \mu x. (a.(3 \dot x))$ and $E_2 = b.(2
  \dot x)$ we have
  \[
  E_1\{E_2/x\} = \mu x.(a. (3 \dot b.(2\dot x))).
  \]
\end{notation}

\begin{theorem}
  \label{thm:final}
  For every $\bar F$-coalgebra $(VS, g)$ with $S$ a finite set there
  exists a unique coalgebra homomorphism from $(VS, g)$ to $(\ED, c)$.
\end{theorem}
\begin{proof}
  Since the coalgebra $(\ED, c)$ is a quotient of the coalgebra $(\E,
  \rcomp)$ we obtain the existence of a homomorphism from Corollary~\ref{cor:weak}.
  It remains to verify its uniqueness.

  So let $m\colon (VS, g) \to (\ED, c)$
  by any $\bar F$-coalgebra homomorphism. Let us assume that $S = \{\,s_1,
  \ldots, s_n\,\}$. It suffices to prove that the $m(s_i)$ are uniquely determined.

  In order to prove this we will first define closed expressions $\expr{s_i}$ and then show
  that these are provably equivalent to $m(s_i)$.

  The expressions $\expr{s_i}$ are defined by an $n$-step
  process. Recall Remark~\ref{rem:vecE}(2), and let $g(s_i) \in \S
  \times (VS)^A$ be
  \begin{equation}
    \label{eq:g}
    g(s_i) = \left(r_i, \left\<\sum\limits_{j = 1}^n r_{ij}^a s_j
      \right\>_{a \in A}\right), \qquad i = 1, \dots, n.
  \end{equation}
  Our expressions will involve the scalars $r_i$, the coefficients $r_{ij}^a$ and $n$
  variables $x_1,\ldots, x_n$. For every $i = 1, \ldots, n$ let
  $$
  E_i^0 = \mu x_i.\left(\ul{r_i} \oplus \bigoplus\limits_{a \in A}  (a.(r_{i1}^a
  \dot x_1) \oplus \cdots \oplus a.(r_{in}^a \dot x_n))\right).
  $$
  Now define for $k = 0, \ldots, n-1$
  $$
  E_i^{k+1} = \left\{
    \begin{array}{cp{2cm}}
      E_i^k\{E_{k+1}^k / x_{k+1}\} & if $i \neq k+1$ \\
      E_i^k & if $i = k+1$.
    \end{array}
  \right.
  $$
  It is easy to see
  that the set of free variables of $E_i^k$ is $\{x_{k+1}, \ldots
  x_n\,\} \setminus \{\,x_i\,\}$, and moreover, every occurrence of
  those variables is free.

  We also see that for every $i$,
  \begin{eqnarray*}
    E_i^n & = & E_i^0 \{E_1^0 / x_1\}\{E_2^1 / x_2\} \cdots
    \{E_{i-1}^{i-2} / x_{i-1}\}\{E_{i+1}^{i} / x_{i+1}\} \cdots
    \{E_n^{n-1} / x_n\} \\
    & = & E_i^{i-1} \{E_{i+1}^i / x_{i+1}\}\cdots\{E_n^{n-1} / x_n\}.
  \end{eqnarray*}
  Observe that $E_i^n$ is a closed term. Moreover, the variable $x_i$
  from $E_i^0$ is never syntactically replaced and it is bound by the
  outermost $\mu x_i$. All other occurrences of $x_i$ in $E_i^n$ are
  not bound by this $\mu$-operator (but by $\mu$-operators further
  inside the term).
  We define
  $$\expr{s_i} = E_i^n.$$
  From now on we will denote equivalence
  classes $[E]$ of expressions in $\ED$ simply by expressions $E$
  representing them.

  It is our goal to prove that $m(s_i) \equiv_D \expr{s_i}$. Let us
  write $m_i$ for (some representative of) $m(s_i)$, for short. We use the fact that $m$ is a coalgebra
  homomorphism, Lemma~\ref{lem:iso} and equation~\refeq{eq:g} to obtain
  \begin{equation}
    \label{eq:comput}
      \begin{array}{rcl}
        m_i
        & = &
        c^{-1} \o Fm \o g(s_i)
        \\
        & = &
        c^{-1} \o Fm \left(r_i, \left\<\sum\limits_{j = 1}^n r_{ij}^a s_j
          \right\>_{a \in A}\right)
        \\
        & = &
        c^{-1} \left(r_i, \left\<\sum\limits_{j = 1}^n r_{ij}^a m_j
          \right\>_{a \in A}\right)
        \\
        & = & \ul{r_i} \oplus \bigoplus\limits_{a \in A}a.\left(1 \dot \sum\limits_{j = 1}^n r_{ij}^a m_j\right).
      \end{array}
  \end{equation}

  For the proof of $m_i \equiv_D \expr{s_i}$, we show the case $n = 3$
  in detail; the general case is completely analogous and is left to
  the reader.

  We start by proving that $m_1 \equiv_D E_1^0[m_2 / x_2][m_3 / x_3]$ by an
  application of the uniqueness rule: from~\refeq{eq:comput}
  we get
  $$
  \begin{array}{rcl}
    m_1
    & \equiv_D &
    \ul{r_1} \oplus \bigoplus\limits_{a \in A} a.(1\dot (r_{11}^a m_1 + r_{12}^a m_2 + r_{13}^a m_3)) \\
    & = &
    (\ul{r_1} \oplus \bigoplus\limits_{a \in A} a.(1\dot (r_{11}^a x_1 +
    r_{12}^a x_2 + r_{13}^a x_3))[m_2 / x_2][m_3/ x_3])[m_1 / x_1].
  \end{array}
  $$

  Next, we prove that $m_2 \equiv_D E_2^1[m_3/x_3]$. Notice that
  $$
  E_1^0[m_2/x_2][m_3/x_3] = E_1^0[m_3 / x_3][m_2 / x_2]
  $$
  since $m_2$ and $m_3$ are closed. Then, applying~\refeq{eq:comput}, we have
  $$
  \begin{array}{rcl}
    m_2 & \equiv_D & \ul{r_2} \oplus \bigoplus\limits_{a \in A} a.(1\dot (r_{21}^a m_1 + r_{22}^a m_2 + r_{23}^a m_3))
    \\
    & \equiv_D &
    \ul{r_2} \oplus \bigoplus\limits_{a \in A} a.(1\dot (r_{21}^a E_1^0[m_2/x_2][m_3/x_3] + r_{22}^a m_2 + r_{23}^a m_3))
    \\
    & = &
    \left(\ul{r_2} \oplus \bigoplus\limits_{a \in A} a.(1\dot (r_{21}^a
      E_1^0[m_3/x_3] + r_{22}^a x_2 + r_{23}^a m_3))\right)[m_2/x_2],
  \end{array}
  $$
  and so we can apply the uniqueness rule to obtain the desired
  equation.

  Now we are able to prove that
  $
  m_1 \equiv_D E_1^0\{E_2^1 / x_2\}[m_3/x_3].
  $
  Notice first that we have $E_1^0\{E_2^1 / x_2\} =
  E_1^0[E_2^1 / x_2]$ since $x_1$ (which is bound in $E_1^0$) is not free
  in $E_2^1$. Now we obtain
  $$
  \begin{array}{rcl}
    E_1^0[E_2^1 / x_2][m_3/x_3]
    & \equiv_D & E_1^0[m_3/x_3] [E_2^1[m_3/x_3] / x_2] \\
    & \equiv_D & E_1^0[m_3/x_3][m_2 / x_2] \\
    & \equiv_D & m_1.
  \end{array}
  $$

  Finally, we show that $m_3 \equiv_D E_3^2$ by another application of the
  uniqueness rule: we have
  $$
  \begin{array}{rcl}
    m_3 & \equiv_D & \ul{r_3} \oplus \bigoplus\limits_{a \in A} a.(1\dot (r_{31}^a
    m_1 + r_{32}^a m_2 + r_{33}^a m_3))
    \\
    & \equiv_D &
    \ul{r_3} \oplus \bigoplus\limits_{a \in A} a.(1\dot (r_{31}^a
    E_1^0\{E_2^1/ x_2\} [m_3/x_3] + r_{32}^a E_2^1[m_3/x_3] + r_{33}^a m_3))
    \\
    & = &
    \left(
    \ul{r_3} \oplus \bigoplus\limits_{a \in A} a.(1\dot (r_{31}^a
    E_1^0\{E_2^1/ x_2\} + r_{32}^a E_2^1 + r_{33}^a x_3))\right)[m_3/x_3].
  \end{array}
  $$
  So we have proved
  $
  m_3 \equiv_D E_3^2 = E_3^3 = \expr{s_3}.
  $
  This implies that
  $$
  m_2 \equiv_D E_2^1[m_3/x_3] \equiv_D E_2^1[E_3^2 / x_3] =  E_2^1\{E_3^2 / x_3\} = E_2^3 = \expr{s_2},
  $$
  where the third step holds since the bound variables
  $x_1$ and $x_2$ of $E_2^1$ are also bound in $E_3^2$.
  Similarly, we have
  $$
  m_1
  \equiv_D
  E_1^0\{E_2^1/ x_2\} [m_3/x_3]
  \equiv_D
  E_1^0\{E_2^1/ x_2\}[E_3^2 / x_3]
  =
  E_1^0\{E_2^1/ x_2\}\{E_3^2 / x_3\} = E_1^3 = \expr{s_1}.
  $$
  This completes the proof.
\qed\end{proof}

\begin{corollary}
  \label{cor:finalexp}
  The coalgebra $(\ED, c)$ is (isomorphic to) the rational fixpoint
  $\rho\bar F$.
\end{corollary}
\begin{proof}
  We prove, equivalently, that $(\ED, c)$ is the final locally finitely presentable
  coalgebra for $\bar F$.
  To see that $(\ED, c)$ is a locally finitely presentable coalgebra
  we use that the coalgebra $(\E, \rcomp)$
  from Remark~\ref{rem:vecE}(4) is locally finitely presentable (see
  Lemma~\ref{lem:lfp}). Since $\ED$ is a quotient coalgebra of $\E$ by
  Corollary~\ref{cor:quotE}, we see that $\ED$ is locally finitely
  presentable, too (apply Lemma~\ref{lem:lfpquot}). The finality of $(\ED, c)$ now follows
  from the previous theorem and Corollary~\ref{cor:final}.
\qed\end{proof}

\begin{theorem}[Completeness]
  \label{thm:complete}
  Whenever we have $\lsem E_1\rsem = \lsem E_2 \rsem$ for two expressions,
  then they are provably equivalent, in symbols: $E_1 \equiv_D E_2$.
\end{theorem}

This is just an application of Theorem~\ref{thm:gencomp} to
$\ED = \rho\bar F$ with the map $q \o q_0\colon \Exp \to \ED$.

\section{Expression Calculus for Non-deterministic Automata}
\label{sec:nondet}

In this section we present some details of an interesting special case
of the work in the previous section---the case of non-deterministic
automata. The calculus becomes somewhat simpler in this case but all
results are just consequences of the more general results of
Section~\ref{sec:lin}.

Here $\S$ is the Boolean semiring, and so the category
$\SMod$ is the category $\Jsl$, of join-semilattices
and join-preserving maps, which is isomorphic to the category of Eilenberg-Moore
algebras for the finite powerset monad $\powf$.

Once again, we work with the functor $FX = 2
\times X^A$, where $A$ is a fixed finite input alphabet and $2$ the two element
join-semilattice.


In~\cite{brs_lmcs}, one considers the language $\Exp$ of closed and guarded expressions defined by the
following grammar
$$
\begin{array}{lcl}
E & ::= & x \mid \zero \mid E \oplus E \mid \ul{1} \mid a.E \mid \mu x.E^g,\\
E^g & :: = & \zero \mid E^g \oplus E^g \mid \ul{1} \mid a.E \mid \mu x. E^g.
\end{array}
$$
Notice that this is just a simplification of the syntax of the
calculus from Section~\ref{sec:lin}. Indeed, $a.E$ corresponds to
$a.(1 \dot E)$, and we do not need the expressions $\ul{0}$ and  $a.(0 \dot E)$
as they are provably equivalent to $\zero$. These syntactic expressions describe
precisely the behaviors of finite non-deterministic automata. Our
set of axioms from Section~\ref{sec:lin} now states that (1) $\mu$ is a
unique fixpoint operator, (2) $\oplus$ is an associative,
commutative and idempotent\footnote{That $\oplus$ is idempotent can be
  derived using the semiring action: $E \oplus E = 1E \oplus 1E =
  (1+1)E = 1E = E$.} binary operation with the neutral element
$\zero$ and that (3) the
$\alpha$-equivalence (i.\,e., renaming of bound variables does not
matter) and the replacement rules are
valid. 
In fact, those are exactly the axioms and rules considered
in~\cite{brs_lmcs}, where they were proven sound and complete with respect to bisimilarity.

To obtain a sound and complete axiomatisation for language equivalence
we only need to add the following two axioms to the above axiomatisation:
\begin{equation}\label{eq:axiom_trace}
a.(E_1 \oplus E_2) \equiv a.E_1\oplus a.E_2
\qquad\textrm{and}\qquad
a.\zero = \zero.
\end{equation}
These new axioms correspond to~\refeq{eq:1} and~\refeq{eq:3}, and the other
axiom~\refeq{eq:2} of the previous section already trivially holds in
the current special case. In this way we recover the result of~\cite{rabinovich} for labelled transition systems (which
are just non-deterministic automata where every state is considered
final). Also note that the result of~\cite{brs_lmcs} coincides precisely
with the results in~\cite{milner} for labelled transition systems and
bisimilarity, which constituted the base of Rabinovich's work.

From Section~\ref{sec:lin}, we get: (1) a Kleene Theorem: every
state of a non-deterministic automaton is language equivalent to an
expression in the calculus and vice-versa; (2) soundness of the
calculus from Theorem~\ref{thm:sound} and (3) completeness of the calculus
from Theorem~\ref{thm:complete}, which uses that the coalgebra $\ED$
of expressions from $\Exp$ modulo \emph{all} the axioms is final
among all locally finite coalgebras for $\bar F$ (cf.~Corollary~\ref{cor:finalexp}).

\section{Conclusions and Future Work}

In this paper, we have presented a general methodology
to extend sound and complete calculi with respect to behavioral
equivalence to sound and complete calculi with respect to coalgebraic
language equivalence. To achieve this goal we have developed a mathematical theory
of finitary coinduction for functors having a lifting to a category of
algebras (satisfying certain finiteness conditions). We illustrated our
general framework by applying it to two concrete instances, non-deterministic
automata and weighted automata. For the former, we recovered the results
of~\cite{rabinovich}, whereas for the latter we presented a new
sound and complete axiomatisation of
weighted language equivalence for automata with weights over a Noetherian semiring.

A key fact to be established in our soundness and
completeness proofs is that expressions modulo proof rules form the final locally finitely
presentable coalgebra. The development of the mathematical theory of
these coalgebras was started in~\cite{m_linexp}, and we continue this
in the current paper.

Even though we did not present the details, our method is generic. For
non-deterministic systems it applies to all coalgebras for $F\powf$ and
for weighted systems we can deal with coalgebras of type $FV$, where $F$
is from an inductively defined class of functors.
However, working out these details is
non-trivial; for example, the generic calculus is syntactically more
involved as it is parametric in the functor $F$. We therefore decided to
treat this generic calculus in a subsequent paper.

In \cite{Myers2011} sound and complete expression calculi for a finitary
endofunctor $F$ over an arbitrary variety $\A$ have been derived from
a presentation of the functor $F$ directly in the variety $\A$. The semantics
of the expressions is given by considering the image of the rational fixpoint
of $F$ inside the final $F$-coalgebra. However, in order to prove completeness,
a full understanding of the interplay between the operators of
the functor presentation and those of the variety is assumed. This is different from our
approach to extend sound and complete calculi with respect to $FT$-equivalence to
sound and complete calculi with respect to $F$-equivalence in the category
of $T$-algebras.  As a consequence our approach is more concrete, and it
requires to explicitly understand the relationship between the final
$FT$-coalgebra and the final $F$-coalgebra and the corresponding
rational fixpoints, respectively.

It is known that language equivalence for automata with weights in a field is 
decidable~\cite{schuetzenberger,Flouret}. This method can be easily 
generalised to weights in a skew field. Language equivalence is also decidable for 
weights in $\Nat$~\cite{Harju}, but it becomes undecidable if
one uses weights in a tropical semiring~\cite{Kro94}. More recently,
a decidability result for automata with weights in a proper and
effectively presentable semiring has been obtained in~\cite{em_2010}.
Using an argument similar to~\cite{em_2010} from our completeness result
it follows that coalgebraic language equivalence for the calculi we have developed
is decidable when weights are from an effectively presentable semiring\footnote{A semiring 
for which its elements have a finitary presentation, and such that sums and products can 
be effectively computed from this presentation (see~\cite{em_2010}).}. On the one hand, 
equivalence is semi-decidable: if two expressions are equivalent one can simply enumerate 
all strings over some finite alphabet, and check which of these strings correspond to 
valid proof chains: because of the completeness, if such a proof chain exists, it will be 
found using this enumeration. On the other hand, non-equivalence is also semi-decidable: if 
two expressions are nonequivalent, there must be a word witnessing this. Thus we can simply 
enumerate all words and for each word compute the weight associated with the languages of the 
two expressions. 
%

We presented the main results of the theory for the base category
$\Set$. In the future we plan to extend this to more general base
categories in order to deal with systems whose state spaces have extra
structure, e.\,g., they form posets, graphs or presheaves.

Unfortunately, our main result on final locally finitely presentable
coalgebras (Theorem~\ref{thm:ratquot}) uses the assumption that
finitely generated objects are closed under kernel pairs. This
assumption is somewhat restrictive, and we intend to study whether
this can be relaxed. This would allow to consider other monads $T$,
i.\,e., other branching types like, for instance, various kinds of
probabilistic systems.

As we saw in our work, the generalised powerset construction lets us
move from systems of type $FT$ to systems of type $F$ (in the category
of $T$-algebras) and hence from bisimilarity to language equivalence.
On the other hand, in coalgebraic trace semantics~\cite{HJS:2007} one considers
functors of the form $TF$ and works with coalgebras for (the lifting of $F$) to
the Kleisli category of $T$ (e.\,g. $TF = \Pow(1+A \times -)$
for non-deterministic automata). This allows to deal with some monads
$T$ that are not finitary, e.\,g.~the full powerset monad or the
subdistribution monad. However, that approach does not allow to
consider the monad $V$ of free semimodules.
It would therefore be desirable to find a common framework that
accommodates both these approaches.

\paragraph{Acknowledgments}
We are very grateful to Zolt\'an \'Esik for his helpful comments and
for pointing out recent work on axiomatizations of rational weighted
languages. We are thankful to the careful reading of Joost Winter,
whose comments improved the presentation and correctness of the some
of the results of this paper, and clarified the question whether the
calculi for coalgebraic language equivalence we have developed are
decidable. Thanks also go to Henning Urbat for pointing out a gap in
Example~\ref{ex:counter}.

In addition, we would like to thank the referees for
the many constructive comments, which greatly helped us improving the
paper. The third author was partially supported by
Funda\c{c}\~ao para a Ci\^encia e a Tecnologia, Portugal, under grant
number \texttt{SFRH/BPD/71956/2010}.
%
%
\nocite{z_2011}
\bibliographystyle{acmsmall}
\bibliography{ourpapers,coalgebra}

%
%
\received{July 20, 2011}{December 15, 2011}{March 10, 2012}

\end{document}